\documentclass[12pt,draftclsnofoot, onecolumn]{IEEEtran}
\usepackage{graphicx,cite,epsfig,amssymb,amsmath,url,stfloats,latexsym,threeparttable,balance}
\usepackage{multirow}
\usepackage{color,epstopdf}
\usepackage{booktabs}
\usepackage{setspace}
\usepackage{amsbsy}
\usepackage{cite}
\usepackage{mathrsfs}
\usepackage{amsfonts}
\usepackage{subfig}
\usepackage{extarrows}
\usepackage{bigints}\usepackage{exscale}\usepackage{relsize}
\usepackage{makecell}

\begin{document}
\title{\mbox{}\vspace{1.5cm}\\
\textsc{Performance Analysis of Joint Transmission Schemes in Ultra-Dense Networks - An Unified Approach} \vspace{1.5cm}}
\author{Shuyi Chen, Xiqing Liu, Tianyu Zhao, Hsiao-Hwa~Chen{$^{^\dagger}$},~\IEEEmembership{Fellow,~IEEE}, and Weixiao Meng,~\IEEEmembership{Member,~IEEE}
\thanks{Shuyi Chen (e-mail: {\tt chenshuyitina@gmail.com}), Tianyu Zhao (e-mail: {\tt tianyuzhao.hit@gmail.com}), and Weixiao Meng (e-mail: {\tt wxmeng@hit.edu.cn}) are with the Communication Research Center, Harbin Institute of Technology, China. Hsiao-Hwa Chen (e-mail: {\tt hshwchen@ieee.org}) and Xiqing Liu (email: hstsliou@gmail.com) are with the Department of Engineering Science, National Cheng Kung University, Tainan, 70101 Taiwan. }
\thanks{The paper was submitted on June 3, 2018, and revised \today.}\\
\vspace{1.5cm}
\underline{{$^{^\dagger}$}Corresponding Author's Address:}\\
$\mbox{Hsiao-Hwa~Chen}$\\
Department of Engineering Science\\
National Cheng Kung University\\
1 Da-Hsueh Road, Tainan City, 70101 Taiwan\\
Tel: +886-6-2757575 ext. 63320\\
Fax: +886-6-2766549 \\
Email: {\tt hshwchen@ieee.org}}

\date{\today}
\renewcommand{\baselinestretch}{1.2}
\thispagestyle{empty} \maketitle \thispagestyle{empty}
\newpage
\setcounter{page}{1}

\begin{abstract}
Ultra-dense network (UDN) is one of the enabling technologies to achieve 1000-fold capacity increase in 5G communication systems, and the application of joint transmission (JT) is an effective method to deal with severe inter-cell interferences in UDNs. However, most works done for performance analysis on JT schemes in the literature were based largely on simulation results due to the difficulties in quantitatively identifying the numbers of desired and interfering transmitters. In this work, we are motivated to propose an analytical approach to investigate the performance of JT schemes with a unified approach based on stochastic geometry, which is in particular useful for studying different JT methods and conventional transmission schemes without JT. Using the proposed approach, we can unveil the statistic characteristics (i.e., expectation, moment generation function, variance) of desired signal and interference powers of a given user equipment (UE), and thus system performances, such as average signal-to-interference-plus-noise ratio (SINR), and area spectral efficiency, can be evaluated analytically. The simulation results are used to verify the effectiveness of the proposed unified approach.
\end{abstract}

\IEEEpeerreviewmaketitle

\section{INTRODUCTION}
A fast growing demand for high-speed data communications has exerted a great pressure on the existing wireless communication systems. We are about to enter the era of the fifth generation (5G) communications. Downlink peak data rate and downlink user experienced data rate of 5G can be as high as 20 Gbps and 100 Mbps, respectively, as specified by IMT 2020 \cite{Shafi20175GCOMST}. In order to meet such a high communication requirement, the current wireless systems should be improved from the following three aspects, i.e., enhancing spectral efficiency, exploring new spectrum resources, and increasing area reuse factor \cite{Wang20145GtechMCON}. The emerging ultra-dense network (UDN) has been viewed as a key enabling technology to improve the area reuse factor. The UDN depicts a network topology, where different types of small cell base stations (SBSs) are densely deployed within the coverage of macro cell base stations (MBSs). The density of SBSs can be even higher than that of user equipments (UEs). For example, the density of SBSs can be as high as $10^3$/km$^2$, while active UEs in a urban scenario are less than 600/km$^2$ \cite{Kamel2016UDNCOMST}. The deployment of UDNs can offload the traffic from MBSs to meet high communication requirements in service-intensive areas, such as apartments, enterprises, and hotspots. It has been proved that with a cell size of less than 10 m$^2$, the capacity target of 1 Gb/s/m$^2$ can be achieved \cite{Zander2017UDNWC}.

However, the inter-cell interference may severely impair the performance of UDNs due to the shortened distances between adjacent cells \cite{Nguyen2017UDNJSAC}. If the density of small cells is higher than a given threshold, network throughput starts to decline and even approaches to zero \cite{Liu2017InterferNetwork}. In order to mitigate inter-cell interferences, coordinated multipoint (CoMP) transmission was proposed, which explores the cooperation among adjacent cells to mitigate the interferences \cite{Bassoy2017CoMPCOMST}. Generally, there are three types of downlink CoMP techniques, i.e., coordinated beamforming (CB), coordinated scheduling (CS), and joint transmission (JT) \cite{TR36814}. In CB, beamformers are designed jointly among cooperative SBSs. In CS, time and frequency resources are allocated jointly to avoid inter-cell interference. In JT, the cooperative SBSs transmit messages collaboratively, and thus the level of inter-cell interference can be reduced. If band-pass signals from cooperative SBSs are aligned at an intended destination, this transmission scheme is called coherent JT. Even though a requirement for phase synchronization in coherent JT may increase its implementation complexity, multiple solutions have been proposed accordingly to make the application of coherent JT possible, such as the use of highly stable GPS-referenced local oscillators or a two-way open-loop carrier synchronization method \cite{Marsch2011CoMPBook}. In this paper, we will focus on the performance analysis of coherent JT in UDNs.

Stochastic geometry has been widely used to analyze large-scale wireless networks for more than three decades, and numerous models have been developed based on different network topologies or application scenarios \cite{ElSawy2013SGCOMST}. In \cite{Yilmaz2012SINRTC}, the ergodic capacity was calculated for $L$-branch coherent diversity combiners, and a unified moment generation function (MGF) of the envelope was derived in a Gamma-shadowed generalized Nakagami-$m$ fading channel. A closed-form expression of the MGF of SNR was derived in \cite{Pena2017MGFTC} in Beckmann fading channels. Apart from MGF, the system performance, such as outage probability and link capacity, can also been obtained with the help of incomplete MGF (IMGF) \cite{Lopez2017SCTC}. In \cite{Schilcher2016SINRTIT}, the authors summarized the mathematical theorems, such as Campbell-Mecke theorem, Campbell's theorem, and probability generating functional (PGF), to calculate different stochastic properties of received power in a homogeneous network, where base stations are distributed following a Poisson point process (PPP). Then, the authors extended the method by proposing a general formula to calculate sum-product functionals for PPP. In \cite{Kamel2017UDNTC}, the performance of JT based on carrier aggregation (CA) in UDNs was analyzed. The involvement of CA guarantees that cooperative SBSs use different frequency resources, and thus the analytical process becomes similar to that in conventional transmission scheme without JT. Another method to analyze the performance of JT was proposed in \cite{Cui2017JTTVT}. The authors ignored the spatial randomness of SBSs and established a relationship between ASE and SINR under a target bit error rate constraint.

To the best of our knowledge, this work is the first effort to analyze the system performance of coherent JT based on stochastic geometry. The main contribution of this paper is to propose a unified analytical approach, which is suitable to analyze  aggregated desired signal and interference powers for different transmission methods with or without coherent JT. Using the proposed method, different stochastic characteristics (i.e., expectation, MGF, and variance) of the aggregated desired signal and interference powers can be obtained, respectively. The other performance indicators, i.e., average SINR, average spectral efficiency, and average area spectral efficiency, can also be calculated easily. Note that the randomness of received power is due mainly to the spatial distribution of SBSs and channel variation. We first analyze the effect of spatial randomness of SBSs, assuming a constant channel coefficient, followed by the results in Rayleigh fading channels and Nagakami-$m$ fading channels.

Major mathematical symbols used in this paper are defined as follows. $\mathbb{E}[X]$, $\mathbb{V}[X]$, $\mathbb{P}[X=x]$, and $\frac{\Delta F}{\Delta x}$ denote the expectation of $X$, the variance of $X$, the probability of $X=x$, and the derivative of function $F$ over variable $x$. $\Gamma(s,a,b)$ is the incomplete Gamma function, which is $\Gamma(s,a,b)=\int_{a}^{b}t^{s-1}e^{-t}dt$. $_2F_1(a,b;c;z)$ is the hypergeometric function, which is $_2F_1(a,b;c;z)=\sum_{n=0}^\infty\frac{(a)_n(b)_n}{(c)_n}\frac{z^n}{n!}$.

The rest of our paper can be outlined as follows. Section II briefly introduces the system model, including the description of different JT methods and the main idea of the proposed unified analytical approach. In Section III, we characterize the properties of aggregated received desired signal and interference powers, such as the expectation, MGF, and variance. Then, with the help of the results obtained in Section III, the system performance parameters are obtained in Section IV, i.e., SINR, spectral efficiency, and area spectral efficiency. Simulation results are presented in Section V, followed by the conclusions given in Section VI.

\section{SYSTEM MODEL}
\subsection{Transmission Schemes}

The very basic idea of JT lies on the fact that a set of base stations transmit the same message cooperatively to one UE simultaneously, and the choice of cooperative set is critical to carry out JT. There are three methods to choose cooperative set, which are summarized as follows.
\begin{enumerate}
\item {2-nearest-SBSs (2NS) method}: The conventional JT method works in a way that a UE chooses the nearest $N$ SBSs, and $N=2$ is the most common situation of this method, which will be analyzed in this paper.
\item {Constant-distance (CD) method}: Each SBS compares its distance to UE with a predetermined value $R_0$. If the distance is no longer than $R_0$, it is a cooperative SBS \cite{Park2016UEassociationTWC}.\cite{Veetil2017UEassociationTWC}. This method can be slightly altered for another JT scheme, which is based on the average received power. More specifically, if the received power from a given SBS is above a given threshold, the SBS is a cooperative SBS \cite{Nie2016UEassociationJSAC}\cite{Skou2017UEassociationTC}.
\item {Fixed-power-difference (FPD) method}: The third method works based on the difference of received powers. To be more specific, if the highest average received power is $P_{max}$ dBm, other SBSs, whose received power are higher than $P_{max}-\eta$ dBm, are regarded as the cooperative SBSs, where $\eta$ dB is a predetermined value \cite{Ge2016UEassociationTVT} \cite{Chen2017UEassociationNetwork}.
\end{enumerate}

There may exist more than two cooperative SBSs in each cooperative set, and thus we can select some or all of SBSs from the set for their participation in JT based on QoS requirement or network traffic. To simplify analytical process, we assume that all SBSs in a cooperative set are involved in transmission process, and the scenario with selective SBSs will be analyzed in our future work. In the rest of this paper, we use the term 2NS, CD, and FPD to represent the aforementioned three methods, respectively. Meanwhile, the traditional transmission method, i.e., a UE is served by its nearest SBS, is also analyzed in this paper as a benchmark, and it is denoted by NoJT method for short.

\subsection{Analytical Approach}
\begin{figure}
\centering
\includegraphics[width=8.8cm]{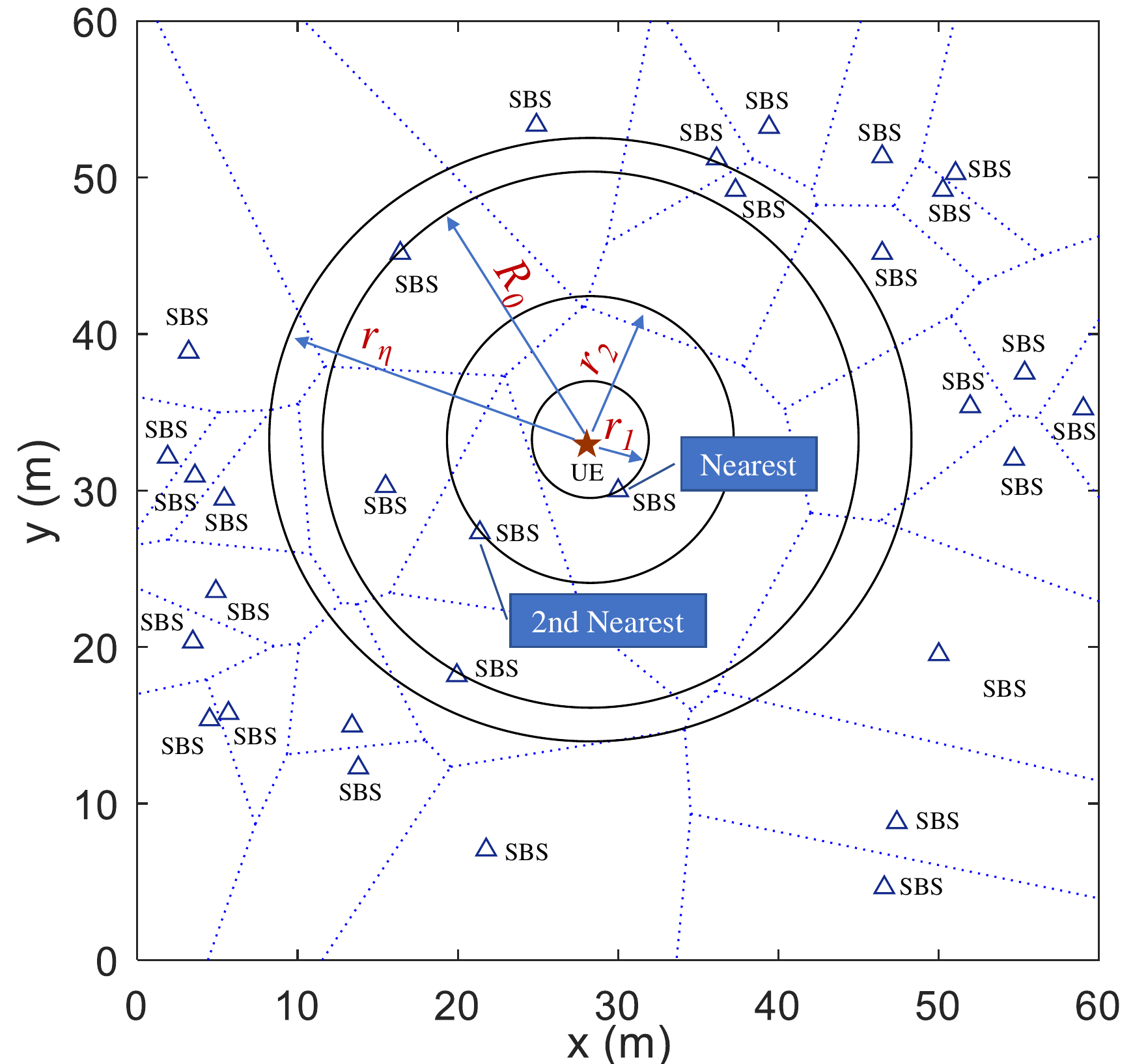}
\caption{{Serving SBSs for different association methods, where the dashed lines represent the Voronoi diagram, and the black concentric circles from inside to outside are the boundaries for NoJT, 2NS, CD, and FPD methods, respectively. The density of SBS is 0.01/m$^2$.}}
\label{distance_topology}
\end{figure}

Even though there are differences in those methods (i.e., three JT methods and one NoJT method), they can be analyzed similarly. As shown in Fig.~\ref{distance_topology}, SBSs are distributed following a Poisson point process (PPP), and a random UE, denoted by a red star, is taken as an example. If NoJT method is applied, the UE is served by its nearest SBS, which can be clearly observed by Voronoi diagram. The distance between UE and its nearest SBS is $r_1$, and a circle is drawn with its radius $r_1$. We can claim that SBSs that are located within the circle are the serving SBSs of the given UE. Similarly, we can determine the circles for the aforementioned three JT methods, and their radii are $r_2$, $R_0$, and $r_{\eta}$ for 2NS, CD, and FPD, respectively. $r_2$ is the distance between UE and the second nearest SBS, and $R_0$ is a predefined constant value. $r_\eta$ can be obtained as follows:
\begin{equation}
\label{Eq.reta}
\eta=10\log\big(K_s\mathbb{E}[h]r_1^{-\alpha_s}P_s\big)-10\log\big(K_s\mathbb{E}[h]r_\eta^{-\alpha_s}P_s\big),
\end{equation}
where $K_s$, $h$, $P_s$, and $\alpha_s$ are the constant factors owing to antenna and average channel attenuation, multipath and shadowing fading, transmit power of SBS, and long-distance fading factor. Thus, $r_\eta=10^{\frac{\eta}{10\alpha_s}}r_1$.

As shown in Fig.~\ref{distance_topology}, the whole area is divided into two regions for each transmission scheme. SBS(s) inside the circle are the desired transmitters, and SBSs outside the circles are the interferers.

Thus, the problem of calculating received desired signal and interference powers can be transformed to the calculation of aggregated powers transmitted from SBSs inside and outside the circle, respectively. Taking NoJT method as an example, the desired and interfering power are $P_1=\sum_{r\leq r_1}K_shr^{-\alpha_s}P_s$ and $I_1=\sum_{r>r_1}K_shr^{-\alpha_s}P_s$, respectively.

Therefore, different mathematical properties of PPP can be directly applied to obtain different statistical characteristics (such as expectation, MGF, and variance) for both desired signal and interference powers. Then, other system performances can be evaluated accordingly.

\subsection{Radius of An Area}
The fundamental step is to determine the radius for different transmission schemes. Assume that SBSs are distributed following a PPP with its density $\lambda_b$, which is defined as the number of SBSs per square meter. The probability distribution function (PDF) of the radius can be calculated as follows.
\begin{enumerate}
\item {NoJT method:} The radius $r_1$ is the distance between UE and its nearest SBS, and thus the PDF of $r_1$ is \cite{ChaHets2014TWC}
\begin{equation}
f_{1}(r) = 2\pi\lambda_b r \exp\;(-\pi\lambda_b r^2).
\end{equation}
\item {2-nearest-SBSs (2NS) method:} The radius $r_2$ is the distance between UE and the second nearest SBS, and similarly, its PDF is  \cite{ChaHets2014TWC}
\begin{equation}
f_{2}(r) =2(\pi\lambda_b)^2 r^3 \exp\;(-\pi\lambda_b r^2).
\end{equation}
\item {Constant-distance (CD) method:} The radius is a pre-defined constant value, which is denoted by $R_0$.
\item {Fixed-power-difference (FPD) method:} The relation between $r_\eta$ and $r_1$ is shown in Eqn.~(\ref{Eq.reta}). Let $\eta_t=10^{-\frac{\eta}{10\alpha_s}}$. We have $f_{\eta}(r)=\eta_t f_1(\eta_t r)$. Thus, the PDF for $r_\eta$ is
\begin{equation}
 f_{\eta}(r)=2\pi\lambda_b\eta_t^2 r \exp\;(-\pi\lambda_b\eta_t^2 r^2).
 \end{equation}
\end{enumerate}

\section{AGGREGATED DESIRED SIGNAL AND INTERFERENCE POWERS}
In this section, we will calculate the expectation, MGF, and variance of both desired signal and interference powers for different transmission schemes. We will first propose a unified analyzing method, and then the results for different transmission schemes are presented accordingly.

There are two factors governing the randomness of the aggregated power, i.e., spacial random distribution of SBSs, and shadowing and multipath fading. Thus, we first present the results when only considering the randomness caused by spacial distribution of SBSs, i.e., the channel parameter $h$ is regarded as a constant value. In other words, the influence of the channel is ignored. Then, the effect of small-scale fading is considered. Since the distance between a transmitter to a receiver is reduced in UNDs, the influence of small-scale fading is significant \cite{Liu2017ShortRangeUDNCM}. The results based on Rayleigh fading and Nakagami-$m$ channels are calculated, respectively. The PDF for Rayleigh fading with a unit mean is $f_{h_R}(h)=\exp(-h)$, and the PDF for Nakagami-$m$ fading is $f_{h_N}(h)=\frac{m^m}{\Omega^m\Gamma(m)}h^{m-1}\exp(-\frac{m}{\Omega}h)$, where $m\geq\frac{1}{2}$ and $\Omega=\mathbb{E}[h]$.

\subsection{Expectation}
The calculation of average aggregated desired signal and interference powers is based on Campbell's theorem of a PPP \cite{Haenggi2013SGbook}, i.e.,
\begin{equation}
\mathbb{E}\Big[\sum_{x\in\Phi}f(x)\Big]=2\pi\lambda\int_\Phi f(x)xdx,
\end{equation}
for any non-negative measurable function $f$ over the PPP $\Phi$. Thus, for all SBSs, whose distance to UE is within the range $[R_1,R_2]$, the average aggregated power received by the UE is
\begin{equation}
\label{Eq.P}
\begin{split}
&\mathbb{E}[P]=2\pi\lambda_b\int_{R_1}^{R_2}K_shP_sr^{-\alpha_s}rdr\\
&=\begin{cases}
\displaystyle
\frac{2\pi\lambda_b K_s h P_s}{2-\alpha}\Big(R_2^{-\alpha_s+2}-R_1^{-\alpha_s+2}\Big),&\alpha_s\neq2,\\
2\pi\lambda_b K_sh P_s\ln\frac{R_2}{R_1},&\alpha_s=2.
\end{cases}
\end{split}
\end{equation}

Eqn.~(\ref{Eq.P}) is useful to calculate both desired signal and interference powers. Assume that the minimum distance between a SBS and a UE is $R_l$, and the radius for a UDN is $R_m$. When calculating the desired power, the lower limit in Eqn.~(\ref{Eq.P}) is $R_l$ (i.e., $R_1=R_l$), and the upper limit is the radius, i.e., $r_1$, $r_2$, $R_0$, and $r_\eta$ for NoJT, 2NS, CD, and FPD methods, respectively. Except for CD method, the radii for the other three methods are random variables with their PDFs given in Section II-C. Thus, the average aggregated desired power can be obtained by integrating the radius (i.e., $r_1$, $r_2$, and $r_\eta$) over the range $[R_l, R_m]$. Similarly, when calculating the interference power, the lower limit is the variable of radius (i.e., $r_1$, $r_2$, $R_0$, and $r_\eta$), and the upper limit is $R_m$. Except for CD method, an integral over $[R_l,R_m]$ is used to obtain the aggregated interference power for the other three transmission methods. Note that when determining the boundary of interfering transmitters, the precise area depends on the location of UE. For UEs that are not located at the center of UDN, the region is not a annulus, and thus Eqn.~(\ref{Eq.P}) is not suitable. In this paper, we assume that the UE is always located at the center of the UDN, and this assumption has been applied in other related works \cite{Jo2012HetnetTWC}.

Meanwhile, $\alpha_s=2$ is the scenario of an infinite space, and $\alpha_s\neq2$ is a more general case. When $h$ is regarded as a constant, the results of $\alpha_s\neq2$ and $\alpha_s=2$ are given in Eqns.~(\ref{EqA1}) and (\ref{EqA2}), respectively, where $P_x$ and $I_x$ are the aggregated received desired signal and interference powers for the $x$ transmission scheme, and the subscripts $1$, $2$, $R_0$, and $\eta$ represent NoJT, 2NS, CD, and FPD methods, respectively.

{
\small
\begin{equation}
\label{EqA1}
\begin{split}
&\mathbb{E}[P_1]=
\mathlarger{\int}_{^{R_l}}^{_{R_m}}\frac{2\pi\lambda_b K_sh P_s}{2-\alpha}\big(r^{2-\alpha_s}-R_l^{2-\alpha_s}\big)f_1(r)dr \\
&{\color{white}{\mathbb{E}[P_1]}}=\frac{2K_s h P_s}{2-\alpha}\Big[\big(\pi\lambda_b\big)^{\frac{\alpha_s}{2}}\Gamma\Big(\frac{4-\alpha_s}{2},\pi\lambda_bR_l^2,\pi\lambda_bR_m^2\Big)+\pi\lambda_bR_l^{2-\alpha_s}\Big(e^{-\pi\lambda_bR_m^2}-e^{-\pi\lambda_bR_l^2}\Big)\Big],\\
&\mathbb{E}[I_1]
=\frac{2K_s h P_s}{2-\alpha}\Big[\pi\lambda_bR_m^{2-\alpha_s}\Big(e^{-\pi\lambda_bR_l^2}-e^{-\pi\lambda_bR_m^2}\Big)-\big(\pi\lambda_b\big)^{\frac{\alpha_s}{2}}\Gamma\Big(\frac{4-\alpha_s}{2},\pi\lambda_bR_l^2,\pi\lambda_bR_m^2\Big)\Big],\\
&\mathbb{E}[P_2]=\mathlarger{\int}_{^{R_l}}^{_{R_m}}\frac{2\pi\lambda_b K_sh P_s}{2-\alpha}\big(r^{2-\alpha_s}-R_l^{2-\alpha_s}\big)f_2(r)dr\\
&{\color{white}{\mathbb{E}[P_2]}}=\frac{2K_s h P_s}{2-\alpha}\bigg\{\big(\pi\lambda_b\big)^{\frac{\alpha_s}{2}}\Gamma\Big(\frac{6-\alpha_s}{2},\pi\lambda_bR_l^2,\pi\lambda_bR_m^2\Big)+\pi\lambda_bR_l^{2-\alpha_s}\Big[(\pi\lambda_bR_m^2+1)e^{-\pi\lambda_bR_m^2}\\
&{\color{white}{\mathbb{E}[P_2]}}-(\pi\lambda_bR_l^2+1)e^{-\pi\lambda_bR_l^2}\Big]\bigg\},\\
&\mathbb{E}[I_2]
=-\frac{2K_s h P_s}{2-\alpha}\bigg\{(\pi\lambda_b)^{\frac{\alpha_s}{2}}\Gamma\Big(\frac{6- \alpha_s}{2} , \pi\lambda_bR_l^2 , \pi\lambda_bR_m^2 \Big) + \pi\lambda_bR_m^{2 - \alpha_s}\Big[(\pi\lambda_bR_m^2 + 1)e^{ -\pi\lambda_bR_m^2} \\
&{\color{white}{\mathbb{E}[P_2]}}- (\pi\lambda_bR_l^2 + 1)e^{ -\pi\lambda_bR_l^2}\Big]\bigg\},\\
&\mathbb{E}[P_{R_0}]  =\frac{2\pi\lambda_b K_s h P_s}{2 - \alpha}\Big(R_0^{2 - \alpha_s}   - R_l^{2 - \alpha_s}\Big),\\
&\mathbb{E}[I_{R_0}]=\frac{2\pi\lambda_b K_s h P_s}{2 - \alpha}\Big(R_m^{2 - \alpha_s}  - R_0^{2 - \alpha_s}\Big),\\
&\mathbb{E}[P_\eta] = \mathlarger{\int}_{^{\frac{R_l}{\eta_t}}}^{_{R_m}} \frac{2\pi\lambda_b K_s h P_s}{2-\alpha} \Big[r^{2 - \alpha_s} -  \Big(\frac{R_l}{\eta_t}\Big)^{2 - \alpha_s} \Big]f_\eta (r)dr\\
 &{\color{white}{\mathbb{E}[P_\eta]}}=\frac{2K_s h P_s}{2-\alpha} \bigg[ \frac{(\pi\lambda_b\eta_t^2)^{\frac{\alpha_s}{2}}}{\eta_t^{2}} \Gamma\Big( \frac{4 - \alpha_s}{2} , \pi\lambda_b\eta_t^2R_l^2 , \pi\lambda_bR_m^2 \Big)+\pi\lambda_b\Big(\frac{R_l}{\eta_t}\Big)^{2 - \alpha_s}\Big(e^{ -\pi\lambda_b\eta_t^2R_m^2} - e^{ - \pi\lambda_bR_l^2}\Big) \bigg],\\
&\mathbb{E}[I_\eta]=\frac{2K_s h P_s}{2-\alpha} \bigg[ \pi\lambda_b\Big( \frac{R_m}{\eta_t} \Big)^{2 - \alpha_s} \Big(e^{ -\pi\lambda_bR_l^2} - e^{ -\pi\lambda_b\eta_t^2R_m^2} \Big)-\frac{(\pi\lambda_b\eta_t^2)^{\frac{\alpha_s}{2}}}{\eta_t^{2}}\Gamma\Big( \frac{4 - \alpha_s}{2} , \pi\lambda_bR_l^2 , \pi\lambda_b\eta_t^2R_m^2\Big) \bigg].\\
\end{split}
\end{equation}

\begin{equation}
\begin{split}
\label{EqA2}
&\mathbb{E}[P_1]=
\mathlarger{\int}_{^{R_l}}^{_{R_m}}2\pi\lambda_b K_sh P_s\Big(\ln r-\ln R_l\Big)f_1(r)dr\\
&{\color{white}{\mathbb{E}[P_1]}}=\pi\lambda_b K_s h P_s\bigg[2\ln \frac{R_l}{R_m} e^{-\pi\lambda_bR_m^2}+\Gamma\Big(0,\pi\lambda_bR_l^2,\pi\lambda_bR_m^2\Big)\bigg],\\
&\mathbb{E}[I_1]=\pi\lambda_b K_s h P_s\bigg[2\ln \frac{R_m}{R_l} e^{-\pi\lambda_bR_l^2}-\Gamma\Big(0,\pi\lambda_bR_l^2,\pi\lambda_bR_m^2\Big)\bigg],\\
&\mathbb{E}[P_2]=\mathlarger{\int}_{^{R_l}}^{_{R_m}}2\pi\lambda_b K_sh P_s\Big(\ln r-\ln R_l\Big)f_2(r)dr\\
&{\color{white}{\mathbb{E}[P_2]}}=4\pi^3\lambda_b^3 K_s h P_s\bigg\{\int_{R_l}^{R_m} \ln rr^3e^{-\pi\lambda_b r^2}dr+\frac{\ln R_l}{2\pi^2\lambda_b^2}\Big[(\pi\lambda_bR_m^2+1)e^{-\pi\lambda_bR_m^2}-(\pi\lambda_bR_l^2+1)e^{-\pi\lambda_bR_l^2}\Big]\bigg\},\\
&\mathbb{E}[I_2]=4\pi^3\lambda_b^3 K_s h P_s \bigg\{\frac{\ln R_m}{2\pi^2\lambda_b^2} \Big[(\pi\lambda_bR_l^2 + 1)e^{-\pi\lambda_bR_l^2} - (\pi\lambda_bR_m^2 + 1)e^{-\pi\lambda_bR_m^2}\Big]-  \int_{R_l}^{R_m}  \ln rr^3e^{-\pi\lambda_b r^2}dr \bigg\},\\
&\mathbb{E}[P_{R_0}]  =2\pi\lambda_b K_s h P_s\ln\frac{R_0}{R_l},\\
&\mathbb{E}[I_{R_0}]=2\pi\lambda_b K_s h P_s\ln\frac{R_m}{R_0},\\
&\mathbb{E}[P_\eta]=\mathlarger{\int}_{^{\frac{R_l}{\eta_t}}}^{_{R_m}} 2\pi\lambda_b K_sh P_s\Big(\ln r-\ln \frac{R_l}{\eta_t}\Big)f_\eta(r)dr\\
&{\color{white}{\mathbb{E}[P_\eta]}}=\pi\lambda_b K_s h P_s\bigg[2\ln \frac{R_l}{\eta_tR_m} e^{-\pi\lambda_b\eta_t^2R_m^2}+\Gamma\Big(0,\pi\lambda_bR_l^2,\pi\lambda_b\eta_t^2R_m^2\Big)\bigg],\\
&\mathbb{E}[I_\eta]=\pi\lambda_b\eta_t^2 K_s h P_s\bigg[2\ln \frac{\eta_tR_m}{R_l} e^{-\pi\lambda_bR_l^2}-\Gamma\Big(0,\pi\lambda_bR_l^2,\pi\lambda_b\eta_t^2R_m^2\Big)\bigg].\\
\end{split}
\end{equation}

}

In Rayleigh and Nakagami-$m$ fading channels, similar results can be obtained by substituting the expectation of the channel (i.e., 1 and $\Omega$, respectively) for $h$ in Eqns.~(\ref{EqA1}) and (\ref{EqA2}). Due to the limited space, the results are not listed in this paper.

\subsection{Moment Generating Functions}
MGF is an alternative representation of probability distribution, which is defined as $\mathcal{M}(z)=\mathbb{E}[e^{-zX}]$ for variable $X$. The relation between MGF and PDF is $\mathcal{M}(z)=\int_{-\infty}^{\infty}e^{-zx}f_X(x)dx$. In other words, MGF is the Laplace transform of PDF. In this paper, MGF is calculated for three purposes, i.e., to obtain the variance in Subsection III-C, the spectral efficiency in Section IV-B, and the numerical value of PDF in Section V-A.

With the help of the Laplace functional of Campbell theorem \cite{Haenggi2013SGbook} (which is also referred to as probability generating functions for a PPP), MGF of the aggregated power in the range $[R_1,R_2]$ is
\begin{equation}
\begin{split}
&\mathcal{M}^{C}(z) = \exp\Big\{\!-2\pi\lambda_b\displaystyle{\int_{R_1}^{R_2}} \Big[1-\exp(-zK_shP_sr^{-\alpha_s})\Big]rdr\Big\}\\
&= \textrm{exp} \bigg\{ 2\pi\lambda_b \Big[ \frac{ (K_shP_s)^{\frac{2}{\alpha_s}}}{\alpha_s}\\
&\times\Gamma \Big( - \frac{2}{\alpha_s}, zK_shP_sR_{2}^{-\alpha_s}, zK_shP_sR_{1}^{-\alpha_s} \Big) - \frac{R_{2}^2}{2} + \frac{R_{1}^2}{2} \Big]  \bigg\},
\end{split}
\end{equation}
where the channel parameter $h$ is assumed to be a constant.

In a Rayleigh fading channel, the result is
\begin{equation}
\begin{split}
\mathcal{M}^{R}(z) &= \mathbb{E}_{r}\Big(\prod_{R_1<r<R_2}\frac{1}{1+zK_sr^{-\alpha_s}P_{s}}\Big)\\
&= \exp\Big( -2\pi\lambda_b\int_{R_1}^{R_2}\frac{rzK_sP_s}{r^{\alpha_s}+zK_sP_s}dr \Big).
\end{split}
\end{equation}
If $\alpha_s$ is a positive integer, a closed form expression can be obtained. For example, if $\alpha_s=2$ for an infinite space, we have
\begin{equation}
\begin{split}
\mathcal{M}^R(z)&=\exp\bigg(-2\pi\lambda_bzK_sP_s\int_{\frac{R_1}{(zK_sP_s)^{1/2}}}^{\frac{R_2}{(zK_sP_s)^{1/2}}}\frac{x}{x^{2}+1}dx\bigg)\\
&\xlongequal[\quad]{u=x^2+1}\Big(\frac{R_2^2+zK_sP_s}{R_1^2+zK_sP_s}\Big)^{-\pi\lambda_bzK_sP_s}.
\end{split}
\end{equation}
If $\alpha_s=4$, we have
\begin{equation}
\begin{split}
\mathcal{M}^R(z)
&=\exp \bigg[ - \pi\lambda_b(zK_sP_s)^{\frac{1}{2}}\\
&\times\arctan \frac{(zK_sP_s)^{\frac{1}{2}}(R_2^2 - R_1^2)}{zK_sP_s + R_1^2R_2^2} \bigg].
\end{split}
\end{equation}
If there is no restriction for $\alpha_s$, a closed form expression can also be obtained when $R_2\rightarrow\infty$ or $R_1\rightarrow0$. If $R_2\rightarrow\infty$, substituting $t=\frac{2zK_sP_sx^{\alpha_s}}{\alpha_s R_2^{\alpha_s}}$, we can get
\begin{equation}
\label{EqMR1}
\begin{split}
\mathcal{M}^R(z) &= \exp \bigg[ - \frac{2\pi\lambda_bzK_sP_sR_1^{2 - \alpha_s}}{\alpha_s-2}\\
&\times_2F_1 \Big( 1 , 1 - \frac{2}{\alpha_s} ;2 - \frac{2}{\alpha_s} ; -\frac{zK_sP_s}{R_1^{\alpha_s}}\Big)\bigg].
\end{split}
\end{equation}
If $R_1\rightarrow0$,
\begin{equation}
\label{EqMR2}
\begin{split}
\mathcal{M}^R(z) &= \exp \bigg[ - \frac{2\pi\lambda_bzK_sP_sR_2^{2 - \alpha_s}}{2}\\
&\times_2F_1 \Big( 1, \frac{2}{\alpha_s} ; 1 + \frac{2}{\alpha_s} ; -\frac{R_2^{\alpha_s}}{zK_sP_s} \Big)\bigg].
\end{split}
\end{equation}
The results of Eqns.~(\ref{EqMR1}) and (\ref{EqMR2}) are obtained based on the equations [3.194] in \cite{Integral7th}.

In a Nakagami-$m$ fading channel, we have
\begin{equation}
\begin{split}
&\quad\mathcal{M}^N (z)
=\mathbb{E}_r\bigg[\prod_{R_1<r<R_2}\displaystyle{\int_0^\infty}\exp(-zK_sP_shr^{-\alpha_s})f_{h_N}(h)dh\bigg]\\
&= \exp \bigg\{ - 2\pi\lambda_b  \displaystyle{\int_{R_1}^{R_2}}   \Big[ 1 - \big( 1 + \frac{\Omega}{m}zK_sP_sr^{-\alpha_s} \big)^{-m} \Big]rdr \bigg\}.
\end{split}
\end{equation}
If $\alpha_s=2$ and $m$ is a positive integer, a closed form expression can be obtained based on the equations [2.117] in \cite{Integral7th}. Here, we take the case of $m=2$ as an example, which is
\begin{equation}
\begin{split}
\mathcal{M}^N (z) &= \exp \Bigg\{ \pi\lambda_b\bigg[ \frac{(\Omega z K_sP_s)^2(R_2^2 - R_1^2)}{(\Omega z K_sP_s+R_1^2)(\Omega z K_sP_s+R_2^2)} \\
&- 2\Omega z K_sP_s\ln\frac{\Omega z K_sP_s+R_2^2}{\Omega z K_sP_s+R_1^2}\bigg] \Bigg\}.
\end{split}
\end{equation}

Thus, the MGFs for different transmission methods in different channels can be calculated and their results are shown in Appendix I.

\subsection{Variance}
The variance measures how far the random results are spread over from the average value, which can be obtained from the MGF. Substitute $s=-z$ to all MGFs to obtain the expression $\mathcal{M}(s)$. Then, we can use the property of MGF to obtain the expectation of the $n$th moment, i.e., $\mathbb{E}[x^n]=\frac{\Delta^n\mathcal{M}(s)}{\Delta s^n}\big|_{s=0}$. Thus, the variance can be calculated as $\mathbb{V}[x]=\mathbb{E}[x^2]-\mathbb{E}[x]^2$.

To apply the aforementioned method to obtain the variance of aggregated desired signal and interference powers, we first need to calculate the expectation of the square, i.e., $\mathbb{E}[P^2]$ and $\mathbb{E}[I^2]$, for different transmission methods. Then, combining with the average value in Subsection III-A gives the variance. An example will be presented below, which is the variance of aggregated desired power for NoJT method when the channel parameter $h$ is a constant. Due to the limited space, the calculation process for other cases will not be included in this paper.

If $h$ is assumed to be a constant, the first order derivative of MGF of the desired power for NoJT method is
\begin{equation}
\begin{split}
&\frac{\Delta\mathcal{M}^C_{P_1}(s)}{\Delta s}\xlongequal{(a)}   \int_{R_l}^{R_m}   \frac{\Delta b(s)}{\Delta s} f_{1}(r)dr\\
&= \int_{R_l}^{R_m} 2\pi\lambda_b \int_{R_l}^{r} K_shP_sx^{1-\alpha_s}a(s)dxb(s)f_{1}(r)dr,
\end{split}
\end{equation}
where (a) is based on Leibniz integral rule $b(s)=\exp\{-2\pi\lambda_b\int_{R_l}^{r}[1-\exp(sK_shP_sx^{-\alpha_s})]xdx\}$, and $a(s)=\exp(sK_shP_sx^{-\alpha_s})$. The second order derivative of MGF is
\begin{equation}
\begin{split}
&\frac{\Delta^2\mathcal{M}_{P_1}(s)}{\Delta s^2}
=2\pi\lambda_bK_shP_s \int_{R_l}^{R_m} \bigg[ \int_{R_l}^r x^{1 - \alpha_s}\frac{\Delta a(s)}{\Delta s}dx b(s)\\
&+\int_{R_l}^rx^{-\alpha_s+1}a(s)dx\frac{\Delta b(s)}{\Delta s}\bigg]f_1(r)dr\\
&=2\pi\lambda_bK_shP_s \int_{R_l}^{R_m} \bigg\{ \int_{R_l}^r
x^{1 - 2\alpha_s}K_shP_sa(s)dxb(s)\\
& + 2\pi\lambda_bK_shP_s\Big[ \int_{R_l}^r x^{1 - \alpha_s}a(s)dx \Big]^2  b(s) \bigg\}f_1(r)dr.
\end{split}
\end{equation}

Thus, the expectation of the second order moment is shown in Eqn.~(\ref{EqP2}),
\begin{equation}
\label{EqP2}
\begin{split}
\qquad\,\mathbb{E}[P_1^2]&= \frac{\Delta^2\mathcal{M}_{P_1}(s)}{\Delta s^2}\Big|_{s=0}\\
&\xlongequal{(a)} 2\pi\lambda_bK_s^2h^2P_s^2 \bigg\{ \Big[ \frac{4\pi\lambda_bR_l^{4 - 2\alpha_s}}{(2 - \alpha_s)^2} - \frac{R_l^{2 - 2\alpha_s}}{2(1 - \alpha_s)} \Big] \Big(e^{-\pi\lambda_bR_l^2}   - e^{-\pi\lambda_bR_m^2} \Big)\\
&+ \Big[\frac{(\pi\lambda_b)^{\alpha_s - 1}}{2(1 - \alpha_s)} + \frac{2(\pi\lambda_b)^{\alpha_s - 1}}{2 - \alpha_s} \Big] \Gamma\big( 2 - \alpha_s , \pi\lambda_bR_l^2 , \pi\lambda_bR_m^2 \big)\\
&- \frac{4R_l^{2 - \alpha_s}(\pi\lambda_b)^{\alpha_s/2}}{(2 - \alpha_s)^2}\Gamma\Big(2-\frac{\alpha_s}{2},R_l^2\pi\lambda_b,R_m^2\pi\lambda_b\Big) \bigg\},
\end{split}
\end{equation}
where (a) is based on the relation $\Gamma(s+1,a,b)=\Gamma(s,a,b)+a^se^{-a}-b^se^{-b}$. The variance of the desired power is calculated in Eqn.~(\ref{EqVP1}).
\begin{equation}
\label{EqVP1}
\begin{split}
\mathbb{V}[P_1]&=\mathbb{E}[P_1^2]-\mathbb{E}[P_1]^2\\
&=2\pi\lambda_bK_s^2h^2P_s^2 \bigg\{ \Big[ \frac{4\pi\lambda_bR_l^{4 - 2\alpha_s}}{(2 - \alpha_s)^2} - \frac{R_l^{2 - 2\alpha_s}}{2(1 - \alpha_s)} \Big] \Big(e^{-\pi\lambda_bR_l^2} - e^{-\pi\lambda_bR_m^2} \Big)+\Big[ \frac{(\pi\lambda_b)^{\alpha_s - 1}}{2(1 - \alpha_s)}\\
 &+ \frac{2(\pi\lambda_b)^{\alpha_s - 1}}{2 - \alpha_s} \Big]\Gamma\big(2 - \alpha_s, \pi\lambda_bR_l^2, \pi\lambda_bR_m^2\big)- \frac{4R_l^{2 - \alpha_s}(\pi\lambda_b)^{\alpha_s/2}}{(2 - \alpha_s)^2}\Gamma\Big(2-\frac{\alpha_s}{2},R_l^2\pi\lambda_b,R_m^2\pi\lambda_b\Big) \bigg\}\\
&- \frac{4K_s^2 h^2 P_s^2}{(2-\alpha)^2} \bigg[ (\pi\lambda_b)^{\frac{\alpha_s}{2}} \Gamma\Big(\frac{4 - \alpha_s}{2},\pi\lambda_bR_l^2,\pi\lambda_bR_m^2\Big)+\pi\lambda_bR_l^{-\alpha_s+2}\big(e^{-\pi\lambda_bR_m^2}-e^{-\pi\lambda_bR_l^2}\big)\bigg]^2.
\end{split}
\end{equation}

\section{SYSTEM PERFORMANCE}
Different statistical characteristics of received power can be used to evaluate system performances, such as SINR, spectral efficiency, and area spectral efficiency. In this section, we will calculate these system performance indicators, respectively.

\subsection{Expectation of SINR}
It is difficult to obtain the exact value of the expectation of SINR, and different approximation methods were proposed accordingly. For example, in \cite{Hamdi2009SINRTC}\cite{Shojaeifard2015SINRTIT}, the expectation of SINR was calculated by
\begin{equation}
\label{EqSINR}
\mathbb{E}[\Upsilon]=\frac{\mathbb{E}[P]}{\mathbb{E}[I]+N_0},
\end{equation}
and the expectation of the $n$th order moment of SINR is
\begin{equation}
\mathbb{E}[\Upsilon^n]= \mathbb{E}[P^n]\int_0^\infty\frac{z^{n-1}}{\Gamma(n)}\mathcal{M}_{I}(z)e^{-zN_0}dz,\quad n>0,
\end{equation}
where $N_0$ is the noise power.

The average (i.e., $\mathbb{E}[P]$ and $\mathbb{E}[I]$), the MGF of interference (i.e., $\mathcal{M}_{I}(z)$), and the expectation of its square (i.e., $\mathbb{E}[P^2]$) have been calculated in Subsection III-A, Subsection III-B, and Subsection III-C, respectively. Thus, we can obtain the expectation of the first and second order moments of SINR for different transmission methods.

\subsection{Spectral Efficiency}
Spectral efficiency is defined as $S=\log(1+\Upsilon)$. In \cite{HamdiIF2010TC}, the authors used the relation $\textrm{ln}(1+x)=\int_0^\infty\frac{1}{s}(1-e^{-sx})e^{-s}ds$ to transform the expression into
\begin{equation}
S=\frac{1}{\textrm{ln}2}\int_0^\infty \frac{1}{s}\big[1-M_\Upsilon(s)\big]e^{-s}ds,
\end{equation}
where $M_\Upsilon(s)=\mathbb{E}[\textrm{exp}(-s\Upsilon)]$, which is the MGF of SINR. Thus, we have
\begin{equation}
\label{Eq.S}
\begin{split}
S&=\frac{1}{\textrm{ln}2}\int_0^\infty \frac{1}{s}\bigg[1-\exp\bigg(-s\frac{P}{I+N_0}\bigg)\bigg]e^{-s}ds\\
& \xlongequal[\quad]{z=\frac{s}{I+N_0}} \frac{1}{\textrm{ln}2} \int_0^\infty \Big[e^{ -zI} - e^{ -z(I + P)}\Big]\frac{e^{ -zN_0}}{z}dz\\
&=\frac{1}{\textrm{ln}2}\int_0^\infty \Big[\mathcal{M}_I(z)-\mathcal{M}_{PI}(z)\Big]\frac{e^{-zN_0}}{z}dz,
\end{split}
\end{equation}
where $\mathcal{M}_{PI}(z)=\mathbb{E}[e^{-z(P+I)}]$ is the MGF of the total received power from both desired and interfering transmitters, which is the same for different types of transmission methods, as shown in Eqn.~(\ref{EqMPI}),
{\small
\begin{equation}
\label{EqMPI}
\mathcal{M}_{PI}(z)=
\begin{cases}
\displaystyle\exp \bigg\{  2\pi\lambda_b \bigg[ \frac{ (K_shP_s)^{\frac{2}{\alpha_s}}}{\alpha_s}\Gamma \Big( - \frac{2}{\alpha_s},zK_shP_sR_{m}^{-\alpha_s},zK_shP_sR_{l}^{-\alpha_s} \Big) - \frac{R_{m}^2}{2} + \frac{R_{l}^2}{2} \bigg] \bigg\},& \textrm{Constant,}\\
\displaystyle\exp\bigg( -2\pi\lambda_b\mathop{\mathlarger{\int}}_{R_l}^{R_m}\frac{rzK_sP_s}{r^{\alpha_s}+zK_sP_s}dr \bigg),&\textrm{Rayleigh,}\\
\displaystyle\exp \bigg\{ - \mathop{\mathlarger{\int}}_{R_l}^{R_m}  2\pi\lambda_b\bigg[ 1 - \big( 1 + \frac{\Omega}{m}zK_sP_sr^{-\alpha_s} \big)^{-m} \bigg]rdr \bigg\},&\textrm{Nakagami-}m.
\end{cases}
\end{equation}
}

Thus, we can obtain the spectral efficiency for different transmission methods using the corresponding MGF of aggregated interference.

\subsection{Area Spectral Efficiency}
Area spectral efficiency was first proposed by Alouini \emph{et al.,} which was defined as the capacity achieved by all users per spectrum in a unit area \cite{AloASE1999TVT}. Comparing to conventional spectral efficiency, ASE can depict the spatial reuse over the area, which is in particular suitable for UDNs \cite{Yunas2015ASECM} \cite{Ding2017ASETWC}.

Assume that all SBSs are serving at least one UE and the same spectral resource is shared by all SBSs. The ASE can be calculated approximatively by
\begin{equation}
C=\sum_{n=1}^{\infty}\mathbb{P}\big[|\Phi|=n\big]\frac{nS}{\pi R_m^2},
\end{equation}
where $\mathbb{P}\big[|\Phi|=n\big]$ is the probability that there are $n$ SBSs distributed by the PPP $\Phi$, which is
\begin{equation}
\mathbb{P}\big[|\Phi|=n\big]=\frac{1}{n!}e^{-\lambda_b\pi R_m^2}\big(\lambda_b\pi R_m^2\big)^{n},\quad n\in\mathbb{Z}^{+}.
\end{equation}

Using spectral efficiency of Eqn.~(\ref{Eq.S}) for different transmission methods, we obtain the approximated ASE.

\section{NUMERICAL RESULTS}
In this section, we will verify the proposed analytical approach and show the performances of different transmission schemes. Assume that we have $R_l=0.1$, $R_m=60$, $K_s=1$ and transmit power of SBS is 17 dBm. The results are presented in Figs.~\ref{average_power} to \ref{other_performance}.

\subsection{Aggregated Received Power}
\subsubsection{Average value}
\begin{figure*}
\centering
\includegraphics[width=15cm]{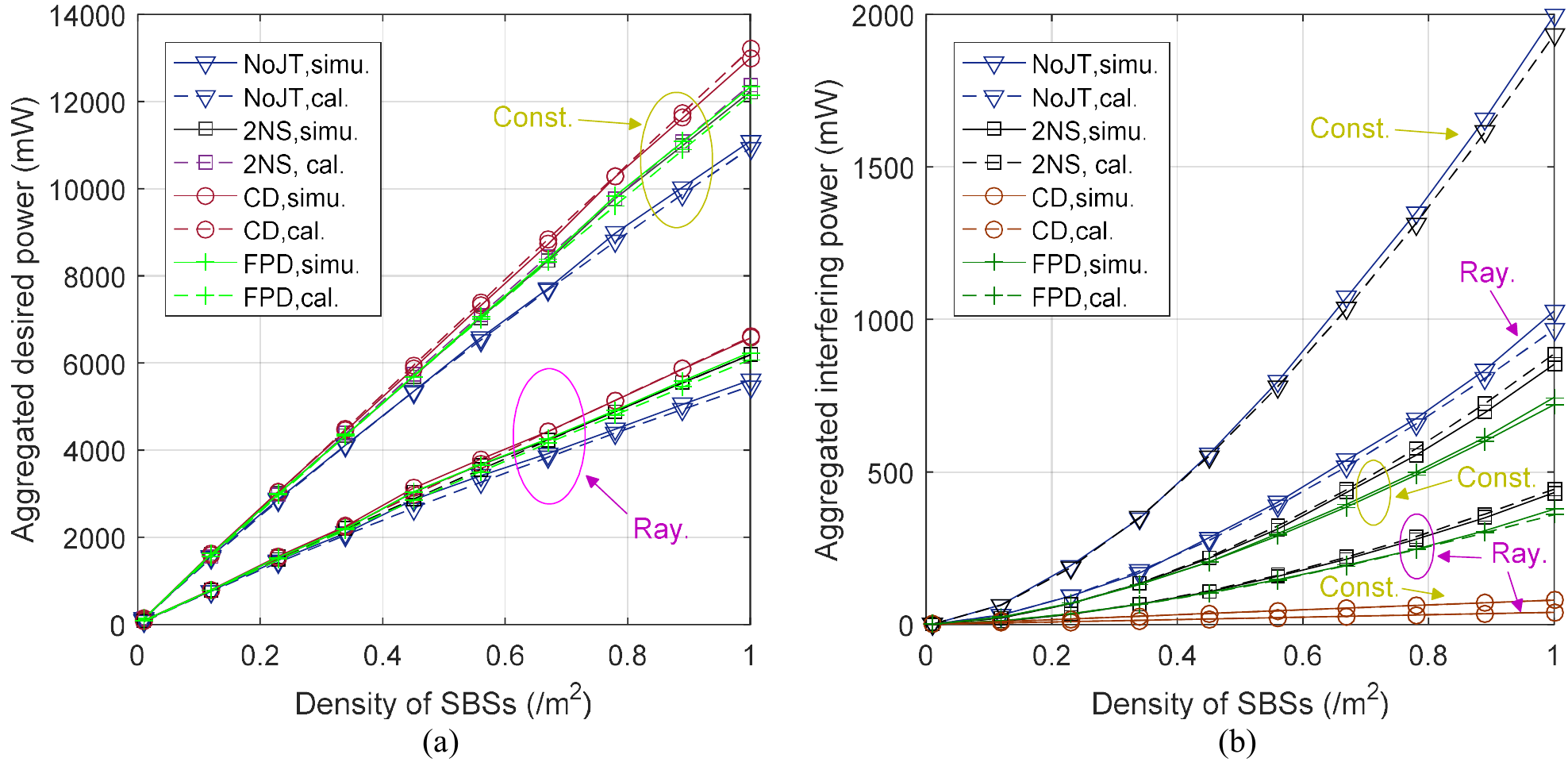}
\caption{{Average desired signal and interference powers for different transmission schemes are shown in (a) and (b), respectively, where $\alpha_s=3.5$, $R_0=3$, $\eta=10$, and $h$ is two if it is a constant. Simulation run for 100,000 times. 'Const.' and 'Ray.' denote the scenarios that channel parameter $h$ is a constant and follows a Rayleigh distribution in fading channel, respectively.}}
 \label{average_power}
\end{figure*}

The average aggregated desired signal and interference powers are shown in Fig.~\ref{average_power}(a) and Fig.~\ref{average_power}(b), respectively. Both desired signal and interference powers increase with SBSs density. A linear increasing trend can be observed for the desired power, and the interference power grows faster with the density, especially for NoJT transmission method. In other words, the use of JT can slow down the growth rate of interference power, and thus JT is proved to be more effective for scenarios with densely distributed SBSs. As proved in Subsection III-A, the same result can be observed when $h=1$ or in Rayleigh fading channel. In other words, the shape of curve depends only on the spatial randomness of SBSs, and the fading channel only affects the amplitude of curve. In order to distinguish these two channel conditions, we assume that $h$ is two if it is a constant in Fig.~\ref{average_power}. Thus, a higher value can be observed if $h=2$.

\subsubsection{Probability Density Functions}

\begin{figure*}
\centering
\includegraphics[width=15cm]{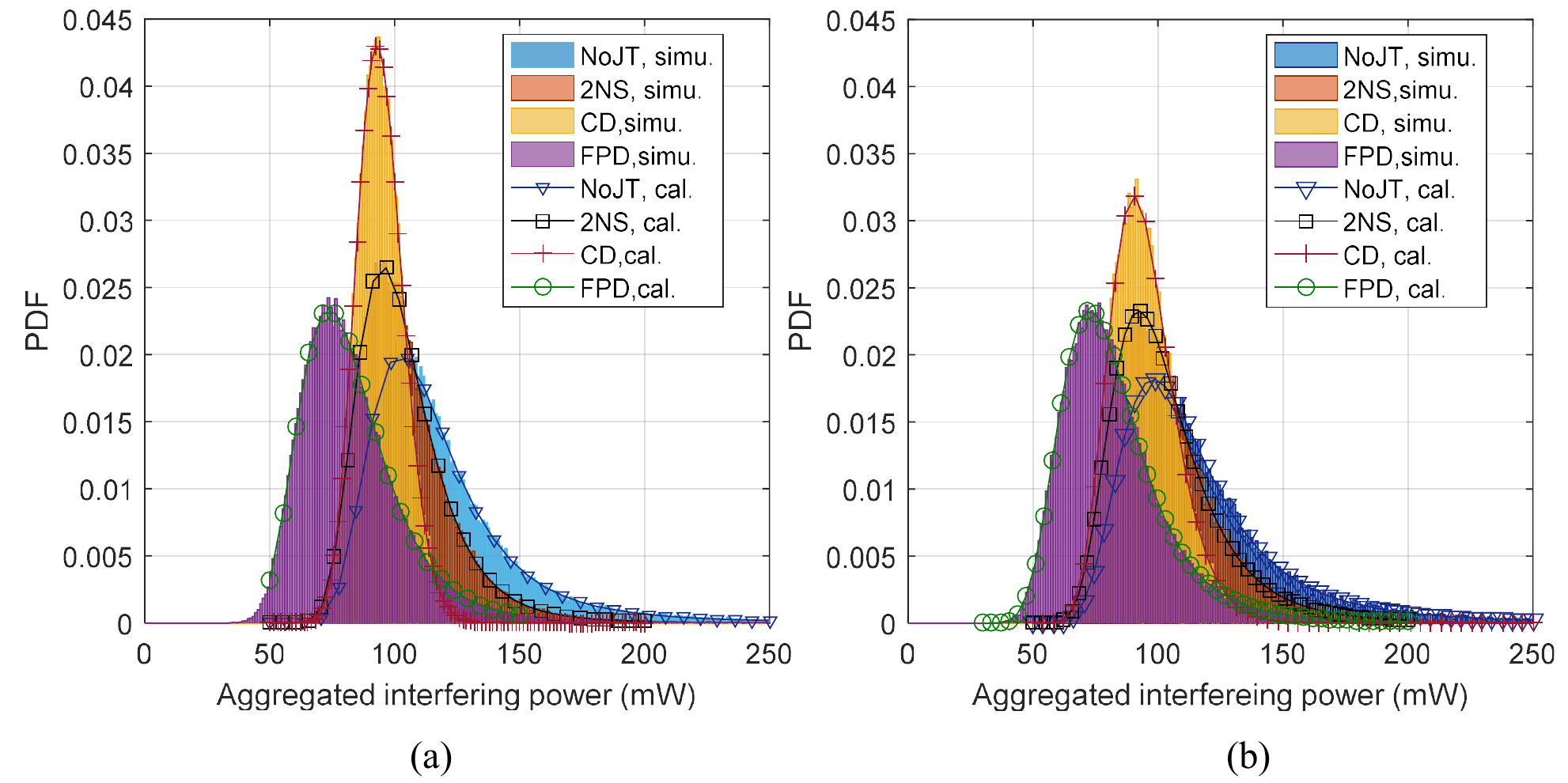}
\caption{{PDFs of interference power for different transmission schemes, where (a) is the scenario that channel parameter $h$ is a constant, and (b) is the scenario in a Rayleigh fading channel. $\alpha_s=2$, $R_0=3$, $\eta=10$, and $h$ is one if it is a constant. Simulation runs for 100,000 times.}}
\label{I_pdf_compare}
\end{figure*}

The theoretical probability density functions (PDFs) are depicted based on the inverse Laplace transform of MGFs, and the PDFs of interference power for different transmission methods are shown in Fig.~\ref{I_pdf_compare}. Fig.~\ref{I_pdf_compare}(a) depicts the scenario that channel parameter $h$ is a constant, and thus the statistical characteristics of aggregated power depend only on the spatial distribution of SBSs. Fig.~\ref{I_pdf_compare}(b) shows the PDFs in a Rayleigh fading channel, and thus both the randomness of spatial distribution and channel condition contribute to the distribution of aggregated power. Comparing Fig.~\ref{I_pdf_compare}(b) with Fig.~\ref{I_pdf_compare}(a), we see that the involvement of channel randomness results in a more scattered distribution of aggregated power for all schemes, and the most obvious example is CD method.

Meanwhile, we also notice that the shape of each curve resembles that of Gamma distribution. In \cite{ElSawy2017StoCOMST} and \cite{Renzo2015EiDTC}, the authors analyzed the distribution of aggregated interference power in NoJT method in Rayleigh fading channel. In a Rayleigh fading channel, the magnitude of received interference can be regarded as the sum of multiple random variables following Rayleigh distributions, resulting in a Gaussian distribution. Thus, they used a Gaussian distribution to represent the aggregated power to simplify the calculation process of other system parameters. The square of Gaussian random variable follows a Gamma distribution, and thus the aggregated interference power follows a Gamma distribution in a Rayleigh fading channel, which is consistent to our results. Meanwhile, we can also conclude that the distribution of aggregated interference power resembles a Gamma distribution, regardless the distribution of  the fading channel, as shown in Fig.~\ref{I_pdf_compare}(a). Meanwhile, we can expect that the distribution of aggregated desired signal power is similar to that of the interference power, and thus a similar approximation method can be used to further simplify the calculation process of system performances.

\subsubsection{Variance}
\begin{figure*}
\centering
\includegraphics[width=15cm]{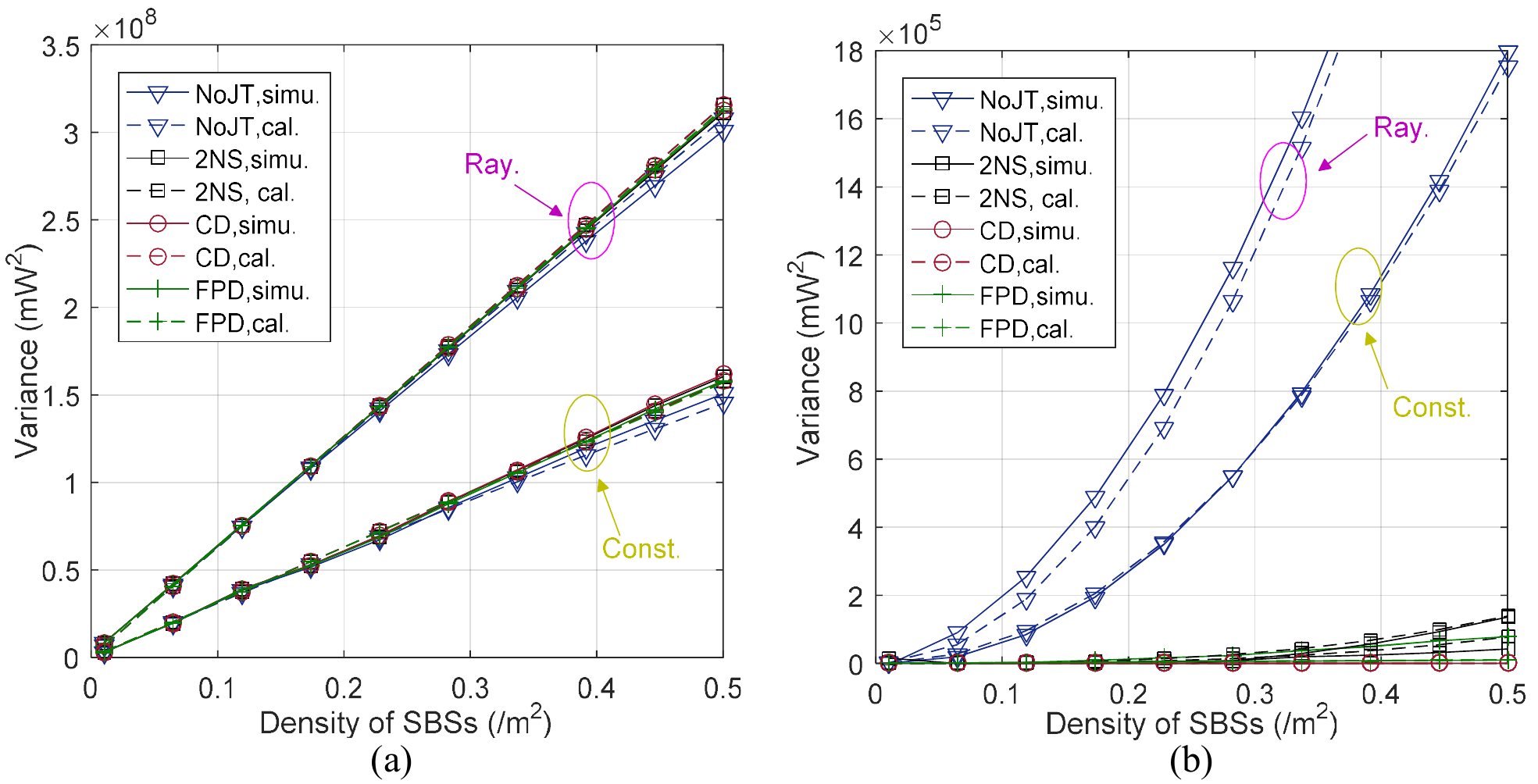}
\caption{{The variances of desired signal and interference power for different transmission schemes are shown in (a) and (b), respectively, where $\alpha_s=3.5$, $R_0=3$, $\eta=10$ and $h$ is one if it is constant. Simulation runs 400,000 times. 'Const.' and 'Ray.' denote the scenarios that channel parameter $h$ is a constant and follows a Rayleigh distribution in fading channel, respectively.}}
\label{variance}
\end{figure*}

The variances of aggregated desired signal and interference powers for different transmission schemes are shown in Fig.~\ref{variance}(a) and (b), respectively. The shape of curve is similar to that in Fig.~\ref{average_power}, and a higher variance can be observed in a Rayleigh channel due to the involvement of extra randomness. Meanwhile, an interesting phenomenon can be observed when comparing Figs.~\ref{average_power}, \ref{I_pdf_compare} and \ref{variance}, that is, the expectation is much larger than the value of the highest PDF, and the variance is even larger. In other words, an extremely large aggregated power can be received with an extremely small probability, and the majority of the received power is much smaller than the average value. Thus, the aggregated power follows a long-tail distribution.

In \cite{Renzo2013alpha-stableTC}, the authors applied an alpha-stable distribution, a representative of long-tail distribution, to depict the aggregated magnitude of interference for NoJT in a Rayleigh fading channel, which is consistent to our results. Moreover, we can also foresee that a similar approach can be applied to other transmission methods as well.

To sum up, both alpha-stable and Gamma distributions can be applied. An alpha-stable distribution can correctly model the tail portion of PDF, while a Gamma distribution is better to depict the head portion of PDF. The problem of determining a better approximation method to depict the aggregated power in UDNs will be our future work.

\begin{figure*}
\centering
\includegraphics[width=15cm]{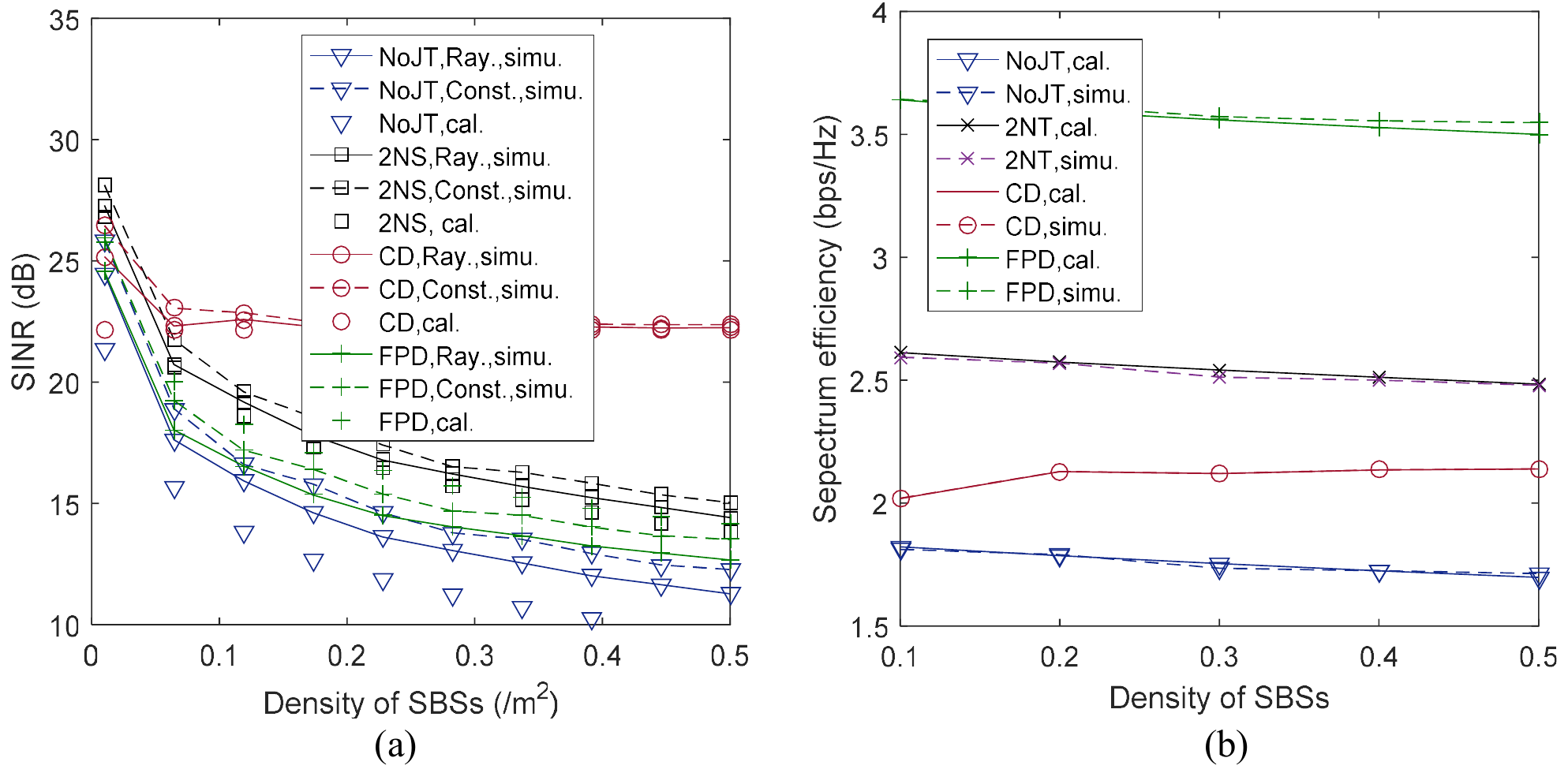}
\caption{{System performances for different transmission schemes, where (a) and (b) are the average SINR and average spectral efficiency, respectively. $\alpha_s=3.5$, $R_0=3$, $\eta=10$ and $h=1$ if it is a constant. Simulation runs 100,000 times. 'Const.' and 'Ray.' denote the scenarios that channel parameter $h$ is a constant and follows a Rayleigh distribution in fading channel, respectively.}}
\label{other_performance}
\end{figure*}

\subsection{System Performance Analysis}

The average SINR for different transmission schemes are shown in Fig.~\ref{other_performance}(a). The analytical average SINR obtained by Eqn.~(\ref{EqSINR}) is the same in both $h=1$ and Rayleigh fading channel. As shown in Fig.~\ref{other_performance}(a), the analytical result can approximately show the actual value in these two channel conditions, and the difference between the analytical and simulation results is within an acceptable range. Comparing to NoJT method, we see that using JT has a better performance, especially in area with high density of SBSs.

The average spectral efficiency is shown in Fig.~\ref{other_performance}(b). The analytical results match to the simulation results, and the correctness of the proposed model is proved. Using JT can improve the average spectral efficiency, and different JT schemes have different performance gains.

The performances of CD and FPD methods are influenced by the value of $R_0$ and $\eta$, and thus we can not jump to the conclusion about which scheme is better based on a single scenario. However, considering the case of $R_0=3$ and $\eta=10$ only, an interesting phenomenon can be observed when comparing Figs.~\ref{average_power} and \ref{other_performance}. If the received power is considered, the performance level can be sorted by CD$>$FPD$>$2NS$>$NoJT. If the SINR is considered, the performance level can be sorted by CD$>$2NS$>$FPD$>$NoJT. If the spectral efficiency is considered, the performance level can be sorted by FPD$>$2NS$>$CD$>$NoJT. Thus, different methods should be chosen based on different requirements.

\section{CONCLUSIONS}
This paper proposed a unified approach to analyze the performance of different joint transmission methods in UDNs. By dividing a UDN into two regions, the calculation of aggregated desired signal and interference powers can be transformed into the integral of received power over each region. Thus, several statistical characteristics of aggregated power, such as expectation, MGF, and variance, can be obtained based on a PPP. Note that the randomness of received power is caused by both spatial distribution of SBSs and multipath or shadowing fading. First, we analyzed the influence of spatial randomness by assuming a constant fading channel. Then, results in Rayleigh and Nagakami-$m$ channel were derived, respectively. As shown in Fig.~\ref{average_power}, the shapes of the average power curves depend only on the spatial randomness of SBSs, and the expectation of the fading channel (i.e., $\mathbb{E}[h]$) affects the scale of the curve. Meanwhile, the involvement of fading channel increases the spreading of aggregated power, resulting in a higher variance and a more decentralized PDF. Moreover, a long-tail distribution feature was observed with or without channel fading. In other words, an extremely high aggregated power can be received at an extremely small probability, and the majority of received power is much lower than the average value. The other system performances, such as SINR, spectral efficiency, and area spectral efficiency, were calculated based on the expectation, MGF, and variance of aggregated power. Simulation results matched to the analytical results very well, which justified the proposed model. Meanwhile, we also note that the proposed unified analytical method can be applied to other scenarios (such as ad hoc networks), as long as transmitters are distributed in a PPP.

\section*{APPENDIX: DERIVATIONS OF MGFs }
If a constant channel parameter $h$ is assumed, the MGFs of desired signal and interference powers for NoJT scheme are derived as
\begin{equation}
\begin{split}
&\mathcal{M}_{P_1}^C(z)=2\pi\lambda_b\int_{{R_l}}^{{R_m}}r\exp \bigg\{  2\pi\lambda_b\bigg[ \frac{ (zK_sh P_{s})^{\frac{2}{\alpha_s}}}{\alpha_s}\\
&\times\Gamma\Big( -\frac{2}{\alpha_s}, zK_sh P_{s}R_m^{-\alpha_s}, zK_sh P_{s}r^{-\alpha_s} \Big) - r^2 + \frac{R_l^2}{2}\bigg]\bigg\}dr,\\
&\mathcal{M}_{I_1}^C(z)=2\pi\lambda_b\!\int_{{R_l}}^{{R_m}}\!r\exp \bigg\{  2\pi\lambda_b\bigg[ \frac{ (zK_sh P_{s})^{\frac{2}{\alpha_s}}}{\alpha_s}\\
&\times \Gamma\Big(\! -\frac{2}{\alpha_s}, zK_sh P_{s}R_m^{-\alpha_s}, zK_sh P_{s}r^{-\alpha_s} \Big) - \frac{R_m^2}{2} \bigg]  \bigg\}dr.
\end{split}
\end{equation}
In a Rayleigh fading channel, the MGFs of NoJT scheme become
\begin{equation}
\begin{split}
\mathcal{M}_{P_1}^R(z)&=2\pi\lambda_b\int_{{R_l}}^{{R_m}}r\exp\Big( -2\pi\lambda_b\times \int_{R_l}^{r}\frac{xzK_sP_s}{x^{\alpha_s}+zK_sP_s}dx \Big)dr,\\
\mathcal{M}_{I_1}^R(z)&=2\pi\lambda_b\int_{R_l}^{{R_m}}r\exp\Big( -2\pi\lambda_b\times\int_{r}^{R_m}\frac{xzK_sP_s}{x^{\alpha_s}+zK_sP_s}dx \Big)dr.
\end{split}
\end{equation}
In a Nakagami-$m$ fading channel, the MGFs of NoJT scheme are
\begin{equation}
\begin{split}
&\mathcal{M}_{P_1}^N(z)=2\pi\lambda_b\int_{^{R_l}}^{_{R_m}}r\exp\Big\{ - 2\pi\lambda_b\\
&\times\int_{R_1}^{r} \Big[1-\Big(1+\frac{\Omega}{m}zK_sP_sx^{-\alpha_s}\Big)^{-m}\Big]xdx-\pi\lambda_br^2 \Big\}dr,\\
&\mathcal{M}_{I_1}^N(z)=2\pi\lambda_b\int_{^{R_l}}^{_{R_m}}r\exp\Big[2\pi\lambda_b\\
&\times\int_{r}^{R_m} \Big(1+\frac{\Omega}{m}zK_sP_sx^{-\alpha_s}\Big)^{-m}xdx-\pi\lambda_bR_m^2\Big]dr.
\end{split}
\end{equation}

Similarly, if a constant channel factor is assumed, the MGFs of desired signal and interference powers of 2NS transmission scheme are
\begin{equation}
\begin{split}
&\mathcal{M}_{P_2}^C(z)=2\pi^2\lambda_b^2\int_{{R_l}}^{{R_m}}\!\!r^3\exp \bigg\{  2\pi\lambda_b\bigg[  - \frac{ (zK_sh P_{s})^{\frac{2}{\alpha_s}}}{\alpha_s}\\
&\times\Gamma\Big(\frac{2}{\alpha_s}, zK_sh P_{s}r^{-\alpha_s},zK_s h P_{s}R_l^{-\alpha_s} \Big) - r^2+ \frac{R_l^2}{2}\bigg]  \bigg\}dr,\\
&\mathcal{M}_{I_2}^C(z)=2\pi^2\lambda_b^2\int_{{R_l}}^{{R_m}}\!\!r^3\exp \bigg\{  2\pi\lambda_b\bigg[ \frac{ (zK_sh P_{s})^{\frac{2}{\alpha_s}}}{\alpha_s}\\
&\times \Gamma\Big( - \frac{2}{\alpha_s}, zK_sh P_{s}R_m^{-\alpha_s},zK_s h P_{s}r^{-\alpha_s} \Big) - \frac{R_m^2}{2} \bigg]  \bigg\}dr.
\end{split}
\end{equation}
In a Rayleigh fading channel, the MGFs of 2NS scheme are
\begin{equation}
\begin{split}
\mathcal{M}_{P_2}^R(z)&=2\pi^2\lambda_b^2\int_{{R_l}}^{{R_m}}r^3\exp\Big( - 2\pi\lambda_b\\
&\times\int_{R_l}^{r}\frac{xzK_sP_s}{x^{\alpha_s}+zK_sP_s}dx-\pi\lambda_br^2 \Big)dr,\\
\mathcal{M}_{I_2}^R(z)&=2\pi^2\lambda_b^2\int_{{R_l}}^{{R_m}}r^3\exp\Big( - 2\pi\lambda_b\\
&\times\int_{r}^{R_m}\frac{xzK_sP_s}{x^{\alpha_s}+zK_sP_s}dx-\pi\lambda_br^2 \Big)dr.
\end{split}
\end{equation}
In a Nakagami-$m$ fading channel, the MGFs of 2NS scheme become
\begin{equation}
\begin{split}
&\mathcal{M}_{P_2}^N(z)=2\pi^2\lambda_b^2\int_{{R_l}}^{{R_m}}r^3\exp\Big\{ - 2\pi\lambda_b\\
&\times \int_{R_1}^{r} \Big[1-\Big(1+\frac{\Omega}{m}zK_sP_sx^{-\alpha_s}\Big)^{-m}\Big]xdx-\pi\lambda_br^2 \Big\}dr,\\
&\mathcal{M}_{I_2}^N(z)=2\pi^2\lambda_b^2\int_{{R_l}}^{{R_m}}r^3\exp\Big[2\pi\lambda_b\\
&\times \int_{r}^{R_m} \Big(1+\frac{\Omega}{m}zK_sP_sx^{-\alpha_s}\Big)^{-m}xdx-\pi\lambda_bR_m^2\Big]dr.
\end{split}
\end{equation}
Similarly, if a deterministic channel is assumed, the MGFs of desired signal and interference powers for the CD transmission scheme are
\begin{equation}
\begin{split}
&\mathcal{M}_{P_{R_0}}^C(z)= \exp  \bigg\{  2\pi\lambda_b\bigg[ \frac{ (zK_s h P_{s})^{\frac{2}{\alpha_s}}}{\alpha_s}\\
&\times \Gamma\Big( - \frac{2}{\alpha_s}, zK_s h P_{s}R_0^{-\alpha_s},zK_s h P_{s}R_l^{-\alpha_s} \Big) - \frac{R_0^2}{2}+ \frac{R_l^2}{2} \bigg]  \bigg\},\\
&\mathcal{M}_{I_{R_0}}^C(z)= \exp \bigg\{ 2\pi\lambda_b\bigg[ \frac{ (zK_s h P_{s})^{\frac{2}{\alpha_s}}}{\alpha_s}\\
&\times \Gamma\Big( - \frac{2}{\alpha_s}, zK_sh P_{s}R_m^{-\alpha_s},zK_sh P_{s}R_0^{-\alpha_s} \Big) - \frac{R_m^2}{2}+ \frac{R_0^2}{2} \bigg]  \bigg\}.
\end{split}
\end{equation}
In a Rayleigh fading channel, the MGFs of CD scheme are
\begin{equation}
\begin{split}
\mathcal{M}_{P_{R_0}}^R(z)=&\exp\Big( - 2\pi\lambda_b\int_{R_l}^{R_0}\frac{xzK_sP_s}{x^{\alpha_s}+zK_sP_s}dx \Big),\\
\mathcal{M}_{I_{R_0}}^R(z)=&\exp\Big( - 2\pi\lambda_b\int_{^{R_0}}^{R_m}\frac{xzK_sP_s}{x^{\alpha_s}+zK_sP_s}dx \Big).
\end{split}
\end{equation}
In a Nakagami-$m$ fading channel, the MGFs of CD scheme are
\begin{equation}
\begin{split}
&\mathcal{M}_{P_{R_0}}^N(z)\\
&=\exp\Big\{ - 2\pi\lambda_b\int_{R_l}^{R_0} \Big[1-\Big(1+
\frac{\Omega}{m}zK_sP_sr^{-\alpha_s}\Big)^{-m}\Big]rdr \Big\},\\
&\mathcal{M}_{I_{R_0}}^N(z)\\
&=\exp\Big\{ - 2\pi\lambda_b\int_{^{R_0}}^{R_m} \Big[1-\Big(1+\frac{\Omega}{m}zK_sP_sr^{-\alpha_s}\Big)^{-m}\Big]rdr \Big\}.
\end{split}
\end{equation}
Similarly, in a deterministic channel the MGFs of desired signal and interference powers for a fixed-power-difference (FPD) transmission scheme are
\begin{equation}
\begin{split}
&\mathcal{M}_{P_{\eta}}^C(z)=2\pi\lambda_b\eta_t^2 \int_{{R_l}}^{{R_m}}\!  r\exp  \bigg\{  2\pi\lambda_b\bigg[ \frac{ (zK_sh P_{s})^{\frac{2}{\alpha_s}}}{\alpha_s}\\
&\times \Gamma\Big(\! - \frac{2}{\alpha_s},\frac{zK_s hP_{s}}{r^{\alpha_s}} ,\frac{ zK_s h P_{s}}{R_l^{\alpha_s}} \Big) - \frac{(1 + \eta_t^2)r^2}{2} + \frac{R_l^2}{2} \bigg]  \bigg\}dr,\\
&\mathcal{M}_{I_{\eta}}^C(z)=2\pi\lambda_b\eta_t^2\int_{{R_l}}^{{R_m}}\! r\exp  \bigg\{  2\pi\lambda_b\bigg[ \frac{ (zK_s h P_{s})^{\frac{2}{\alpha_s}}}{\alpha_s}\\
&\times \Gamma\Big(\! - \frac{2}{\alpha_s}, zK_sh P_{s}R_m^{-\alpha_s},zK_s h P_{s}r^{-\alpha_s} \Big) - \frac{R_m^2}{2} \bigg] \bigg\}dr.
\end{split}
\end{equation}
In a Rayleigh fading channel, the MGFs of FPD scheme are
\begin{equation}
\begin{split}
\mathcal{M}_{P_{\eta}}^R(z)&=2\pi\lambda_b\eta_t^2\int_{{R_l}}^{{R_m}} r\exp\Big( - 2\pi\lambda_b\\
&\times\int_{R_l}^{r}\frac{xzK_sP_s}{x^{\alpha_s}+zK_sP_s}dx-\pi\lambda_b\eta_t^2r^2 \Big)dr,\\
\mathcal{M}_{I_{\eta}}^R(z)&=2\pi\lambda_b\eta_t^2\int_{{R_l}}^{{R_m}}r\exp\Big( - 2\pi\lambda_b\\
&\times\int_{r}^{R_m}\frac{xzK_sP_s}{x^{\alpha_s}+zK_sP_s}dx-\pi\lambda_b\eta_t^2r^2 \Big)dr.
\end{split}
\end{equation}
In a Nakagami-$m$ fading channel, the MGFs of FPD scheme are
\begin{equation}
\begin{split}
&\mathcal{M}_{P_{\eta}}^N(z)=2\pi\lambda_b\eta_t^2\int_{^{R_l}}^{_{R_m}} r\exp\Big\{  - 2\pi\lambda_b\\
&\times\int_{R_1}^{r} \Big[1-\Big(1+\frac{\Omega}{m}zK_sP_sx^{-\alpha_s}\Big)^{-m}\Big]xdx - \pi\lambda_b\eta_t^2r^2 \Big\}dr,\\
&\mathcal{M}_{I_{\eta}}^N(z)=2\pi\lambda_b\eta_t^2 \int_{^{R_l}}^{_{R_m}} r\exp\Big\{ - 2\pi\lambda_b\\
&\times\int_{r}^{R_m} \Big[1 - \Big(1 +\frac{\Omega}{m}zK_sP_sx^{-\alpha_s}\Big)^{-m} \Big]xdx - \pi\lambda_b\eta_t^2r^2\Big\}dr.
\end{split}
\end{equation}

\section*{ACKNOWLEDGMENTS}
This work was supported in part by the Natural Science Foundation General Program of China (U1764263 and 61671186) and Taiwan Ministry of Science and Technology (106-2221-E-006-028-MY3 and 106-2221-E-006-021-MY3).

\balance

\vfill

\newpage
\begin{center}
{\Huge Supplement for ``Performance Analysis of Joint Transmission Schemes in Ultra-Dense Networks - An Unified Approach''}
\end{center}

\vspace{0.32in}
\begin{abstract}
This document will present the detailed calculation process for the submitted journal paper ``Performance Analysis of Joint Transmission Schemes in Ultra-Dense Networks -  An Unified Approach''.
\end{abstract}

\setcounter{section}{0}
\section{Brief Introduction of the Journal Paper}
Ultra-dense network (UDN) is one of the enabling technologies to achieve 1000-fold capacity increase in 5G communication systems, and the application of joint transmission (JT) is an effective method to deal with severe inter-cell interferences in UDNs. However, most works done for performance analysis on JT schemes in the literature were based largely on simulation results due to the difficulties in quantitatively identifying the numbers of desired and interfering transmitters. In this work, we are motivated to propose an analytical approach to investigate the performance of JT schemes with a unified approach based on stochastic geometry, which is in particular useful for studying different JT methods and conventional transmission schemes without JT. Using the proposed approach, we can unveil the statistical characteristics (i.e., expectation, moment generation function, variance) of desired signal and interference powers of a given user equipment (UE), and thus system performances, such as average signal-to-interference-plus-noise ratio (SINR), and area spectral efficiency, can be evaluated analytically. The simulation results are used to verify the effectiveness of the proposed unified approach.

\section{Detailed Calculation Process}
In this section, we will calculate the expectation, MGF, and variance of both desired signal and interference powers for different transmission schemes.

\subsection{Expectation}

\begin{itemize}
\item \textbf{NoJT:} if no CoMP method is applied, i.e., UE is served by the nearest BS. The average received desired power can be calculated based one either traditional method or our proposed method. The result shows that these two methods are equivalent, which validates the effectiveness of the proposed method.

    In traditional method, the average desired power is
\begin{equation}
\begin{split}
\mathbb{E}[P_1]
& = \int_0^\infty\int_{R_l}^{R_m}{K_sh P_s r^{-\alpha_s}f_1(r)drf_h(h)dh} \\
& = \int_{R_l}^{R_m}{K_s P_s\mathbb{E}[h] r^{-\alpha_s} 2\pi\lambda_b r \exp{(-\pi \lambda_b r^2)}dr} \\
& = 2\pi\lambda_b K_s\mathbb{E}[h] P_s \times \frac{1}{2} \int_{R_l}^{R_m}{ {(r^2)}^{-\alpha_s/2}  \exp{(-\pi \lambda_b r^2)}dr^2} \\
& = \pi\lambda_b K_s\mathbb{E}[h] P_s \int_{\pi\lambda_bR_l^2}^{\pi\lambda_bR_m^2} {(s/{\pi\lambda_b})}^{-\alpha_s/2}  \exp{(-s)}d(s/{\pi\lambda_b)} \\
& = {(\pi\lambda_b)}^{\alpha_s/2} K_s\mathbb{E}[h] P_s \int_{\pi\lambda_bR_l^2}^{\pi\lambda_bR_m^2} {(s)}^{-\alpha_s/2}  \exp{(-s)}ds \\
& = {(\pi\lambda_b)}^{\alpha_s/2} K_s\mathbb{E}[h] P_s \left[ \gamma(1-\alpha_s/2, \pi\lambda_bR_m^2) -\gamma(1-\alpha_s/2, \pi\lambda_bR_l^2) \right],
\end{split}
\end{equation}
where $\gamma(s,x)$ is the lower incomplete Gamma function.

Using the proposed method, if $\alpha_s\neq2$, the average received power is
\begin{equation}
\begin{split}
\mathbb{E}[P_1]=&\int_0^\infty\int_{R_l}^{R_m}\frac{2\pi\lambda_b K_sh P_s}{2-\alpha_s}(r^{-\alpha_s+2}-R_l^{-\alpha_s+2})f_1(r)drf_h(h)dh\\
=&\frac{4\pi^2\lambda_b^2 K_s\mathbb{E}[h]  P_s}{2-\alpha_s}\bigg(\int_{R_l}^{R_m} r^{-\alpha_s+3}e^{-\pi\lambda_b r^2}dr-R_l^{-\alpha_s+2}\int_{R_l}^{R_m} re^{-\pi\lambda_b r^2}dr\bigg)\\
=&\frac{4\pi^2\lambda_b^2 K_s \mathbb{E}[h]  P_s}{2-\alpha_s}\bigg[\frac{\gamma(\frac{4-\alpha_s}{2},\pi\lambda_bR_m^2)-\gamma(\frac{4-\alpha_s}{2},\pi\lambda_bR_l^2)}{2(\pi\lambda_b)^{\frac{4-\alpha_s}{2}}}\\
&+R_l^{-\alpha_s+2}\frac{1}{2\pi\lambda_b}(e^{-\pi\lambda_bR_m^2}-e^{-\pi\lambda_bR_0^2})\bigg]\\
=&\frac{2K_s \mathbb{E}[h]  P_s}{2-\alpha_s}\bigg\{(\pi\lambda_b)^{\alpha_s/2}\bigg[\gamma(\frac{4-\alpha_s}{2},\pi\lambda_bR_m^2)-\gamma(\frac{4-\alpha_s}{2},\pi\lambda_bR_l^2)\bigg]\\
&+\pi\lambda_bR_l^{-\alpha_s+2}(e^{-\pi\lambda_bR_m^2}-e^{-\pi\lambda_bR_l^2})\bigg\}.\\
\end{split}
\end{equation}

Based on the following equation
\begin{equation}
\begin{split}
\int_a^b x\ln x e^{-cx^2}dx&=-\frac{\ln x}{2c}e^{-cx^2}|_a^b
-\int_a^b\frac{1}{x}-\frac{e^{-cx^2}}{2c}dx\\
&=\frac{\ln a e^{-ca^2}-\ln b e^{-cb^2}}{2c}+\frac{1}{4c}\int_{ca^2}^{cb^2}t^{-1}e^{-t}dt\\
&=\frac{\ln a e^{-ca^2}-\ln b e^{-cb^2}}{2c}+\frac{1}{4c}\Gamma(0,ca^2,cb^2),\\
\end{split}
\end{equation}

we can obtain the result when $\alpha_s=2$, which is
\begin{equation}
\begin{split}
\mathbb{E}[P_1]=&\int_0^\infty\int_{R_l}^{R_m}2\pi\lambda_b K_sh P_s(\ln r-\ln R_l)f_1(r)drf_h(h)dh\\
=&4\pi^2\lambda_b^2 K_s\mathbb{E}[h] P_s\bigg(\int_{R_l}^{R_m} \ln rre^{-\pi\lambda_b r^2}dr-\ln R_l\int_{R_l}^{R_m} re^{-\pi\lambda_b r^2}dr\bigg)\\
=&4\pi^2\lambda_b^2 K_s \mathbb{E}[h] P_s\bigg[\frac{\ln R_l e^{-\pi\lambda_bR_l^2}-\ln R_m e^{-\pi\lambda_bR_m^2}}{2\pi\lambda_b}+\frac{1}{4\pi\lambda_b}\Gamma(0,\pi\lambda_bR_l^2,\pi\lambda_bR_m^2)\\
&+\ln R_l\frac{1}{2\pi\lambda_b}(e^{-\pi\lambda_bR_m^2}-e^{-\pi\lambda_bR_l^2})\bigg]\\
=&4\pi^2\lambda_b^2 K_s \mathbb{E}[h] P_s\bigg[\frac{\ln \frac{R_l}{R_m} e^{-\pi\lambda_bR_m^2}}{2\pi\lambda_b}+\frac{1}{4\pi\lambda_b}\Gamma(0,\pi\lambda_bR_l^2,\pi\lambda_bR_m^2)\bigg]\\
=&\pi\lambda_b K_s \mathbb{E}[h] P_s\Big[2\ln \frac{R_l}{R_m} e^{-\pi\lambda_bR_m^2}+\Gamma(0,\pi\lambda_bR_l^2,\pi\lambda_bR_m^2)\Big].\\
\end{split}
\end{equation}

Same result can be obtained for both of these methods based on the relation $\gamma(s+1,x)=s\gamma(s,x)-x^se^{-x}$.

Similarly, if $\alpha_s\neq2$, the average interfering power is

\begin{equation}
\begin{split}
\mathbb{E}[I_1]=&\int_0^\infty\int_{R_l}^{R_m}\frac{2\pi\lambda_b K_s hP_s}{2-\alpha_s}(R_m^{-\alpha_s+2}-r^{-\alpha_s+2})f_1(r)drf_h(h)dh\\
=&\frac{4\pi^2\lambda_b^2 K_s \mathbb{E}[h] P_s}{2-\alpha_s}\bigg(R_m^{-\alpha_s+2}\int_{R_l}^{R_m} re^{-\pi\lambda_b r^2}dr-\int_{R_l}^{R_m} r^{-\alpha_s+3}e^{-\pi\lambda_b r^2}dr\\
&-R_l^{-\alpha_s+2}\int_{R_l}^{R_m} re^{-\pi\lambda_b r^2}dr\bigg)\\
=&\frac{2K_s \mathbb{E}[h] P_s}{2-\alpha_s}\bigg\{\pi\lambda_bR_m^{-\alpha_s+2}(e^{-\pi\lambda_bR_l^2}-e^{-\pi\lambda_bR_m^2})\\
&+(\pi\lambda_b)^{\alpha_s/2}\bigg[\gamma(\frac{4-\alpha_s}{2},\pi\lambda_bR_l^2)-\gamma(\frac{4-\alpha_s}{2},\pi\lambda_bR_m^2)\bigg]\bigg\}.\\
\end{split}
\end{equation}
If $\alpha_s=2$,
\begin{equation}
\begin{split}
\mathbb{E}[I_1]=&\int_0^\infty\int_{R_l}^{R_m}2\pi\lambda_b K_sh P_s(\ln R_m-\ln r)f_1(r)drf_h(h)dh\\
=&4\pi^2\lambda_b^2 K_s\mathbb{E}[h] P_s\bigg(\ln R_m\int_{R_l}^{R_m}re^{-\pi\lambda_b r^2}dr-\int_{R_l}^{R_m}\ln r re^{-\pi\lambda_b r^2}dr\bigg)\\
=&4\pi^2\lambda_b^2 K_s \mathbb{E}[h] P_s\bigg[\ln R_m\frac{1}{2\pi\lambda_b}(e^{-\pi\lambda_bR_l^2}-e^{-\pi\lambda_bR_m^2})\\
&-\frac{\ln R_l e^{-\pi\lambda_bR_l^2}-\ln R_m e^{-\pi\lambda_bR_m^2}}{2\pi\lambda_b}-\frac{1}{4\pi\lambda_b}\Gamma(0,\pi\lambda_bR_l^2,\pi\lambda_bR_m^2)\bigg]\\
=&4\pi^2\lambda_b^2 K_s \mathbb{E}[h] P_s\bigg[\frac{\ln \frac{R_m}{R_l} e^{-\pi\lambda_bR_l^2}}{2\pi\lambda_b}-\frac{1}{4\pi\lambda_b}\Gamma(0,\pi\lambda_bR_l^2,\pi\lambda_bR_m^2)\bigg]\\
=&\pi\lambda_b K_s \mathbb{E}[h] P_s\Big[2\ln \frac{R_m}{R_l} e^{-\pi\lambda_bR_l^2}-\Gamma(0,\pi\lambda_bR_l^2,\pi\lambda_bR_m^2)\Big].\\
\end{split}
\end{equation}

\item\textbf{2NS:} a UE is served by two nearest SBSs. If $\alpha_s\neq 2$, the average aggregate desired power is
\begin{equation}
\begin{split}
\mathbb{E}[P_2]=&\int_0^\infty\int_{R_l}^{R_m}\frac{2\pi\lambda_b K_s hP_s}{2-\alpha_s}(r^{-\alpha_s+2}-R_l^{-\alpha_s+2})f_2(r)drf_h(h)dh\\
=&\frac{4\pi^3\lambda_b^3 K_s \mathbb{E}[h] P_s}{2-\alpha_s}\bigg(\int_{R_l}^{R_m} r^{-\alpha_s+5}e^{-\pi\lambda_b r^2}dr-R_l^{-\alpha_s+2}\int_{R_l}^{R_m} r^3e^{-\pi\lambda_b r^2}dr\bigg)\\
=&\frac{4\pi^3\lambda_b^3 K_s \mathbb{E}[h] P_s}{2-\alpha_s}\bigg\{\frac{\gamma(\frac{6-\alpha_s}{2},\pi\lambda_bR_m^2)-\gamma(\frac{6-\alpha_s}{2},\pi\lambda_bR_l^2)}{2(\pi\lambda_b)^{\frac{6-\alpha_s}{2}}}\\
&+R_l^{-\alpha_s+2}\frac{1}{2\pi^2\lambda_b^2}\Big[(\pi\lambda_bR_m^2+1)e^{-\pi\lambda_bR_m^2}-(\pi\lambda_bR_l^2+1)e^{-\pi\lambda_bR_l^2}\Big]\bigg\}\\
=&\frac{2K_s \mathbb{E}[h] P_s}{2-\alpha_s}\bigg\{(\pi\lambda_b)^{\alpha_s/2}\Big[\gamma(\frac{6-\alpha_s}{2},\pi\lambda_bR_m^2)-\gamma(\frac{6-\alpha_s}{2},\pi\lambda_bR_l^2)\Big]\\
&+\pi\lambda_bR_l^{-\alpha_s+2}\Big[(\pi\lambda_bR_m^2+1)e^{-\pi\lambda_bR_m^2}-(\pi\lambda_bR_l^2+1)e^{-\pi\lambda_bR_l^2}\Big]\bigg\}.\\
\end{split}
\end{equation}

If $\alpha_s=2$,
\begin{equation}
\begin{split}
\mathbb{E}[P_2]=&\int_0^\infty\int_{R_l}^{R_m}2\pi\lambda_b K_sh P_s(\ln r-\ln R_l)f_2(r)drf_h(h)dh\\
=&4\pi^3\lambda_b^3 K_s\mathbb{E}[h] P_s\bigg(\int_{R_l}^{R_m} \ln rr^3e^{-\pi\lambda_b r^2}dr-\ln R_l\int_{R_l}^{R_m} r^3e^{-\pi\lambda_b r^2}dr\bigg)\\
=&4\pi^3\lambda_b^3 K_s \mathbb{E}[h] P_s\bigg\{\frac{\ln R_l}{2\pi^2\lambda_b^2}\Big[(\pi\lambda_bR_m^2\!+\!1)e^{-\pi\lambda_bR_m^2}\\
&-(\pi\lambda_bR_l^2+1)e^{-\pi\lambda_bR_l^2}\Big]+\int_{R_l}^{R_m} \ln rr^3e^{-\pi\lambda_b r^2}dr\bigg\}.\\
\end{split}
\end{equation}

If $\alpha_s\neq 2$, the average aggregate interfering power is
\begin{equation}
\begin{split}
\mathbb{E}[I_2]=&\int_0^\infty\int_{R_l}^{R_m}\frac{2\pi\lambda_b K_s hP_s}{2-\alpha_s}(R_m^{-\alpha_s+2}-r^{-\alpha_s+2})f_2(r)drf_h(h)dh\\
=&\frac{2K_s H P_s}{2-\alpha_s}\bigg\{(\pi\lambda_b)^{\alpha_s/2}\Big[\gamma(\frac{6-\alpha_s}{2},\pi\lambda_bR_l^2)-\gamma(\frac{6-\alpha_s}{2},\pi\lambda_bR_m^2)\Big]\\
&+\pi\lambda_bR_m^{-\alpha_s+2}\Big[(\pi\lambda_bR_l^2+1)e^{-\pi\lambda_bR_l^2}-(\pi\lambda_bR_m^2+1)e^{-\pi\lambda_bR_m^2}\Big]\bigg\}.\\
\end{split}
\end{equation}

If $\alpha_s=2$,
\begin{equation}
\begin{split}
\mathbb{E}[I_2]=&\int_0^\infty\int_{R_l}^{R_m}2\pi\lambda_b K_sh P_s(\ln R_m-\ln r)f_2(r)drf_h(h)dh\\
=&4\pi^3\lambda_b^3 K_sh P_s\bigg(\ln R_m\int_{R_l}^{R_m} r^3e^{-\pi\lambda_b r^2}dr-\int_{R_l}^{R_m}\ln r r^3e^{-\pi\lambda_b r^2}dr\bigg)\\
=&4\pi^3\lambda_b^3 K_s h P_s\bigg\{\frac{\ln R_m}{2\pi^2\lambda_b^2}\Big[(\pi\lambda_bR_l^2+1)e^{-\pi\lambda_bR_l^2}-(\pi\lambda_bR_m^2+1)e^{-\pi\lambda_bR_m^2}\Big]\\
&-\int_{R_l}^{R_m} \ln rr^3e^{-\pi\lambda_b r^2}dr\bigg\}.\\
\end{split}
\end{equation}

\item \textbf{CD:} each SBS compares its distance to UE with a predetermined value $R_0$. If the distance is no longer than $R_0$, it is a cooperative SBS. The aggregate desired and interfering power are
\begin{equation}
\begin{split}
\mathbb{E}[P_{R_0}]=\left\{\begin{array}{ll}
\frac{2\pi\lambda_b K_s \mathbb{E}[h] P_s}{2-\alpha_s}(R_0^{-\alpha_s+2}-R_l^{-\alpha_s+2}),&\alpha_s\neq 2,\\
2\pi\lambda_b K_s \mathbb{E}[h] P_s\ln\frac{R_0}{R_l},&\alpha_s=2,\\
\end{array}\right.
\end{split}
\end{equation}
\begin{equation}
\mathbb{E}[I_{R_0}]=\left\{\begin{array}{ll}
\frac{2\pi\lambda_b K_s \mathbb{E}[h] P_s}{2-\alpha_s}(R_m^{-\alpha_s+2}-R_0^{-\alpha_s+2}),&\alpha_s\neq 2,\\
2\pi\lambda_b K_s \mathbb{E}[h] P_s\ln\frac{R_m}{R_0},&\alpha_s=2,\\
\end{array}\right.
\end{equation}
respectively.

\item \textbf{FPD:} if the highest average received power is $P_{max}$ dBm, other SBSs, whose received power are higher than $P_{max}-\eta$ dBm, are regarded as the cooperative SBSs, where $\eta$ dB is a predetermined value. If $\alpha_s\neq 2$, the average aggregate desired power is
\begin{equation}
\begin{split}
\mathbb{E}[P_\eta]=&\int_0^\infty\int_{\frac{R_l}{\eta_t}}^{R_m}\frac{2\pi\lambda_b K_s h P_s}{2-\alpha_s}\Big[r^{-\alpha_s+2}-(\frac{R_l}{\eta_t})^{-\alpha_s+2}\Big]f_\eta(r)drf_h(h)dh\\
=&\frac{4\pi^2\lambda_b^2\eta_t^2 K_s \mathbb{E}[h] P_s}{2-\alpha_s}\bigg[\int_{R_l/\eta_t}^{R_m} r^{-\alpha_s+3}e^{-\pi\lambda_b\eta_t^2 r^2}dr-(\frac{R_l}{\eta_t})^{-\alpha_s+2}\int_{R_l/\eta_t}^{R_m} re^{-\pi\lambda_b\eta_t^2 r^2}dr\bigg]\\
=&\frac{4\pi^2\lambda_b^2\eta_t^2 K_s \mathbb{E}[h] P_s}{2-\alpha_s}\bigg[\frac{\gamma(\frac{4-\alpha_s}{2},\pi\lambda_b\eta_t^2R_m^2)-\gamma(\frac{4-\alpha_s}{2},\pi\lambda_bR_l^2)}{2(\pi\lambda_b\eta_t^2)^{\frac{4-\alpha_s}{2}}}\\
&+R_l^{-\alpha_s+2}\frac{1}{2\pi\lambda_b\eta_t^{-\alpha_s+4}}(e^{-\pi\lambda_b\eta_t^2R_m^2}-e^{-\pi\lambda_bR_l^2})\bigg]\\
=&\frac{2K_s \mathbb{E}[h] P_s}{2-\alpha_s}\bigg\{\eta_t^{-2}(\pi\lambda_b\eta_t^2)^{\alpha_s/2}\Big[\gamma(\frac{4-\alpha_s}{2},\pi\lambda_b\eta_t^2R_m^2)-\gamma(\frac{4-\alpha_s}{2},\pi\lambda_bR_l^2)\Big]\\
&+\pi\lambda_b(\frac{R_l}{\eta_t})^{-\alpha_s+2}(e^{-\pi\lambda_b\eta_t^2R_m^2}-e^{-\pi\lambda_bR_l^2})\bigg\}.\\
\end{split}
\end{equation}

If $\alpha_s=2$,
\begin{equation}
\begin{split}
\mathbb{E}[P_\eta]=&\int_0^\infty\int_{\frac{R_l}{\eta_t}}^{R_m}2\pi\lambda_b K_sh P_s(\ln r-\ln \frac{R_l}{\eta_t})f_\eta(r)drf_h(h)dh\\
=&4\pi^2\lambda_b^2 \eta_t^2K_s\mathbb{E}[h] P_s\bigg(\int_{\frac{R_l}{\eta_t}}^{R_m} \ln rre^{-\pi\lambda_b \eta_t^2r^2}dr-\ln \frac{R_l}{\eta_t}\int_{\frac{R_l}{\eta_t}}^{R_m} re^{-\pi\lambda_b\eta_t^2 r^2}dr\bigg)\\
=&4\pi^2\lambda_b^2\eta_t^2 K_s \mathbb{E}[h] P_s\bigg[\frac{\ln\frac{R_l}{\eta_t} e^{-\pi\lambda_bR_l^2}-\ln R_m e^{-\pi\lambda_bR_m^2}}{2\pi\lambda_b}+\frac{1}{4\pi\lambda_b}\Gamma(0,\pi\lambda_bR_l^2,\pi\lambda_b\eta_t^2R_m^2)\\
&+\ln \frac{R_l}{\eta_t}\frac{1}{2\pi\lambda_b}(e^{-\pi\lambda_b\eta_t^2R_m^2}-e^{-\pi\lambda_bR_l^2})\bigg]\\
=&4\pi^2\lambda_b^2 K_s \mathbb{E}[h] P_s\bigg[\frac{\ln \frac{R_l}{\eta_tR_m} e^{-\pi\lambda_b\eta_t^2R_m^2}}{2\pi\lambda_b}+\frac{1}{4\pi\lambda_b}\Gamma(0,\pi\lambda_bR_l^2,\pi\lambda_b\eta_t^2R_m^2)\bigg]\\
=&\pi\lambda_b K_s \mathbb{E}[h] P_s\bigg[2\ln \frac{R_l}{\eta_tR_m} e^{-\pi\lambda_b\eta_t^2R_m^2}+\Gamma(0,\pi\lambda_bR_l^2,\pi\lambda_b\eta_t^2R_m^2)\bigg].\\
\end{split}
\end{equation}

If $\alpha_s\neq 2$, the average interfering power is
\begin{equation}
\begin{split}
\mathbb{E}[I_\eta]=&\int_0^\infty\int_{R_l/\eta_t}^{R_m}\frac{2\pi\lambda_b K_s h P_s}{2-\alpha_s}(R_m^{-\alpha_s+2}-r^{-\alpha_s+2})f_\eta(r)drf_h(h)dh\\
=&\frac{2K_s \mathbb{E}[h] P_s}{2-\alpha_s}\bigg\{\eta_t^{-2}(\pi\lambda_b\eta_t^2)^{\alpha_s/2}\Big[\gamma(\frac{4-\alpha_s}{2},\pi\lambda_bR_l^2)-\gamma(\frac{4-\alpha_s}{2},\pi\lambda_b\eta_t^2R_m^2)\Big]\\
&+\pi\lambda_b(\frac{R_m}{\eta_t})^{-\alpha_s+2}(e^{-\pi\lambda_bR_l^2}-e^{-\pi\lambda_b\eta_t^2R_m^2})\bigg\}\\
=&\frac{2K_s \mathbb{E}[h] P_s}{2-\alpha_s}\bigg[\pi\lambda_b(\frac{R_m}{\eta_t})^{-\alpha_s+2}(e^{-\pi\lambda_bR_l^2}-e^{-\pi\lambda_b\eta_t^2R_m^2})\\
&-\eta_t^{-2}(\pi\lambda_b\eta_t^2)^{\alpha_s/2}\Gamma(\frac{4-\alpha_s}{2},\pi\lambda_bR_l^2,\pi\lambda_b\eta_t^2R_m^2)\bigg].\\
\end{split}
\end{equation}

If $\alpha_s=2$,
\begin{equation}
\begin{split}
\mathbb{E}[I_\eta]=&\int_0^\infty\int_{\frac{R_l}{\eta_t}}^{R_m}2\pi\lambda_b K_sh P_s(\ln R_m-\ln r )f_\eta(r)drf_h(h)dh\\
=&4\pi^2\lambda_b^2 \eta_t^2K_s\mathbb{E}[h] P_s\bigg(\int_{\frac{R_l}{\eta_t}}^{R_m} \ln R_mre^{-\pi\lambda_b \eta_t^2r^2}dr-\int_{\frac{R_l}{\eta_t}}^{R_m} \ln r re^{-\pi\lambda_b\eta_t^2 r^2}dr\bigg)\\
=&4\pi^2\lambda_b^2 K_s \mathbb{E}[h] P_s\bigg[\ln R_m\frac{1}{2\pi\lambda_b}(e^{-\pi\lambda_bR_l^2}-e^{-\pi\lambda_b\eta_t^2R_m^2})\\
&-\frac{\ln\frac{R_l}{\eta_t} e^{-\pi\lambda_bR_l^2}-\ln R_m e^{-\pi\lambda_b\eta_t^2R_m^2}}{2\pi\lambda_b}-\frac{1}{4\pi\lambda_b}\Gamma(0,\pi\lambda_bR_l^2,\pi\lambda_b\eta_t^2R_m^2)\bigg]\\
=&4\pi^2\lambda_b^2 K_s \mathbb{E}[h] P_s\Big[\frac{\ln \frac{\eta_tR_m}{R_l} e^{-\pi\lambda_bR_l^2}}{2\pi\lambda_b}-\frac{1}{4\pi\lambda_b}\Gamma(0,\pi\lambda_bR_l^2,\pi\lambda_b\eta_t^2R_m^2)\Big]\\
=&\pi\lambda_b\eta_t^2 K_s \mathbb{E}[h] P_s\Big[2\ln \frac{\eta_tR_m}{R_l} e^{-\pi\lambda_bR_l^2}-\Gamma(0,\pi\lambda_bR_l^2,\pi\lambda_b\eta_t^2R_m^2)\Big].\\
\end{split}
\end{equation}
\end{itemize}
\subsection{Moment generating function}
The moment-generating function (MGF) of a real-valued random variable is an alternative specification of its probability distribution. Thus, it provides the basis of an alternative route to analytical results compared with working directly with probability density functions or cumulative distribution functions. There are particularly simple results for the moment-generating functions of distributions defined by the weighted sums of random variables. MGF is defined as $\mathcal{M}_{X}(z)=\mathbb{E}[e^{-zX}]$, and its relation with the PDF is $\mathcal{M}_{X}(z)=\int_{-\infty}^{\infty}e^{-zX}f_X(x)dx$. In other words, MGF is the Laplace transform of PDF.

For BSs distributed following PPP, we can obtain the MGF of aggregated desired or interfering power by using the property
\begin{equation}
\mathbb{E}[\prod_{x\in\Phi}f(x)]=\exp\Big\{-2\pi\lambda\int_\Phi [1-f(x)]xdx\Big\}.
\end{equation}

If $h$ is constant, applying the Campbell theorem of PPP, the MGF for the aggregated power in the range $[R_1,R_2]$ is
\begin{equation}
\begin{split}
\mathcal{M}(z)&=\exp\Big\{-2\pi\lambda_b\int_{R_1}^{R_2} \big[1-\exp(-zK_shP_sr^{-\alpha_s})\big]rdr\Big\}\\
&=\textrm{exp}\bigg\{2\pi\lambda_b\Big[\frac{ (K_shP_s)^{2/\alpha_s}}{\alpha_s}\Gamma\Big(-\frac{2}{\alpha_s},zK_shP_sR_{2}^{-\alpha_s},zK_shP_sR_{1}^{-\alpha_s}\Big)\!-\!\frac{R_{2}^2}{2}+\frac{R_{1}^2}{2}\Big]\bigg\}.
\end{split}
\end{equation}

In a Rayleigh fading channel, the result is
\begin{equation}
\begin{split}
\mathcal{M}(z) &=\mathbb{E}_{r}\Big[\prod\frac{1}{1+zK_sr^{-\alpha_s}P_{s}}\Big]\\
&=\exp\bigg[-2\pi\lambda_b\int_{R_1}^{R_2} \Big(1-\frac{1}{1+zK_sr^{-\alpha_s}P_{s}}\Big)rdr\bigg]\\
&=\exp\bigg[-2\pi\lambda_b\int_{R_1}^{R_2}\frac{zK_sP_s}{r^{\alpha_s}+zK_sP_s}dr\bigg]\\
&=\exp\bigg[-2\pi\lambda_b(zK_sP_s)^{2/\alpha_s}\int_{\frac{R_1}{(zK_sP_s)^{1/\alpha_s}}}^{\frac{R_2}{(zK_sP_s)^{1/\alpha_s}}}\frac{x}{x^{\alpha_s}+1}dx\bigg].\\
\end{split}
\end{equation}
If $\alpha_s$ is a positive integer, a closed form expression can be obtained. For example, if $\alpha_s=2$, it is
\begin{equation}
\begin{split}
\mathcal{M}(z)&=\exp\bigg[-2\pi\lambda_bzK_sP_s\int_{\frac{R_1}{(zK_sP_s)^{1/2}}}^{\frac{R_2}{(zK_sP_s)^{1/2}}}\frac{x}{x^{2}+1}dx\bigg]\\
&\xlongequal[\quad]{u=x^2+1}\exp\bigg(-\pi\lambda_bzK_sP_s\int_{\frac{R_1^2}{zK_sP_s}+1}^{\frac{R_2^2}{zK_sP_s}+1}\frac{1}{u}du\bigg)\\
&=\exp\bigg(-\pi\lambda_bzK_sP_s\ln\frac{R_2^2+zK_sP_s}{R_1^2+zK_sP_s}\bigg)\\
&=\Big(\frac{R_2^2+zK_sP_s}{R_1^2+zK_sP_s}\Big)^{-\pi\lambda_bzK_sP_s}.
\end{split}
\end{equation}
If $\alpha_s=4$, it is
\begin{equation}
\begin{split}
\mathcal{M}(z) &=\exp\Big[-2\pi\lambda_b(zK_sP_s)^{1/2}\int_{\frac{R_1}{(zK_sP_s)^{1/4}}}^{\frac{R_2}{(zK_sP_s)^{1/4}}}\frac{x}{x^4+1}dxBig]\\
&\xlongequal[\quad]{u=x^2}\exp\Big[-\pi\lambda_b(zK_sP_s)^{1/2}\int_{\frac{R_1^2}{(zK_sP_s)^{1/2}}}^{\frac{R_2^2}{(zK_sP_s)^{1/2}}}\frac{1}{x^2+1}dx\Big]\\
&=\exp\bigg\{-\pi\lambda_b(zK_sP_s)^{1/2}\Big[\arctan\frac{R_2^2}{(zK_sP_s)^{1/2}}-\arctan\frac{R_1^2}{(zK_sP_s)^{1/2}}\Big]\bigg\}.\\
\end{split}
\end{equation}

If $R_2\rightarrow\infty$, substituting $t=\frac{2zK_sP_sx^{\alpha_s}}{\alpha_s R_2^{\alpha_s}}$, we can get
\begin{equation}
\mathcal{M}(z)=\exp\Big[-\frac{2\pi\lambda_bzK_sP_sR_1^{2-\alpha_s}}{\alpha_s-2}_2F_1(1,1-\frac{2}{\alpha_s};2-\frac{2}{\alpha_s};-\frac{zK_sP_s}{R_1^{\alpha_s}})\Big].
\end{equation}
If $R_1\rightarrow0$,
\begin{equation}
\mathcal{M}(z)=\exp\Big[-\frac{2\pi\lambda_bzR_2^{2-\alpha_s}}{2}_2F_1(1,\frac{2}{\alpha_s};1+\frac{2}{\alpha_s};-\frac{R_2^{\alpha_s}}{zK_sP_s})\Big].
\end{equation}
The relation is based on the equations [3.194] in \cite{Integral7th}.

In a Nakagami-m fading channel, the result is
\begin{equation}
\begin{split}
\mathcal{M}(z) &=\mathbb{E}_{r}\Big[\prod\int_0^\infty\exp(-zK_sP_shr^{-\alpha_s})f_{h_N}(h)dh\Big]\\
&=\mathbb{E}_{r}\Big[(1+\frac{\Omega}{m}zK_sP_sr^{-\alpha_s})^{-m}\Big]\\
&=\exp\bigg\{-2\pi\lambda_b\int_{R_1}^{R_2} \Big[1-(1+\frac{\Omega}{m}zK_sP_sr^{-\alpha_s})^{-m}\Big]rdr\bigg\}.\\
\end{split}
\end{equation}
If $\alpha_s=2$ and $m$ is a positive integer, a closed form expression can be obtained based on the equations [2.117] in \cite{Integral7th}. Here, we present the case of $m=2$ as an example, which is
\begin{equation}
\begin{split}
\mathcal{M}(z)=&\exp\bigg[-\pi\lambda_b\int_{R_1^{-2}}^{R_2^{-2}} \big(1+\frac{\Omega}{2}zK_sP_su\big)^{-2}u^{-2}du+\pi\lambda_b(R_1^2-R_2^2)\bigg]\\
=&\exp\bigg\{-\pi\lambda_b\Big[-\Big(\frac{1}{u}+\Omega zK_sP_s\Big)\Big(1+\frac{\Omega zK_sP_s}{2}u\Big)^{-1}\\
&+\Omega zK_sP_s\ln\frac{1+\frac{\Omega zK_sP_s}{2}u}{u}\Big]\Big|_{R_1^{-2}}^{R_2^{-2}}
+\pi\lambda_b(R_1^2-R_2^2)\bigg\}\\
=&\exp\bigg\{-\pi\lambda\Big[2G\ln\frac{R_1^{-2}(GR_2^{-2}+1)}{R_2^{-2}(GR_1^{-2}+1)}-\frac{2GR_2^{-2}+1}{R_2^{-2}(GR_2^{-2}+1)}+\frac{2GR_1^{-2}+1}{R_1^{-2}(GR_1^{-2}+1)}\\
&+R_2^2-R_1^2\Big]\bigg\}\\
=&\exp\bigg[-\pi\lambda\Big(2G\ln\frac{G+R_2^2}{G+R_1^2}-\frac{2G+R_2^{2}}{GR_2^{-2}+1}+\frac{2G+R_1^{2}}{GR_1^{-2}+1}+R_2^2-R_1^2\Big)\bigg]\\
=&\exp\bigg[\pi\lambda\Big(\frac{G}{GR_2^{-2}+1}-\frac{G}{GR_1^{-2}+1}-2G\ln\frac{G+R_2^2}{G+R_1^2}\Big)\bigg]\\
=&\exp\bigg\{\pi\lambda\Big[\frac{G^2(R_2^2-R_1^2)}{(G+R_1^2)(G+R_2^2)}-2G\ln\frac{G+R_2^2}{G+R_1^2}\Big]\bigg\},\\
\end{split}
\end{equation}
where $G=\Omega z K_sP_s$.

Thus, the MGF for different transmission schemes can be obtained as follows.
\begin{itemize}
\item \textbf{NoJT:} if $h$ is a constant, using traditional calculation method, the MGF for aggregate desired power is
\begin{equation}
\begin{split}
\mathcal{M}_{P_1}(z)&=\int_{R_l}^{R_m}\exp(-zK_shP_sr^{-\alpha_s})f_1(r)dr\\
&=2\pi\lambda_b \int_{R_l}^{R_m}r\exp(-zK_shP_sr^{-\alpha_s}-\pi\lambda_br^2)dr.\\
\end{split}
\end{equation}

In a Rayleigh fading channel, the result is
\begin{equation}
\begin{split}
\mathcal{M}_{P_1}(z)&=\int_{R_l}^{R_m}\frac{1}{1+zK_sP_sr^{-\alpha_s}}f_1(r)dr\\
&=2\pi\lambda_b \int_{R_l}^{R_m}\frac{r}{1+zK_sP_sr^{-\alpha_s}}\exp(-\pi\lambda_br^2)dr.\\
\end{split}
\end{equation}
If $\alpha_s=2$, a closed-form expression can be obtained, which is
\begin{equation}
\begin{split}
\mathcal{M}_{P_1}(z)&=\mathbb{E}_{r}\bigg[\frac{1}{1+zK_shP_{s}r^{-2}}\bigg]\\
&=2c\int_{R_l}^{R_m}\frac{x^3e^{-cx^2}}{x^2+b}dx\\
&\xlongequal[\quad]{u=x^2}c\int_{R_l^2}^{R_m^2}\frac{ue^{-cu}}{u+b}du\\
&=c\bigg[-e^{bc}u\textrm{E}_{\textrm{i}}(cu+bc)\Big|_{R_l^2}^{R_m^2}-\int_{R_l^2}^{R_m^2}\!\!-e^{bc}\textrm{E}_\textrm{i}(cu+bc)du\bigg]\\
&\!\xlongequal[\quad]{v=cu+bc}e^{bc}\bigg[-cR_m^2\textrm{E}_\textrm{i}(cR_m^2+bc)+cR_l^2\textrm{E}_\textrm{i}(cR_l^2+bc)+\int_{cR_l^2+bc}^{cR_m^2+bc}\textrm{E}_\textrm{i}(v)dv\bigg]\\
&=bc e^{bc}[\textrm{E}_\textrm{i}(cR_m^2+bc)-\textrm{E}_\textrm{i}(cR_l^2+bc)]-e^{-cR_m^2}+e^{-cR_l^2},\\
\end{split}
\end{equation}
where $b=zK_sP_s$, $c=\pi\lambda_b$ and $\textrm{E}_\textrm{i}(x)$ is the exponential integral function.

Using our proposed method, the result is
\begin{equation}
\begin{split}
\mathcal{M}_{P_{1}}(z)=&\int_{R_l}^{R_m}\exp\bigg\{-2\pi\lambda_b\int_{R_l}^{r}[1-\exp(-zK_shP_sx^{-\alpha_s})]xdx\bigg\}f_{1}(r)dr\\
=&2\pi\lambda_b\int_{R_l}^{R_m}\exp\Bigg\{2\pi\lambda_b\bigg\{\frac{ (zK_s P_{s})^{\frac{2}{\alpha_s}}}{\alpha_s}\Gamma\Big(-\frac{2}{\alpha_s},zK_sh P_{s}r^{-\alpha_s},\\
&zK_sh P_{s}R_l^{-\alpha_s}\Big)-\frac{r^2}{2}+\frac{R_l^2}{2}\bigg\}\Bigg\}r\exp(-\pi\lambda_br^2)dr\\
=&2\pi\lambda_b\int_{R_l}^{R_m}r\exp\Bigg\{2\pi\lambda_b\bigg[\frac{ (zK_sh P_{s})^{\frac{2}{\alpha_s}}}{\alpha_s}\Gamma\Big(\!-\frac{2}{\alpha_s},zK_sh P_{s}R_m^{-\alpha_s},zK_sh P_{s}r^{-\alpha_s}\Big)\\
&-r^2+\frac{R_l^2}{2}\bigg]\Bigg\}dr.\\
\end{split}
\end{equation}

In a Rayleigh fading channel, the result is
\begin{equation}
\begin{split}
\mathcal{M}_{P_1}(z)&=\int_{R_l}^{R_m}\exp\bigg[-2\pi\lambda_b\int_{R_l}^{r} \Big(1-\frac{1}{1+zK_sx^{-2}P_{s}}\Big)xdx\bigg]f_1(r)dr\\
&=2\pi\lambda_b\int_{R_l}^{R_m}r\exp\Bigg[-2\pi\lambda_b(zK_sP_s)^{2/\alpha_s}\int_{\frac{R_l}{(zK_sP_s)^{1/\alpha_s}}}^{\frac{r}{(zK_sP_s)^{1/\alpha_s}}}\frac{x}{x^{\alpha_s}+1}dx-\pi\lambda_br^2\Bigg]dr.\\
\end{split}
\end{equation}

In a Nakagami-m fading channel, the result is
\begin{equation}
\begin{split}
\mathcal{M}_{P_1}(z) &=\int_{R_l}^{R_m}\exp\bigg\{-2\pi\lambda_b\int_{R_1}^{r} \Big[1-(1+\frac{\Omega}{m}zK_sP_sx^{-\alpha_s})^{-m}\Big]xdx\bigg\}f_1(r)dr\\
&=\int_{R_l}^{R_m}2\pi\lambda_br\exp\bigg\{-2\pi\lambda_b\int_{R_1}^{r} \Big[1-(1+\frac{\Omega}{m}zK_sP_sx^{-\alpha_s})^{-m}\Big]xdx-\pi\lambda_br^2\bigg\}dr.\\
\end{split}
\end{equation}

If $h$ is a constant, MGF for interfering power is
\begin{equation}
\begin{split}
\mathcal{M}_{I_{1}}(z)=&\int_{R_l}^{R_m}\exp\bigg\{-2\pi\lambda_b\int_{r}^{R_m}[1-\exp(-zK_shP_sx^{-\alpha_s})]xdx\bigg\}f_{1}(r)dr\\
=&2\pi\lambda_b\int_{R_l}^{R_m}\exp\Bigg\{2\pi\lambda_b\bigg\{\frac{ (zK_s h P_{s})^{\frac{2}{\alpha_s}}}{\alpha_s}\Big[\Gamma\Big(-\frac{2}{\alpha_s},zK_s hP_{s}R_m^{-\alpha_s}\Big)-\\
&\Gamma\Big(-\frac{2}{\alpha_s},zK_sh P_{s}r^{-\alpha_s}\Big)\Big]-\frac{R_m^2}{2}+\frac{r^2}{2}\bigg\}\Bigg\}r\exp(-\pi\lambda_br^2)dr\\
=&2\pi\lambda_b\int_{R_l}^{R_m}r\exp\Bigg\{2\pi\lambda_b\bigg\{\frac{ (zK_sh P_{s})^{\frac{2}{\alpha_s}}}{\alpha_s}\Big[\Gamma\Big(-\frac{2}{\alpha_s},zK_sh P_{s}R_m^{-\alpha_s}\Big)-\\
&\Gamma\Big(-\frac{2}{\alpha_s},zK_sh P_{s}r^{-\alpha_s}\Big)\Big]-\frac{R_m^2}{2}\bigg\}\Bigg\}dr\\
=&2\pi\lambda_b\int_{R_l}^{R_m}r\exp\Bigg\{2\pi\lambda_b\bigg[\frac{ (zK_sh P_{s})^{\frac{2}{\alpha_s}}}{\alpha_s}\Gamma\Big(-\frac{2}{\alpha_s},zK_sh P_{s}R_m^{-\alpha_s},\\
&zK_sh P_{s}r^{-\alpha_s}\Big)-\frac{R_m^2}{2}\bigg]\Bigg\}dr.\\
\end{split}
\end{equation}

In a Rayleigh fading channel, the result is
\begin{equation}
\begin{split}
\mathcal{M}_{I_1}(z)&=\int_{R_l}^{R_m}\exp\bigg[-2\pi\lambda_b\int_{r}^{R_m} \Big(1-\frac{1}{1+zK_sx^{-2}P_{s}}\Big)xdx\bigg]f_1(r)dr\\
&=2\pi\lambda_b\int_{R_l}^{R_m}r\exp\Bigg[-2\pi\lambda_b(zK_sP_s)^{2/\alpha_s}\int_{\frac{r}{(zK_sP_s)^{1/\alpha_s}}}^{\frac{R_m}{(zK_sP_s)^{1/\alpha_s}}}\frac{x}{x^{\alpha_s}+1}dx-\pi\lambda_br^2\Bigg]dr.\\
\end{split}
\end{equation}

In a Nakagami-m fading channel, the result is
\begin{equation}
\begin{split}
\mathcal{M}_{I_1}(z) &=\int_{R_l}^{R_m}\exp\bigg\{-2\pi\lambda_b\int_{r}^{R_m} \Big[1-(1+\frac{\Omega}{m}zK_sP_sx^{-\alpha_s})^{-m}\Big]xdx\bigg\}f_1(r)dr\\
&=2\pi\lambda_b\int_{R_l}^{R_m}r\exp\bigg[2\pi\lambda_b\int_{r}^{R_m} (1+\frac{\Omega}{m}zK_sP_sx^{-\alpha_s})^{-m}xdx-\pi\lambda_bR_m^2\bigg]dr.\\
\end{split}
\end{equation}
\item \textbf{2NS:} if $h$ is a constant, MGF for desired power is

\begin{equation}
\begin{split}
\mathcal{M}_{P_{2}}(z)=&\int_{R_l}^{R_m}\exp\bigg\{-2\pi\lambda_b\int_{R_l}^{r}[1-\exp(-zK_shP_sx^{-\alpha_s})]xdx\bigg\}f_{2}(r)dr\\
=&2\pi^2\lambda_b^2\int_{R_l}^{R_m}\exp\Bigg\{2\pi\lambda_b\bigg\{\frac{ (zK_sh P_{s})^{\frac{2}{\alpha_s}}}{\alpha_s}\Big[\Gamma\Big(-\frac{2}{\alpha_s},zK_sh P_{s}r^{-\alpha_s}\Big)-\\
&\Gamma\Big(-\frac{2}{\alpha_s},zK_s h P_{s}R_l^{-\alpha_s}\Big)\Big]-\frac{r^2}{2}+\frac{R_l^2}{2}\bigg\}\Bigg\}r^3\exp(-\pi\lambda_br^2)dr\\
=&2\pi^2\lambda_b^2\exp\bigg\{\pi\lambda_b\bigg[R_l^2-\frac{ 2(zK_s h P_{s})^{\frac{2}{\alpha_s}}}{\alpha_s}\Gamma\Big(-\frac{2}{\alpha_s},zK_s hP_{s}R_l^{-\alpha_s}\Big)\bigg]\bigg\}\\
&\times\int_{R_l}^{R_m}r^3\exp\bigg\{2\pi\lambda_b\bigg\{\frac{ (zK_sh P_{s})^{\frac{2}{\alpha_s}}}{\alpha_s}\Gamma\Big(-\frac{2}{\alpha_s},zK_s hP_{s}r^{-\alpha_s}\Big)-r^2\bigg]\bigg\}dr\\
=&2\pi^2\lambda_b^2\int_{R_l}^{R_m}r^3\exp\Bigg\{2\pi\lambda_b\bigg[-\frac{ (zK_sh P_{s})^{\frac{2}{\alpha_s}}}{\alpha_s}\Gamma\Big(\frac{2}{\alpha_s},zK_sh P_{s}r^{-\alpha_s},\\
&zK_s h P_{s}R_l^{-\alpha_s}\Big)-r^2+\frac{R_l^2}{2}\bigg]\Bigg\}dr.\\
\end{split}
\end{equation}

In a Rayleigh fading channel, the result is
\begin{equation}
\begin{split}
\mathcal{M}_{P_2}(z)&=\int_{R_l}^{R_m}\exp\bigg[-2\pi\lambda_b\int_{R_l}^{r} \Big(1-\frac{1}{1+zK_sx^{-2}P_{s}}\Big)xdx\bigg]f_2(r)dr\\
&=2\pi^2\lambda_b^2\int_{R_l}^{R_m}r^3\exp\Bigg[-2\pi\lambda_b(zK_sP_s)^{2/\alpha_s}\int_{\frac{R_l}{(zK_sP_s)^{1/\alpha_s}}}^{\frac{r}{(zK_sP_s)^{1/\alpha_s}}}\frac{x}{x^{\alpha_s}+1}dx-\pi\lambda_br^2\Bigg]dr.\\
\end{split}
\end{equation}

In a Nakagami-m fading channel, the result is
\begin{equation}
\begin{split}
\mathcal{M}_{P_2}(z) &=\int_{R_l}^{R_m}\exp\bigg\{-2\pi\lambda_b\int_{R_1}^{r} \Big[1-(1+\frac{\Omega}{m}zK_sP_sx^{-\alpha_s})^{-m}\Big]xdx\bigg\}f_2(r)dr\\
&=\int_{R_l}^{R_m}2\pi^2\lambda_b^2r^3\exp\bigg\{-2\pi\lambda_b\int_{R_1}^{r} \Big[1-(1+\frac{\Omega}{m}zK_sP_sx^{-\alpha_s})^{-m}\Big]xdx-\pi\lambda_br^2\bigg\}dr.\\
\end{split}
\end{equation}
If $h$ is a constant, MGF for interfering power is
\begin{equation}
\begin{split}
\mathcal{M}_{I_{2}}(z)=&2\pi^2\lambda_b^2\int_{R_l}^{R_m}r^3\exp\Bigg\{2\pi\lambda_b\bigg[\frac{ (zK_sh P_{s})^{\frac{2}{\alpha_s}}}{\alpha_s}\Gamma\Big(-\frac{2}{\alpha_s},\\
&zK_sh P_{s}R_m^{-\alpha_s},zK_s h P_{s}r^{-\alpha_s}\Big)-\frac{R_m^2}{2}\bigg]\Bigg\}dr.\\
\end{split}
\end{equation}

In a Rayleigh fading channel, the result is
\begin{equation}
\begin{split}
\mathcal{M}_{I_2}(z)&=\int_{R_l}^{R_m}\exp\bigg[-2\pi\lambda_b\int_{r}^{R_m} \Big(1-\frac{1}{1+zK_sx^{-2}P_{s}}\Big)xdx\bigg]f_2(r)dr\\
&=2\pi^2\lambda_b^2\int_{R_l}^{R_m}r^3\exp\bigg[\!\!-\!2\pi\lambda_b(zK_sP_s)^{2/\alpha_s}\int_{\frac{r}{(zK_sP_s)^{1/\alpha_s}}}^{\frac{R_m}{(zK_sP_s)^{1/\alpha_s}}}\frac{x}{x^{\alpha_s}+1}dx-\pi\lambda_br^2\bigg]dr.\\
\end{split}
\end{equation}

In a Nakagami-m fading channel, the result is
\begin{equation}
\begin{split}
\mathcal{M}_{I_1}(z) &=\int_{R_l}^{R_m}\exp\bigg\{-2\pi\lambda_b\int_{r}^{R_m} \Big[1-(1+\frac{\Omega}{m}zK_sP_sx^{-\alpha_s})^{-m}\Big]xdx\bigg\}f_2(r)dr\\
&=2\pi^2\lambda_b^2\int_{R_l}^{R_m}r^3\exp\bigg[2\pi\lambda_b\int_{r}^{R_m} (1+\frac{\Omega}{m}zK_sP_sx^{-\alpha_s})^{-m}xdx-\pi\lambda_bR_m^2\bigg]dr.\\
\end{split}
\end{equation}
\item \textbf{CD:} if $h$ is a constant, MGF for desired power is
\begin{equation}
\begin{split}
\mathcal{M}_{P_{R_0}}(z)= \textrm{exp}\bigg\{2\pi\lambda_b\Big[\frac{ (K_shP_s)^{2/\alpha_s}}{\alpha_s}\Gamma\Big(-\frac{2}{\alpha_s},zK_shP_sR_{0}^{-\alpha_s},zK_shP_sR_{l}^{-\alpha_s}\Big)\!-\!\frac{R_{0}^2}{2}+\frac{R_{l}^2}{2}\Big]\bigg\}.
\end{split}
\end{equation}
In a Rayleigh fading channel, the result is
\begin{equation}
\begin{split}
\mathcal{M}_{P_{R_0}}(z) =\exp\bigg[-2\pi\lambda_b(zK_sP_s)^{2/\alpha_s}\int_{\frac{R_l}{(zK_sP_s)^{1/\alpha_s}}}^{\frac{R_0}{(zK_sP_s)^{1/\alpha_s}}}\frac{x}{x^{\alpha_s}+1}dx\bigg].\\
\end{split}
\end{equation}

In a Nakagami-m fading channel, the result is
\begin{equation}
\begin{split}
\mathcal{M}_{P_{R_0}}(z)
&=\exp\bigg\{-2\pi\lambda_b\int_{R_l}^{R_0} \Big[1-(1+\frac{\Omega}{m}zK_sP_sr^{-\alpha_s})^{-m}\Big]rdr\bigg\}.\\
\end{split}
\end{equation}

If $h$ is a constant, MGF for interfering power is
\begin{equation}
\begin{split}
\mathcal{M}_{I_{R_0}}(z)=& \textrm{exp}\bigg\{2\pi\lambda_b\Big[\frac{ (K_shP_s)^{2/\alpha_s}}{\alpha_s}\Gamma\Big(-\frac{2}{\alpha_s},zK_shP_sR_{m}^{-\alpha_s},zK_shP_sR_{0}^{-\alpha_s}\Big)\!-\!\frac{R_{m}^2}{2}+\frac{R_{l}^2}{2}\Big]\bigg\}.
\end{split}
\end{equation}
In a Rayleigh fading channel, the result is
\begin{equation}
\begin{split}
\mathcal{M}_{I_{R_0}}(z) =\exp\bigg[-2\pi\lambda_b(zK_sP_s)^{2/\alpha_s}\int_{\frac{R_0}{(zK_sP_s)^{1/\alpha_s}}}^{\frac{R_m}{(zK_sP_s)^{1/\alpha_s}}}\frac{x}{x^{\alpha_s}+1}dx\bigg].\\
\end{split}
\end{equation}

In a Nakagami-m fading channel, the result is
\begin{equation}
\begin{split}
\mathcal{M}_{I_{R_0}}(z)
&=\exp\bigg\{-2\pi\lambda_b\int_{R_0}^{R_m} \Big[1-(1+\frac{\Omega}{m}zK_sP_sr^{-\alpha_s})^{-m}\Big]rdr\bigg\}.\\
\end{split}
\end{equation}
\item \textbf{FPD:} If $h$ is a constant, MGF for desired power is
\begin{equation}
\begin{split}
\mathcal{M}_{P_{\eta}}(z)=&\int_{R_l}^{R_m}\exp\Big\{-2\pi\lambda_b\int_{R_l}^{r}[1-\exp(-zK_shP_sx^{-\alpha_s})]xdx\Big\}f_{\eta}(r)dr\\
=&2\pi\lambda_b\eta_t^2\int_{R_l}^{R_m}\exp\Bigg\{2\pi\lambda_b\bigg\{\frac{ (zK_sh P_{s})^{\frac{2}{\alpha_s}}}{\alpha_s}\Big[\Gamma\Big(-\frac{2}{\alpha_s},zK_sh P_{s}r^{-\alpha_s}\Big)-\\
&\Gamma\Big(-\frac{2}{\alpha_s},zK_s h P_{s}R_l^{-\alpha_s}\Big)\Big]-\frac{r^2}{2}+\frac{R_l^2}{2}\bigg\}\Bigg\}r\exp(-\pi\lambda_b\eta_t^2r^2)dr\\
=&2\pi\lambda_b\exp\bigg\{\pi\lambda_b\Big[R_l^2-\frac{ 2(zK_sh P_{s})^{\frac{2}{\alpha_s}}}{\alpha_s}]\Gamma\Big(-\frac{2}{\alpha_s},zK_sh P_{s}R_l^{-\alpha_s}\Big)\bigg\}\\
&\times\int_{R_l}^{R_m}r\exp\bigg\{\pi\lambda_b\Big[\frac{ 2(zK_sh P_{s})^{\frac{2}{\alpha_s}}}{\alpha_s}\Gamma\Big(-\frac{2}{\alpha_s},zK_s h P_{s}r^{-\alpha_s}\Big)-(1+\eta_t^2)r^2\Big]\bigg\}dr\\
=&2\pi\lambda_b\eta_t^2\int_{R_l}^{R_m}r\exp\Bigg\{2\pi\lambda_b\bigg[\frac{ (zK_sh P_{s})^{\frac{2}{\alpha_s}}}{\alpha_s}\Gamma\Big(-\frac{2}{\alpha_s},zK_s hP_{s}r^{-\alpha_s},zK_s h P_{s}R_l^{-\alpha_s}\Big)\\
&-\frac{(1+\eta_t^2)r^2}{2}+\frac{R_l^2}{2}\bigg]\Bigg\}dr.\\
\end{split}
\end{equation}

In a Rayleigh fading channel, the result is
\begin{equation}
\begin{split}
\mathcal{M}_{P_{\eta}}(z)&=\int_{R_l}^{R_m}\exp\bigg[-2\pi\lambda_b\int_{R_l}^{r} \Big(1-\frac{1}{1+zK_sx^{-2}P_{s}}\Big)xdx\bigg]f_{\eta}(r)dr\\
&=2\pi\lambda_b\eta_t^2\int_{R_l}^{R_m}r\exp\bigg[-2\pi\lambda_b(zK_sP_s)^{2/\alpha_s}\int_{\frac{R_l}{(zK_sP_s)^{1/\alpha_s}}}^{\frac{r}{(zK_sP_s)^{1/\alpha_s}}}\frac{x}{x^{\alpha_s}+1}dx-\pi\lambda_b\eta_t^2r^2\bigg]dr.\\
\end{split}
\end{equation}

In a Nakagami-m fading channel, the result is
\begin{equation}
\begin{split}
\mathcal{M}_{P_{\eta}}(z) &=\int_{R_l}^{R_m}\exp\bigg\{-2\pi\lambda_b\int_{R_1}^{r} \Big[1-(1+\frac{\Omega}{m}zK_sP_sx^{-\alpha_s})^{-m}\Big]xdx\bigg\}f_\eta(r)dr\\
&=2\pi\lambda_b\eta_t^2\int_{R_l}^{R_m}r\exp\bigg\{-2\pi\lambda_b\int_{R_1}^{r} \Big[1-(1+\frac{\Omega}{m}zK_sP_sx^{-\alpha_s})^{-m}\Big]xdx\\
&-\pi\lambda_b\eta_t^2r^2\bigg\}dr.\\
\end{split}
\end{equation}

If $h$ is a constant, MGF for interfering power is
\begin{equation}
\begin{split}
\mathcal{M}_{I_{\eta}}(z)=&2\pi\lambda_b\eta_t^2\int_{R_l}^{R_m}\exp\Bigg\{2\pi\lambda_b\bigg\{\frac{ (zK_sh P_{s})^{\frac{2}{\alpha_s}}}{\alpha_s}\Big[\Gamma\Big(-\frac{2}{\alpha_s},zK_sh P_{s}R_m^{-\alpha_s}\Big)-\\
&\Gamma\Big(-\frac{2}{\alpha_s},zK_sh P_{s}r^{-\alpha_s}\Big)\Big]-\frac{R_m^2}{2}+\frac{r^2}{2}\bigg\}\Bigg\}r\exp(-\pi\lambda_b\eta_t^2r^2)dr\\
=&2\pi\lambda_b\eta_t^2\int_{R_l}^{R_m}r\exp\Bigg\{2\pi\lambda_b\bigg[\!\frac{ (zK_s h P_{s})^{\frac{2}{\alpha_s}}}{\alpha_s}\Gamma\Big(-\frac{2}{\alpha_s},zK_sh P_{s}R_m^{-\alpha_s},\\
&zK_s h P_{s}r^{-\alpha_s}\Big)-\frac{R_m^2}{2}\bigg]\Bigg\}dr.\\
\end{split}
\end{equation}
\end{itemize}

In a Rayleigh fading channel, the result is
\begin{equation}
\begin{split}
\mathcal{M}_{I_{\eta}}(z)&=\int_{R_l}^{R_m}\exp\bigg[-2\pi\lambda_b\int_{r}^{R_m} \Big(1-\frac{1}{1+zK_sx^{-2}P_{s}}\Big)xdx\bigg]f_{\eta}(r)dr\\
&=2\pi\lambda_b\eta_t^2\int_{R_l}^{R_m}r\exp\bigg[-2\pi\lambda_b(zK_sP_s)^{2/\alpha_s}\int_{\frac{r}{(zK_sP_s)^{1/\alpha_s}}}^{\frac{R_m}{(zK_sP_s)^{1/\alpha_s}}}\frac{x}{x^{\alpha_s}+1}dx-\pi\lambda_b\eta_t^2r^2\bigg]dr.\\
\end{split}
\end{equation}

In a Nakagami-m fading channel, the result is
\begin{equation}
\begin{split}
\mathcal{M}_{I_{\eta}}(z) &=\int_{R_l}^{R_m}\exp\bigg\{-2\pi\lambda_b\int_{r}^{R_m} \Big[1-(1+\frac{\Omega}{m}zK_sP_sx^{-\alpha_s})^{-m}\Big]xdx\bigg\}f_\eta(r)dr\\
&=2\pi\lambda_b\eta_t^2\int_{R_l}^{R_m}r\exp\bigg\{-2\pi\lambda_b\int_{r}^{R_m} \Big[1-(1+\frac{\Omega}{m}zK_sP_sx^{-\alpha_s})^{-m}\Big]xdx-\pi\lambda_b\eta_t^2r^2\bigg\}dr.\\
\end{split}
\end{equation}

\subsection{Variance}
Substitute $s=-z$ into the all MGFs. Then we can use the property of MGF to obtain the expect of different moments of the variable, i.e., $\mathbb{E}[x^n]=\frac{d^n\mathcal{M}_x(s)}{ds^n}\big|_{s=0}$. Thus, the variance can be calculated be $\mathbb{V}[x]=\mathbb{E}[x^2]-\mathbb{E}[x]^2$.
\begin{itemize}
\item \textbf{NoJT:} if $h$ is a constant, using the conventional calculation method, the first and second derivatives of MGF of the desired power are
    \begin{equation}
    \begin{split}
    \mathcal{M}'_{P_1}(s)\xlongequal{(a)}&2\pi\lambda_b \int_{R_l}^{R_m}r\frac{d\exp(sK_shP_sr^{-\alpha_s})}{ds}\exp(-\pi\lambda_br^2)dr\\
    =&2\pi\lambda_b \int_{R_l}^{R_m}K_shP_sr^{-\alpha_s+1}\exp(sK_shP_sr^{-\alpha_s})\exp(-\pi\lambda_br^2)dr,\\
    \end{split}
    \end{equation}
     \begin{equation}
    \begin{split}
    \mathcal{M}''_{P_1}(s)=&2\pi\lambda_b \int_{R_l}^{R_m}K_s^2h^2P_s^2r^{-2\alpha_s+1}\exp(sK_shP_sr^{-\alpha_s})\exp(-\pi\lambda_br^2)dr,\\
    \end{split}
    \end{equation}
    where (a) is based on Leibniz integral rule. Thus, the expectation of the second moment of received power is
    \begin{equation}
    \begin{split}
    \mathbb{E}[P_1^2]=&\mathcal{M}''_{P_1}(0)\\
    =&2\pi\lambda_b \int_{R_l}^{R_m}K_s^2h^2P_s^2r^{-2\alpha_s+1}\exp(-\pi\lambda_br^2)dr\\
    =&(\pi\lambda_b)^{\alpha_s}K_s^2h^2P_s^2[\gamma(1-\alpha_s,\pi\lambda_bR_m^2)-\gamma(1-\alpha_s,\pi\lambda_bR_l^2)].
    \end{split}
    \end{equation}
    The variance of the desired power is
    \begin{equation}
    \begin{split}
    \mathbb{V}[P_1]=&\mathbb{E}[P_1^2]-\mathbb{E}[P_1]^2\\
    =&(\pi\lambda_b)^{\alpha_s/2}K_s^2h^2P_s^2\Big\{\gamma(1-\alpha_s,\pi\lambda_bR_m^2)-\gamma(1-\alpha_s,\pi\lambda_bR_l^2)\\
    &-(\pi\lambda_b)^{\alpha_s/2}\big[\gamma(1-\alpha_s/2,\pi\lambda_bR_m^2)-\gamma(1-\alpha_s/2,\pi\lambda_bR_l^2)\big]^2\Big\}.
    \end{split}
    \end{equation}

    Using our proposed method, the first and second derivatives of MGF are
    \begin{equation}
    \begin{split}
    \mathcal{M}'_{P_1}(s)=&\int_{R_l}^{R_m}\frac{\Delta \exp\{-2\pi\lambda_b\int_{R_l}^{r}[1-\exp(sK_shP_sx^{-\alpha_s})]xdx\}}{\Delta s}f_{1}(r)dr\\
    =&\int_{R_l}^{R_m}\frac{\Delta -2\pi\lambda_b\int_{R_l}^{r}[1-\exp(sK_shP_sx^{-\alpha_s})]xdx}{\Delta s}\\
    &\times\exp\Big\{-2\pi\lambda_b\int_{R_l}^{r}[1-\exp(sK_shP_sx^{-\alpha_s})]xdx\Big\}f_{1}(r)dr\\
    =&\int_{R_l}^{R_m}2\pi\lambda_b\int_{R_l}^{r}K_shP_sx^{-\alpha_s+1}\exp(sK_shP_sx^{-\alpha_s})dx\\
    &\times\exp\Big\{-2\pi\lambda_b\int_{R_l}^{r}[1-\exp(sK_shP_sx^{-\alpha_s})]xdx\Big\}f_{1}(r)dr,\\
    \end{split}
    \end{equation}
     \begin{equation}
    \begin{split}
    \mathcal{M}''_{P_1}(s)
    =&2\pi\lambda_bK_shP_s\int_{R_l}^{R_m}\Big[\int_{R_l}^rx^{-\alpha_s+1}\frac{\Delta a(s)}{\Delta s}dx b(s)\\
    &+\int_{R_l}^rx^{-\alpha_s+1}a(s)dx\frac{\Delta b(s)}{\Delta s}\Big]f_1(r)dr\\
    =&2\pi\lambda_bK_shP_s\int_{R_l}^{R_m}\bigg\{\int_{R_l}^rx^{-2\alpha_s+1}K_shP_sa(s)dxb(s)\\
    &+2\pi\lambda_bK_shP_s\Big[\int_{R_l}^rx^{-\alpha_s+1}a(s)dx\Big]^2b(s)\bigg\}f_1(r)dr,\\
    \end{split}
    \end{equation}
    where $a(s)=\exp(sK_shP_sx^{-\alpha_s})$ and $b(s)=\exp\{-2\pi\lambda_b\int_{R_l}^{r}[1-\exp(sK_shP_sx^{-\alpha_s})]xdx\}$.
    Thus, the expectation of the second moment of received power is
    \begin{equation}
    \begin{split}
    \mathbb{E}[P_1^2]=&\mathcal{M}''_{P_1}(0)\\
    =&2\pi\lambda_bK_s^2h^2P_s^2\int_{R_l}^{R_m}\bigg[\int_{R_l}^rx^{-2\alpha_s+1}dx+2\pi\lambda_b\Big(\int_{R_l}^rx^{-\alpha_s+1}dx\Big)^2\bigg]f_1(r)dr\\
    =&K_s^2h^2P_s^2\int_{R_l}^{R_m}\bigg[\frac{\pi\lambda_b}{1-\alpha_s}(r^{-2\alpha_s+2}-R_l^{-2\alpha_s+2})+\frac{4\pi^2\lambda_b^2}{(2-\alpha_s)^2}(r^{2-\alpha_s}-R_l^{2-\alpha_s})^2\bigg]f_1(r)dr\\
    =&K_s^2h^2P_s^2\int_{R_l}^{R_m}\bigg[\frac{2\pi^2\lambda_b^2r}{1-\alpha_s}(r^{-2\alpha_s+2}-R_l^{-2\alpha_s+2})+\frac{8\pi^3\lambda_b^3r}{(2-\alpha_s)^2}(r^{2-\alpha_s}-R_l^{2-\alpha_s})^2\bigg]\\
    &\times\exp(-\pi\lambda_br^2)dr\\
    =&2\pi\lambda_bK_s^2h^2P_s^2\Bigg\{\frac{(\pi\lambda_b)^{\alpha_s-1}}{2(1-\alpha_s)}\Gamma(2-\alpha_s,R_l^2\pi\lambda_b,R_m^2\pi\lambda_b)\\
    &+\Big[-\frac{R_l^{2-2\alpha_s}}{2(1-\alpha_s)}+\frac{2\pi\lambda_bR_l^{4-2\alpha_s}}{(2-\alpha_s)^2}\Big](e^{-R_l^2\pi\lambda_b}-e^{-R_m^2\pi\lambda_b})\\
    &+\frac{2(\pi\lambda_b)^{\alpha_s-1}}{(2-\alpha_s)^2}\Gamma(3-\alpha_s,R_l^2\pi\lambda_b,R_m^2\pi\lambda_b)\\
    &-\frac{4R_l^{2-\alpha_s}(\pi\lambda_b)^{\alpha_s/2}}{(2-\alpha_s)^2}\Gamma(2-\alpha_s/2,R_l^2\pi\lambda_b,R_m^2\pi\lambda_b)\Bigg\}\\
    =&2\pi\lambda_bK_s^2h^2P_s^2\Bigg\{\frac{(\pi\lambda_b)^{\alpha_s-1}}{2(1-\alpha_s)}\Gamma(2-\alpha_s,R_l^2\pi\lambda_b,R_m^2\pi\lambda_b)\\
    &+\Big[-\frac{R_l^{2-2\alpha_s}}{2(1-\alpha_s)}+\frac{2\pi\lambda_bR_l^{4-2\alpha_s}}{(2-\alpha_s)^2}\Big](e^{-R_l^2\pi\lambda_b}-e^{-R_m^2\pi\lambda_b})\\
    &+\frac{2(\pi\lambda_b)^{\alpha_s-1}}{(2-\alpha_s)^2}\Big[(2-\alpha_s)\Gamma(2-\alpha_s,R_l^2\pi\lambda_b,R_m^2\pi\lambda_b)+(R_l^2\pi\lambda_b)^{2-\alpha_s}e^{-R_l^2\pi\lambda_b}\\
    &-(R_m^2\pi\lambda_b)^{2-\alpha_s}e^{-R_m^2\pi\lambda_b}\Big]-\frac{4R_l^{2-\alpha_s}(\pi\lambda_b)^{\alpha_s/2}}{(2-\alpha_s)^2}\Gamma(2-\alpha_s/2,R_l^2\pi\lambda_b,R_m^2\pi\lambda_b)\Bigg\}\\
    =&2\pi\lambda_bK_s^2h^2P_s^2\Bigg\{\Big[\frac{(\pi\lambda_b)^{\alpha_s-1}}{2(1-\alpha_s)}+\frac{2(\pi\lambda_b)^{\alpha_s-1}}{2-\alpha_s}\Big]\Gamma(2-\alpha_s,R_l^2\pi\lambda_b,R_m^2\pi\lambda_b)\\
    &+\Big[\frac{4\pi\lambda_bR_l^{4-2\alpha_s}}{(2-\alpha_s)^2}-\frac{R_l^{2-2\alpha_s}}{2(1-\alpha_s)}\Big](e^{-R_l^2\pi\lambda_b}-e^{-R_m^2\pi\lambda_b})\\
    &-\frac{4R_l^{2-\alpha_s}(\pi\lambda_b)^{\alpha_s/2}}{(2-\alpha_s)^2}\Gamma(2-\alpha_s/2,R_l^2\pi\lambda_b,R_m^2\pi\lambda_b)\Bigg\}.\\
    \end{split}
    \end{equation}
    Due to $\Gamma(s+1,a,b)=\Gamma(s,a,b)+a^se^{-a}-b^se{-b}$.
    The variance of the desired power is
    \begin{equation}
    \begin{split}
    \mathbb{V}[P_1]=&\mathbb{E}[P_1^2]-\mathbb{E}[P_1]^2\\
    =&2\pi\lambda_bK_s^2h^2P_s^2\Bigg\{\Big[\frac{(\pi\lambda_b)^{\alpha_s-1}}{2(1-\alpha_s)}+\frac{2(\pi\lambda_b)^{\alpha_s-1}}{2-\alpha_s}\Big]\Gamma(2-\alpha_s,R_l^2\pi\lambda_b,R_m^2\pi\lambda_b)\\
    &+\Big[\frac{4\pi\lambda_bR_l^{4-2\alpha_s}}{(2-\alpha_s)^2}-\frac{R_l^{2-2\alpha_s}}{2(1-\alpha_s)}\Big](e^{-R_l^2\pi\lambda_b}-e^{-R_m^2\pi\lambda_b})\\
    &-\frac{4R_l^{2-\alpha_s}(\pi\lambda_b)^{\alpha_s/2}}{(2-\alpha_s)^2}\Gamma(2-\alpha_s/2,R_l^2\pi\lambda_b,R_m^2\pi\lambda_b)\Bigg\}\\
    &-\Bigg\{\frac{2K_s h P_s}{2-\alpha_s}\Big[(\pi\lambda_b)^{\alpha_s/2}\Gamma(\frac{4-\alpha_s}{2},\pi\lambda_bR_l^2,\pi\lambda_bR_m^2)\\
    &+\pi\lambda_bR_l^{-\alpha_s+2}(e^{-\pi\lambda_bR_m^2}-e^{-\pi\lambda_bR_l^2})\Big]\Bigg\}^2.
    \end{split}
    \end{equation}

    In a Rayleigh fading channel, the first and second derivatives of the MGF of the desired power are
    \begin{equation}
    \begin{split}
    \mathcal{M}'_{P_1}(s)=&\int_{R_l}^{R_m}\frac{\Delta \exp\{-2\pi\lambda_b\int_{R_l}^{r}[1-(1-sK_sP_sx^{-\alpha_s})^{-1}]xdx\}}{\Delta s}f_{1}(r)dr\\
    =&\int_{R_l}^{R_m}\frac{\Delta -2\pi\lambda_b\int_{R_l}^{r}[1-(1-sK_sP_sx^{-\alpha_s})^{-1}]xdx}{\Delta s}\\
    &\times\exp\bigg\{-2\pi\lambda_b\int_{R_l}^{r}\Big[1-(1-sK_sP_sx^{-\alpha_s})^{-1}\Big]xdx\bigg\}f_{1}(r)dr\\
    =&\int_{R_l}^{R_m}2\pi\lambda_b\int_{R_l}^{r}K_sP_sx^{-\alpha_s+1}(1-sK_sP_sx^{-\alpha_s})^{-2}dx\\
    &\times\exp\bigg\{-2\pi\lambda_b\int_{R_l}^{r}\Big[1-(1-sK_sP_sx^{-\alpha_s})^{-1}\Big]xdx\bigg\}f_{1}(r)dr\\
    \end{split}
    \end{equation}
     \begin{equation}
    \begin{split}
    \mathcal{M}''_{P_1}(s)
    =&2\pi\lambda_bK_sP_s\int_{R_l}^{R_m}\Big[\int_{R_l}^rx^{-\alpha_s+1}\frac{\Delta a(s)^{-2}}{\Delta s}dx b(s)\\
    &+\int_{R_l}^rx^{-\alpha_s+1}a(s)^{-2}dx\frac{\Delta b(s)}{\Delta s}\Big]f_1(r)dr\\
    =&2\pi\lambda_bK_sP_s\int_{R_l}^{R_m}\bigg\{2K_sP_s\int_{R_l}^rx^{-2\alpha_s+1}a(s)^{-3}dxb(s)\\
    &+2\pi\lambda_bK_sP_s\Big[\int_{R_l}^rx^{-\alpha_s+1}a(s)^{-2}dx\Big]^2b(s)\bigg\}f_1(r)dr,\\
    \end{split}
    \end{equation}
    where $a(s)=1-sK_sP_sx^{-\alpha_s}$ and $b(s)=\exp\{-2\pi\lambda_b\int_{R_l}^{r}[1-(1-sK_sP_sx^{-\alpha_s})^{-1}]xdx\}$.
    Thus, the expectation of the second moment of received power is
    \begin{equation}
    \begin{split}
    \mathbb{E}[P_1^2]=&\mathcal{M}''_{P_1}(0)\\
    =&4\pi\lambda_bK_s^2P_s^2\int_{R_l}^{R_m}\bigg[\int_{R_l}^rx^{-2\alpha_s+1}dx+\pi\lambda_b\Big(\int_{R_l}^rx^{-\alpha_s+1}dx\Big)^2\bigg]f_1(r)dr\\
    =&K_s^2P_s^2\int_{R_l}^{R_m}\bigg[\frac{2\pi\lambda_b}{1-\alpha_s}(r^{-2\alpha_s+2}-R_l^{-2\alpha_s+2})+\frac{4\pi^2\lambda_b^2}{(2-\alpha_s)^2}(r^{2-\alpha_s}-R_l^{2-\alpha_s})^2\bigg]f_1(r)dr\\
    =&K_s^2P_s^2\int_{R_l}^{R_m}\bigg[\frac{4\pi^2\lambda_b^2r}{1-\alpha_s}(r^{-2\alpha_s+2}-R_l^{-2\alpha_s+2})+\frac{8\pi^3\lambda_b^3r}{(2-\alpha_s)^2}(r^{2-\alpha_s}-R_l^{2-\alpha_s})^2\bigg]\\
    &\times\exp(-\pi\lambda_br^2)dr\\
    =&2\pi\lambda_bK_s^2P_s^2\Bigg\{\frac{(\pi\lambda_b)^{\alpha_s-1}}{1-\alpha_s}\Gamma(2-\alpha_s,R_l^2\pi\lambda_b,R_m^2\pi\lambda_b)\\
    &+\Big[-\frac{R_l^{2-2\alpha_s}}{1-\alpha_s}+\frac{2\pi\lambda_bR_l^{4-2\alpha_s}}{(2-\alpha_s)^2}\Big](e^{-R_l^2\pi\lambda_b}-e^{-R_m^2\pi\lambda_b})\\
    &+\frac{2(\pi\lambda_b)^{\alpha_s-1}}{(2-\alpha_s)^2}\Gamma(3-\alpha_s,R_l^2\pi\lambda_b,R_m^2\pi\lambda_b)\\
    &-\frac{4R_l^{2-\alpha_s}(\pi\lambda_b)^{\alpha_s/2}}{(2-\alpha_s)^2}\Gamma(2-\alpha_s/2,R_l^2\pi\lambda_b,R_m^2\pi\lambda_b)\Bigg\}\\
    =&2\pi\lambda_bK_s^2P_s^2\Bigg\{\frac{(\pi\lambda_b)^{\alpha_s-1}}{(1-\alpha_s)}\Gamma(2-\alpha_s,R_l^2\pi\lambda_b,R_m^2\pi\lambda_b)\\
    &+\Big[-\frac{R_l^{2-2\alpha_s}}{(1-\alpha_s)}+\frac{2\pi\lambda_bR_l^{4-2\alpha_s}}{(2-\alpha_s)^2}\Big](e^{-R_l^2\pi\lambda_b}-e^{-R_m^2\pi\lambda_b})\\
    &+\frac{2(\pi\lambda_b)^{\alpha_s-1}}{(2-\alpha_s)^2}\Big[(2-\alpha_s)\Gamma(2-\alpha_s,R_l^2\pi\lambda_b,R_m^2\pi\lambda_b)+(R_l^2\pi\lambda_b)^{2-\alpha_s}e^{-R_l^2\pi\lambda_b}\\
    &-(R_m^2\pi\lambda_b)^{2-\alpha_s}e^{-R_m^2\pi\lambda_b}\Big]-\frac{4R_l^{2-\alpha_s}(\pi\lambda_b)^{\alpha_s/2}}{(2-\alpha_s)^2}\Gamma(2-\alpha_s/2,R_l^2\pi\lambda_b,R_m^2\pi\lambda_b)\Bigg\}\\
    =&2\pi\lambda_bK_s^2P_s^2\Bigg\{\Big[\frac{(\pi\lambda_b)^{\alpha_s-1}}{(1-\alpha_s)}+\frac{2(\pi\lambda_b)^{\alpha_s-1}}{2-\alpha_s}\Big]\Gamma(2-\alpha_s,R_l^2\pi\lambda_b,R_m^2\pi\lambda_b)\\
    &+\Big[\frac{4\pi\lambda_bR_l^{4-2\alpha_s}}{(2-\alpha_s)^2}-\frac{R_l^{2-2\alpha_s}}{(1-\alpha_s)}\Big](e^{-R_l^2\pi\lambda_b}-e^{-R_m^2\pi\lambda_b})\\
    &-\frac{4R_l^{2-\alpha_s}(\pi\lambda_b)^{\alpha_s/2}}{(2-\alpha_s)^2}\Gamma(2-\alpha_s/2,R_l^2\pi\lambda_b,R_m^2\pi\lambda_b)\Bigg\}.\\
    \end{split}
    \end{equation}

    The variance of the desired power is
    \begin{equation}
    \begin{split}
    \mathbb{V}[P_1]=&\mathbb{E}[P_1^2]-\mathbb{E}[P_1]^2\\
    =&2\pi\lambda_bK_s^2P_s^2\Bigg\{\Big[\frac{(\pi\lambda_b)^{\alpha_s-1}}{(1-\alpha_s)}+\frac{2(\pi\lambda_b)^{\alpha_s-1}}{2-\alpha_s}\Big]\Gamma(2-\alpha_s,R_l^2\pi\lambda_b,R_m^2\pi\lambda_b)\\
    &+\Big[\frac{4\pi\lambda_bR_l^{4-2\alpha_s}}{(2-\alpha_s)^2}-\frac{R_l^{2-2\alpha_s}}{(1-\alpha_s)}\Big](e^{-R_l^2\pi\lambda_b}-e^{-R_m^2\pi\lambda_b})\\
    &-\frac{4R_l^{2-\alpha_s}(\pi\lambda_b)^{\alpha_s/2}}{(2-\alpha_s)^2}\Gamma(2-\alpha_s/2,R_l^2\pi\lambda_b,R_m^2\pi\lambda_b)\Bigg\}\\
    &-\bigg\{\frac{2K_s P_s}{2-\alpha_s}\Big[(\pi\lambda_b)^{\alpha_s/2}\Gamma(\frac{4-\alpha_s}{2},\pi\lambda_bR_l^2,\pi\lambda_bR_m^2)\\
    &+\pi\lambda_bR_l^{-\alpha_s+2}(e^{-\pi\lambda_bR_m^2}-e^{-\pi\lambda_bR_l^2})\Big]\bigg\}^2.
    \end{split}
    \end{equation}

   In a Nakagami-m fading channel, the first and second derivatives of the MGF of the desired power are
   \begin{equation}
    \begin{split}
    \mathcal{M}'_{P_1}(s)=&\int_{R_l}^{R_m}\frac{\Delta \exp\{-2\pi\lambda_b\int_{R_l}^{r}[1-(1-s\frac{\Omega}{m}K_sP_sx^{-\alpha_s})^{-m}]xdx\}}{\Delta s}f_{1}(r)dr\\
    =&\int_{R_l}^{R_m}\frac{\Delta -2\pi\lambda_b\int_{R_l}^{r}[1-(1-s\frac{\Omega}{m}K_sP_sx^{-\alpha_s})^{-m}]xdx}{\Delta s}\\
    &\times\exp\bigg\{-2\pi\lambda_b\int_{R_l}^{r}\Big[1-(1-s\frac{\Omega}{m}K_sP_sx^{-\alpha_s})^{-m}\Big]xdx\bigg\}f_{1}(r)dr\\
    =&\int_{R_l}^{R_m}2\pi\lambda_b\int_{R_l}^{r}\Omega K_sP_sx^{-\alpha_s+1}(1-s\frac{\Omega}{m}K_sP_sx^{-\alpha_s})^{-m-1}dx\\
    &\times\exp\bigg\{-2\pi\lambda_b\int_{R_l}^{r}\Big[1-(1-s\frac{\Omega}{m}K_sP_sx^{-\alpha_s})^{-m}\Big]xdx\bigg\}f_{1}(r)dr\\
    \end{split}
    \end{equation}
     \begin{equation}
    \begin{split}
    \mathcal{M}''_{P_1}(s)
    =&2\pi\lambda_b\Omega K_sP_s\int_{R_l}^{R_m}\Big[\int_{R_l}^rx^{-\alpha_s+1}\frac{\Delta a(s)^{-m-1}}{\Delta s}dx b(s)\\
    &+\int_{R_l}^rx^{-\alpha_s+1}a(s)^{-m-1}dx\frac{\Delta b(s)}{\Delta s}\Big]f_1(r)dr\\
    =&2\pi\lambda_b\Omega K_sP_s\int_{R_l}^{R_m}\bigg\{\frac{(m+1)\Omega}{m}K_sP_s\int_{R_l}^rx^{-2\alpha_s+1}a(s)^{-m-2}dxb(s)\\
    &+2\pi\lambda_b\Omega K_sP_s\Big[\int_{R_l}^rx^{-\alpha_s+1}a(s)^{-m-1}dx\Big]^2b(s)\bigg\}f_1(r)dr,\\
    \end{split}
    \end{equation}
    where $a(s)=1-\frac{\Omega}{m}sK_sP_sx^{-\alpha_s}$ and $b(s)=\exp\{-2\pi\lambda_b\int_{R_l}^{r}[1-(1-s\frac{\Omega}{m}K_sP_sx^{-\alpha_s})^{-m}]xdx\}$.
    Thus, the expectation of the second moment of received power is
    \begin{equation}
    \begin{split}
    \mathbb{E}[P_1^2]=&\mathcal{M}''_{P_1}(0)\\
    =&2\pi\lambda_b\Omega^2K_s^2P_s^2\int_{R_l}^{R_m}\bigg[\frac{m+1}{m}\int_{R_l}^rx^{-2\alpha_s+1}dx+2\pi\lambda_b\Big(\int_{R_l}^rx^{-\alpha_s+1}dx\Big)^2\bigg]f_1(r)dr\\
    =&K_s^2\Omega^2P_s^2\int_{R_l}^{R_m}\bigg[\frac{\pi\lambda_b(m+1)}{m(1-\alpha_s)}(r^{-2\alpha_s+2}-R_l^{-2\alpha_s+2})+\frac{4\pi^2\lambda_b^2}{(2-\alpha_s)^2}(r^{2-\alpha_s}\\
    &-R_l^{2-\alpha_s})^2\bigg]f_1(r)dr\\
    =&K_s^2\Omega^2P_s^2\int_{R_l}^{R_m}\bigg[\frac{2\pi^2\lambda_b^2(m+1)r}{m(1-\alpha_s)}(r^{-2\alpha_s+2}-R_l^{-2\alpha_s+2})+\frac{8\pi^3\lambda_b^3r}{(2-\alpha_s)^2}(r^{2-\alpha_s}\\
    &-R_l^{2-\alpha_s})^2\bigg]\exp(-\pi\lambda_br^2)dr\\
    =&2\pi\lambda_bK_s^2\Omega^2P_s^2\Bigg\{\frac{(\pi\lambda_b)^{\alpha_s-1}(m+1)}{2m(1-\alpha_s)}\Gamma(2-\alpha_s,R_l^2\pi\lambda_b,R_m^2\pi\lambda_b)+\Big[-\frac{(m+1)R_l^{2-2\alpha_s}}{2m(1-\alpha_s)}\\
    &+\frac{2\pi\lambda_bR_l^{4-2\alpha_s}}{(2-\alpha_s)^2}\Big](e^{-R_l^2\pi\lambda_b}-e^{-R_m^2\pi\lambda_b})+\frac{2(\pi\lambda_b)^{\alpha_s-1}}{(2-\alpha_s)^2}\Gamma(3-\alpha_s,R_l^2\pi\lambda_b,R_m^2\pi\lambda_b)\\
    &-\frac{4R_l^{2-\alpha_s}(\pi\lambda_b)^{\alpha_s/2}}{(2-\alpha_s)^2}\Gamma(2-\alpha_s/2,R_l^2\pi\lambda_b,R_m^2\pi\lambda_b)\Bigg\}\\
    =&2\pi\lambda_bK_s^2\Omega^2P_s^2\Bigg\{\frac{(\pi\lambda_b)^{\alpha_s-1}(m+1)}{2m(1-\alpha_s)}\Gamma(2-\alpha_s,R_l^2\pi\lambda_b,R_m^2\pi\lambda_b)\\
    &+\Big[-\frac{R_l^{2-2\alpha_s}(m+1)}{2m(1-\alpha_s)}+\frac{2\pi\lambda_bR_l^{4-2\alpha_s}}{(2-\alpha_s)^2})(e^{-R_l^2\pi\lambda_b}-e^{-R_m^2\pi\lambda_b})\\
    &+\frac{2(\pi\lambda_b)^{\alpha_s-1}}{(2-\alpha_s)^2}\Big[(2-\alpha_s)\Gamma(2-\alpha_s,R_l^2\pi\lambda_b,R_m^2\pi\lambda_b)+(R_l^2\pi\lambda_b)^{2-\alpha_s}e^{-R_l^2\pi\lambda_b}\\
    &-(R_m^2\pi\lambda_b)^{2-\alpha_s}e^{-R_m^2\pi\lambda_b}\Big]-\frac{4R_l^{2-\alpha_s}(\pi\lambda_b)^{\alpha_s/2}}{(2-\alpha_s)^2}\Gamma(2-\alpha_s/2,R_l^2\pi\lambda_b,R_m^2\pi\lambda_b)\Bigg\}\\
    =&2\pi\lambda_bK_s^2\Omega^2P_s^2\Bigg\{\Big[\frac{(\pi\lambda_b)^{\alpha_s-1}(m+1)}{2m(1-\alpha_s)}+\frac{2(\pi\lambda_b)^{\alpha_s-1}}{2-\alpha_s}\Big]\Gamma(2-\alpha_s,R_l^2\pi\lambda_b,R_m^2\pi\lambda_b)\\
    &+\Big[\frac{4\pi\lambda_bR_l^{4-2\alpha_s}}{(2-\alpha_s)^2}-\frac{R_l^{2-2\alpha_s}(m+1)}{2m(1-\alpha_s)}\Big](e^{-R_l^2\pi\lambda_b}-e^{-R_m^2\pi\lambda_b})\\
    &-\frac{4R_l^{2-\alpha_s}(\pi\lambda_b)^{\alpha_s/2}}{(2-\alpha_s)^2}\Gamma(2-\alpha_s/2,R_l^2\pi\lambda_b,R_m^2\pi\lambda_b)\Bigg\}.\\
    \end{split}
    \end{equation}

    The variance of the desired power is
    \begin{equation}
    \begin{split}
    \mathbb{V}[P_1]=&\mathbb{E}[P_1^2]-\mathbb{E}[P_1]^2\\
    =&2\pi\lambda_bK_s^2\Omega^2P_s^2\bigg\{\Big[\frac{(\pi\lambda_b)^{\alpha_s-1}(m+1)}{2m(1-\alpha_s)}+\frac{2(\pi\lambda_b)^{\alpha_s-1}}{2-\alpha_s}\Big]\Gamma(2-\alpha_s,R_l^2\pi\lambda_b,R_m^2\pi\lambda_b)\\
    &+\Big[\frac{4\pi\lambda_bR_l^{4-2\alpha_s}}{(2-\alpha_s)^2}-\frac{R_l^{2-2\alpha_s}(m+1)}{2m(1-\alpha_s)}\Big](e^{-R_l^2\pi\lambda_b}-e^{-R_m^2\pi\lambda_b})\\
    &-\frac{4R_l^{2-\alpha_s}(\pi\lambda_b)^{\alpha_s/2}}{(2-\alpha_s)^2}\Gamma(2-\alpha_s/2,R_l^2\pi\lambda_b,R_m^2\pi\lambda_b)\bigg\}-\bigg\{\frac{2K_s\Omega P_s}{2-\alpha_s}\Big[(\pi\lambda_b)^{\alpha_s/2}\\
    &\Gamma(\frac{4-\alpha_s}{2},\pi\lambda_bR_l^2,\pi\lambda_bR_m^2)+\pi\lambda_bR_l^{-\alpha_s+2}(e^{-\pi\lambda_bR_m^2}-e^{-\pi\lambda_bR_l^2})\Big]\bigg\}^2.
    \end{split}
    \end{equation}
Similarly, we can obtain the variance of the interfering power. For example, if $h$ is a constant, the expectation of the second moment of interference is
    \begin{equation}
    \begin{split}
    \mathbb{E}[I_1^2]=&\mathcal{M}''_{I_1}(0)\\
    =&K_s^2h^2P_s^2\int_{R_l}^{R_m}\bigg[\frac{2\pi^2\lambda_b^2r}{1-\alpha_s}(Rm^{-2\alpha_s+2}-r^{-2\alpha_s+2})+\frac{8\pi^3\lambda_b^3r}{(2-\alpha_s)^2}(r^{2-\alpha_s}-R_m^{2-\alpha_s})^2\bigg]\\
    &\times\exp(-\pi\lambda_br^2)dr\\
    =&2\pi\lambda_bK_s^2h^2P_s^2\Bigg\{\Big[-\frac{(\pi\lambda_b)^{\alpha_s-1}}{2(1-\alpha_s)}+\frac{2(\pi\lambda_b)^{\alpha_s-1}}{2-\alpha_s}\Big]\Gamma(2-\alpha_s,R_l^2\pi\lambda_b,R_m^2\pi\lambda_b)\\
    &+\Big[\frac{4\pi\lambda_bR_m^{4-2\alpha_s}}{(2-\alpha_s)^2}+\frac{R_m^{2-2\alpha_s}}{2(1-\alpha_s)}\Big](e^{-R_l^2\pi\lambda_b}-e^{-R_m^2\pi\lambda_b})\\
    &-\frac{4R_m^{2-\alpha_s}(\pi\lambda_b)^{\alpha_s/2}}{(2-\alpha_s)^2}\Gamma(2-\alpha_s/2,R_l^2\pi\lambda_b,R_m^2\pi\lambda_b)\Bigg\}.\\
    \end{split}
    \end{equation}
Thus, the variance is
    \begin{equation}
    \begin{split}
    \mathbb{V}[I_1]
        =&2\pi\lambda_bK_s^2h^2P_s^2\bigg\{\Big[-\frac{(\pi\lambda_b)^{\alpha_s-1}}{2(1-\alpha_s)}+\frac{2(\pi\lambda_b)^{\alpha_s-1}}{2-\alpha_s}\Big]\Gamma(2-\alpha_s,R_l^2\pi\lambda_b,R_m^2\pi\lambda_b)\\
    &+\Big[\frac{4\pi\lambda_bR_m^{4-2\alpha_s}}{(2-\alpha_s)^2}+\frac{R_m^{2-2\alpha_s}}{2(1-\alpha_s)})(e^{-R_l^2\pi\lambda_b}-e^{-R_m^2\pi\lambda_b})-\frac{4R_m^{2-\alpha_s}(\pi\lambda_b)^{\alpha_s/2}}{(2-\alpha_s)^2}\\
    &\times\Gamma(2-\alpha_s/2,R_l^2\pi\lambda_b,R_m^2\pi\lambda_b)\bigg\}-\frac{4K_s^2 h^2P_s^2}{(2-\alpha_s)^2}\bigg\{\pi\lambda_bR_m^{-\alpha_s+2}(e^{-\pi\lambda_bR_l^2}-e^{-\pi\lambda_bR_m^2})\\
&+(\pi\lambda_b)^{\alpha_s/2}\Big[\gamma(\frac{4-\alpha_s}{2},\pi\lambda_bR_l^2)-\gamma(\frac{4-\alpha_s}{2},\pi\lambda_bR_m^2)\Big]\bigg\}^2.\\
    \end{split}
    \end{equation}
    In a Rayleigh fading channel, the expectation of the second moment of interference is
     \begin{equation}
    \begin{split}
    \mathbb{E}[I_1^2]=&\mathcal{M}''_{I_1}(0)\\
    =&K_s^2P_s^2\int_{R_l}^{R_m}\bigg[\frac{4\pi^2\lambda_b^2r}{1-\alpha_s}(r^{-2\alpha_s+2}-R_l^{-2\alpha_s+2})+\frac{8\pi^3\lambda_b^3r}{(2-\alpha_s)^2}(r^{2-\alpha_s}-R_l^{2-\alpha_s})^2\bigg]\\
    &\times\exp(-\pi\lambda_br^2)dr\\
    =&2\pi\lambda_bK_s^2P_s^2\Bigg\{\Big[-\frac{(\pi\lambda_b)^{\alpha_s-1}}{(1-\alpha_s)}+\frac{2(\pi\lambda_b)^{\alpha_s-1}}{2-\alpha_s}\Big]\Gamma(2-\alpha_s,R_l^2\pi\lambda_b,R_m^2\pi\lambda_b)\\
    &+\Big[\frac{4\pi\lambda_bR_m^{4-2\alpha_s}}{(2-\alpha_s)^2}-\frac{R_m^{2-2\alpha_s}}{(1-\alpha_s)}\Big](e^{-R_l^2\pi\lambda_b}-e^{-R_m^2\pi\lambda_b})\\
    &-\frac{4R_m^{2-\alpha_s}(\pi\lambda_b)^{\alpha_s/2}}{(2-\alpha_s)^2}\Gamma(2-\alpha_s/2,R_l^2\pi\lambda_b,R_m^2\pi\lambda_b)\Bigg\}.\\
    \end{split}
    \end{equation}

    The variance of the interfering power is
    \begin{equation}
    \begin{split}
    \mathbb{V}[I_1]=&\mathbb{E}[I_1^2]-\mathbb{E}[I_1]^2\\
    =&2\pi\lambda_bK_s^2P_s^2\Bigg\{\Big[-\frac{(\pi\lambda_b)^{\alpha_s-1}}{(1-\alpha_s)}+\frac{2(\pi\lambda_b)^{\alpha_s-1}}{2-\alpha_s}\Big]\Gamma(2-\alpha_s,R_l^2\pi\lambda_b,R_m^2\pi\lambda_b)\\
    &+\Big[\frac{4\pi\lambda_bR_m^{4-2\alpha_s}}{(2-\alpha_s)^2}-\frac{R_m^{2-2\alpha_s}}{(1-\alpha_s)}\Big](e^{-R_l^2\pi\lambda_b}-e^{-R_m^2\pi\lambda_b})\\
    &-\frac{4R_m^{2-\alpha_s}(\pi\lambda_b)^{\alpha_s/2}}{(2-\alpha_s)^2}\Gamma(2-\alpha_s/2,R_l^2\pi\lambda_b,R_m^2\pi\lambda_b)\Bigg\}-\bigg\{\frac{2K_s P_s}{2-\alpha_s}\Big[(\pi\lambda_b)^{\alpha_s/2}\\
    &\Gamma(\frac{4-\alpha_s}{2},\pi\lambda_bR_l^2,\pi\lambda_bR_m^2)+\pi\lambda_bR_m^{-\alpha_s+2}(e^{-\pi\lambda_bR_m^2}-e^{-\pi\lambda_bR_l^2})\Big]\bigg\}^2.
    \end{split}
    \end{equation}
    In a Nakagami-m fading channel, the expectation of the second moment of interference is
\begin{equation}
    \begin{split}
    \mathbb{E}[I_1^2]=&\mathcal{M}''_{I_1}(0)\\
    =&K_s^2\Omega^2P_s^2\int_{R_l}^{R_m}\bigg[\frac{2\pi^2\lambda_b^2(m+1)r}{m(1-\alpha_s)}(R_m^{-2\alpha_s+2}-r^{-2\alpha_s+2})+\frac{8\pi^3\lambda_b^3r}{(2-\alpha_s)^2}(R_m^{2-\alpha_s}\\
    &-r^{2-\alpha_s})^2\bigg]\exp(-\pi\lambda_br^2)dr\\
    =&2\pi\lambda_bK_s^2\Omega^2P_s^2\Bigg\{\Big[-\frac{(\pi\lambda_b)^{\alpha_s-1}(m+1)}{2m(1-\alpha_s)}+\frac{2(\pi\lambda_b)^{\alpha_s-1}}{2-\alpha_s}\Big]\Gamma(2-\alpha_s,R_l^2\pi\lambda_b,R_m^2\pi\lambda_b)\\
    &+\Big[\frac{4\pi\lambda_bR_m^{4-2\alpha_s}}{(2-\alpha_s)^2}-\frac{R_m^{2-2\alpha_s}(m+1)}{2m(1-\alpha_s)}\Big](e^{-R_l^2\pi\lambda_b}-e^{-R_m^2\pi\lambda_b})\\
    &-\frac{4R_m^{2-\alpha_s}(\pi\lambda_b)^{\alpha_s/2}}{(2-\alpha_s)^2}\Gamma(2-\alpha_s/2,R_l^2\pi\lambda_b,R_m^2\pi\lambda_b)\Bigg\}.\\
    \end{split}
    \end{equation}

    The variance of the interfering power is
    \begin{equation}
    \begin{split}
    \mathbb{V}[I_1]=&\mathbb{E}[I_1^2]-\mathbb{E}[I_1]^2\\
    =&2\pi\lambda_bK_s^2\Omega^2P_s^2\Bigg\{\Big[-\frac{(\pi\lambda_b)^{\alpha_s-1}(m+1)}{2m(1-\alpha_s)}+\frac{2(\pi\lambda_b)^{\alpha_s-1}}{2-\alpha_s}\Big]\Gamma(2-\alpha_s,R_l^2\pi\lambda_b,R_m^2\pi\lambda_b)\\
    &+\Big[\frac{4\pi\lambda_bR_m^{4-2\alpha_s}}{(2-\alpha_s)^2}-\frac{R_m^{2-2\alpha_s}(m+1)}{2m(1-\alpha_s)}\Big](e^{-R_l^2\pi\lambda_b}-e^{-R_m^2\pi\lambda_b})\\
    &-\frac{4R_m^{2-\alpha_s}(\pi\lambda_b)^{\alpha_s/2}}{(2-\alpha_s)^2}\Gamma(2-\alpha_s/2,R_l^2\pi\lambda_b,R_m^2\pi\lambda_b)\Bigg\}-\bigg\{\frac{2K_s\Omega P_s}{2-\alpha_s}\Big[(\pi\lambda_b)^{\alpha_s/2}\\
    &\Gamma(\frac{4-\alpha_s}{2},\pi\lambda_bR_l^2,\pi\lambda_bR_m^2)+\pi\lambda_bR_m^{-\alpha_s+2}(e^{-\pi\lambda_bR_m^2}-e^{-\pi\lambda_bR_l^2})\Big]\bigg\}^2.
    \end{split}
    \end{equation}
\item \textbf{2NS:} if $h$ is a constant, the first and second derivatives of MGF of desired power are
    \begin{equation}
    \begin{split}
    \mathcal{M}'_{P_2}(s)    =&\int_{R_l}^{R_m}2\pi\lambda_b\int_{R_l}^{r}K_shP_sx^{-\alpha_s+1}\exp(sK_shP_sx^{-\alpha_s})dx\\
    &\times\exp\bigg\{-2\pi\lambda_b\int_{R_l}^{r}\Big[1-\exp(sK_shP_sx^{-\alpha_s})\Big]xdx\bigg\}f_{2}(r)dr,\\
    \end{split}
    \end{equation}
     \begin{equation}
    \begin{split}
    \mathcal{M}''_{P_2}(s)    =&2\pi\lambda_bK_shP_s\int_{R_l}^{R_m}\Big\{\int_{R_l}^rx^{-2\alpha_s+1}K_shP_sa(s)dxb(s)\\
    &+2\pi\lambda_bK_shP_s\Big[\int_{R_l}^rx^{-\alpha_s+1}a(s)dx\Big]^2b(s)\Big\}f_2(r)dr,\\
    \end{split}
    \end{equation}
    where $a(s)=\exp(sK_shP_sx^{-\alpha_s})$ and $b(s)=\exp\{-2\pi\lambda_b\int_{R_l}^{r}[1-\exp(sK_shP_sx^{-\alpha_s})]xdx\}$.
    Thus, the expectation of the second moment of received power is
    \begin{equation}
    \begin{split}
    \mathbb{E}[P_2^2]=&\mathcal{M}''_{P_2}(0)\\
    =&2\pi\lambda_bK_s^2h^2P_s^2\int_{R_l}^{R_m}\bigg[\int_{R_l}^rx^{-2\alpha_s+1}dx+2\pi\lambda_b\Big(\int_{R_l}^rx^{-\alpha_s+1}dx\Big)^2\bigg]f_2(r)dr\\
    =&K_s^2h^2P_s^2\int_{R_l}^{R_m}\bigg[\frac{2\pi^3\lambda_b^3r^3}{1-\alpha_s}(r^{-2\alpha_s+2}-R_l^{-2\alpha_s+2})\\
    &+\frac{8\pi^4\lambda_b^4r^3}{(2-\alpha_s)^2}(r^{2-\alpha_s}-R_l^{2-\alpha_s})^2\bigg]\exp(-\pi\lambda_br^2)dr.
    \end{split}
    \end{equation}
    The variance of the desired power is
    \begin{equation}
    \begin{split}
    \mathbb{V}[P_2]=&\mathbb{E}[P_2^2]-\mathbb{E}[P_2]^2\\
    =&K_s^2h^2P_s^2\int_{R_l}^{R_m}\bigg[\frac{\pi\lambda_b}{1-\alpha_s}(r^{-2\alpha_s+2}-R_l^{-2\alpha_s+2})+\frac{4\pi^2\lambda_b^2}{(2-\alpha_s)^2}(r^{2-\alpha_s}-R_l^{2-\alpha_s})^2\bigg]f_2(r)dr\\
    &-\frac{4K_s^2 h^2 P_s^2}{(2-\alpha_s)^2}\Bigg\{(\pi\lambda_b)^{\alpha_s/2}\Gamma(\frac{6-\alpha_s}{2},\pi\lambda_bR_l^2,\pi\lambda_bR_m^2)+\pi\lambda_bR_l^{-\alpha_s+2}\\
    &\times\Big[(\pi\lambda_bR_m^2+1)e^{-\pi\lambda_bR_m^2}-(\pi\lambda_bR_l^2+1)e^{-\pi\lambda_bR_l^2}\Big]\Bigg\}^2.\\
    \end{split}
    \end{equation}
    In a Rayleigh fading channel, the expectation of the second moment of received power is
    \begin{equation}
    \begin{split}
    \mathbb{E}[P_2^2]=&\mathcal{M}''_{P_2}(0)\\
    =&K_s^2P_s^2\int_{R_l}^{R_m}\bigg[\frac{2\pi\lambda_b}{1-\alpha_s}(r^{-2\alpha_s+2}-R_l^{-2\alpha_s+2})+\frac{4\pi^2\lambda_b^2}{(2-\alpha_s)^2}(r^{2-\alpha_s}-R_l^{2-\alpha_s})^2\bigg]f_2(r)dr\\
    =&K_s^2P_s^2\int_{R_l}^{R_m}\bigg[\frac{4\pi^3\lambda_b^3r^3}{1-\alpha_s}(r^{-2\alpha_s+2}-R_l^{-2\alpha_s+2})+\frac{8\pi^4\lambda_b^4r^3}{(2-\alpha_s)^2}(r^{2-\alpha_s}\\
    &-R_l^{2-\alpha_s})^2\bigg]\exp(-\pi\lambda_br^2)dr.\\
    \end{split}
    \end{equation}
    The variance of the desired power is
    \begin{equation}
    \begin{split}
    \mathbb{V}[P_2]=&K_s^2P_s^2\int_{R_l}^{R_m}\bigg[\frac{4\pi^3\lambda_b^3r^3}{1-\alpha_s}(r^{-2\alpha_s+2}-R_l^{-2\alpha_s+2})+\frac{8\pi^4\lambda_b^4r^3}{(2-\alpha_s)^2}(r^{2-\alpha_s}-R_l^{2-\alpha_s})^2\bigg]\\
    &\times\exp(-\pi\lambda_br^2)dr-\frac{4K_s^2 P_s^2}{(2-\alpha_s)^2}\Bigg\{(\pi\lambda_b)^{\alpha_s/2}\Gamma(\frac{6-\alpha_s}{2},\pi\lambda_bR_l^2,\pi\lambda_bR_m^2)\\
    &+\pi\lambda_bR_l^{-\alpha_s+2}\Big[(\pi\lambda_bR_m^2+1)e^{-\pi\lambda_bR_m^2}-(\pi\lambda_bR_l^2+1)e^{-\pi\lambda_bR_l^2}\Big]\Bigg\}^2.\\
    \end{split}
    \end{equation}
    In a Nakagami-m fading channel, the expectation of the second moment of the desired power is
    \begin{equation}
    \begin{split}
    \mathbb{E}[P_2^2]=&\mathcal{M}''_{P_2}(0)\\
    =&K_s^2\Omega^2P_s^2\int_{R_l}^{R_m}\bigg[\frac{2\pi^3\lambda_b^3(m+1)r^3}{m(1-\alpha_s)}(r^{-2\alpha_s+2}-R_l^{-2\alpha_s+2})+\frac{8\pi^4\lambda_b^4r^3}{(2-\alpha_s)^2}(r^{2-\alpha_s}\\
    &-R_l^{2-\alpha_s})^2\bigg]\exp(-\pi\lambda_br^2)dr.\\
    \end{split}
    \end{equation}
    The variance of the desired power is
    \begin{equation}
    \begin{split}
    \mathbb{V}[P_2]=&K_s^2\Omega^2P_s^2\int_{R_l}^{R_m}\bigg[\frac{2\pi^3\lambda_b^3(m+1)r^3}{m(1-\alpha_s)}(r^{-2\alpha_s+2}-R_l^{-2\alpha_s+2})+\frac{8\pi^4\lambda_b^4r^3}{(2-\alpha_s)^2}(r^{2-\alpha_s}\\
    &-R_l^{2-\alpha_s})^2\bigg]\exp(-\pi\lambda_br^2)dr-\frac{4K_s^2 \Omega^2P_s^2}{(2-\alpha_s)^2}\Bigg\{(\pi\lambda_b)^{\alpha_s/2}\Gamma(\frac{6-\alpha_s}{2},\pi\lambda_bR_l^2,\pi\lambda_bR_m^2)\\
    &+\pi\lambda_bR_l^{-\alpha_s+2}\Big[(\pi\lambda_bR_m^2+1)e^{-\pi\lambda_bR_m^2}-(\pi\lambda_bR_l^2+1)e^{-\pi\lambda_bR_l^2}\Big]\Bigg\}^2.\\
    \end{split}
    \end{equation}
    Similarly, we can obtain the variance of the interfering power. For example, if $h$ is a constant, the variance of interfering power is
    \begin{equation}
    \begin{split}
    \mathbb{V}[I_2]
    =&K_s^2h^2P_s^2\int_{R_l}^{R_m}\bigg[\frac{2\pi^3\lambda_b^3r^3}{1-\alpha_s}(R_m^{-2\alpha_s+2}-r^{-2\alpha_s+2})+\frac{8\pi^4\lambda_b^4r^3}{(2-\alpha_s)^2}(R_m^{2-\alpha_s}-r^{2-\alpha_s})^2\bigg]\\
    &\times \exp(-\pi\lambda_br^2)dr-\frac{4K_s^2 h^2P_s^2}{(2-\alpha_s)^2}\bigg\{(\pi\lambda_b)^{\alpha_s/2}\Gamma\Big(\frac{6-\alpha_s}{2},\pi\lambda_bR_l^2,\pi\lambda_bR_m^2\Big)\\
    &-\pi\lambda_bR_m^{-\alpha_s+2}\Big[(\pi\lambda_bR_l^2+1)e^{-\pi\lambda_bR_l^2}-(\pi\lambda_bR_m^2+1)e^{-\pi\lambda_bR_m^2}\Big]\bigg\}^2.\\
    \end{split}
    \end{equation}
    In a Rayleigh fading channel, the variance of interfering power is
    \begin{equation}
    \begin{split}
    \mathbb{V}[I_2]
    =&K_s^2P_s^2\int_{R_l}^{R_m}\bigg[\frac{4\pi^3\lambda_b^3r^3}{1-\alpha_s}(R_m^{-2\alpha_s+2}-r^{-2\alpha_s+2})+\frac{8\pi^4\lambda_b^4r^3}{(2-\alpha_s)^2}(R_m^{2-\alpha_s}-r^{2-\alpha_s})^2\bigg]\\
    &\times \exp(-\pi\lambda_br^2)dr-\frac{4K_s^2 P_s^2}{(2-\alpha_s)^2}\bigg\{(\pi\lambda_b)^{\alpha_s/2}\Gamma\Big(\frac{6-\alpha_s}{2},\pi\lambda_bR_l^2,\pi\lambda_bR_m^2\Big)\\
    &-\pi\lambda_bR_m^{-\alpha_s+2}\Big[(\pi\lambda_bR_l^2+1)e^{-\pi\lambda_bR_l^2}-(\pi\lambda_bR_m^2+1)e^{-\pi\lambda_bR_m^2}\Big]\bigg\}^2.\\
    \end{split}
    \end{equation}
    In a Nakagami-m fading channel, the variance of interfering power is
    \begin{equation}
    \begin{split}
    \mathbb{V}[I_2]
    =&K_s^2\Omega^2P_s^2\int_{R_l}^{R_m}\bigg[\frac{2\pi^3(m+1)\lambda_b^3r^3}{m(1-\alpha_s)}(R_m^{-2\alpha_s+2}-r^{-2\alpha_s+2})+\frac{8\pi^4\lambda_b^4r^3}{(2-\alpha_s)^2}(R_m^{2-\alpha_s}\\
    &-r^{2-\alpha_s})^2\bigg]\exp(-\pi\lambda_br^2)dr-\frac{4K_s^2\Omega^2 P_s^2}{(2-\alpha_s)^2}\bigg\{(\pi\lambda_b)^{\alpha_s/2}\Gamma\Big(\frac{6-\alpha_s}{2},\pi\lambda_bR_l^2,\pi\lambda_bR_m^2\Big)\\
    &-\pi\lambda_bR_m^{-\alpha_s+2}\Big[(\pi\lambda_bR_l^2+1)e^{-\pi\lambda_bR_l^2}-(\pi\lambda_bR_m^2+1)e^{-\pi\lambda_bR_m^2}\Big]\bigg\}^2.\\
    \end{split}
    \end{equation}
 \item \textbf{CD:} if $h$ is a constant, the first and second derivatives of MGF of desired power are
     \begin{equation}
    \begin{split}
    \mathcal{M}'_{P_{R_0}}(s)    =&2\pi\lambda_b\int_{R_l}^{R_0}K_shP_sx^{-\alpha_s+1}\exp(sK_shP_sx^{-\alpha_s})dx\\
    &\times\exp\Big\{-2\pi\lambda_b\int_{R_l}^{R_0}[1-\exp(sK_shP_sx^{-\alpha_s})]xdx\Big\},\\
    \end{split}
    \end{equation}
     \begin{equation}
    \begin{split}
    \mathcal{M}''_{P_{R_0}}(s)    =&2\pi\lambda_bK_shP_s\bigg\{\int_{R_l}^{R_0}x^{-2\alpha_s+1}K_shP_sa(s)dxb(s)\\
    &+2\pi\lambda_bK_shP_s\Big[\int_{R_l}^{R_0}x^{-\alpha_s+1}a(s)dx\Big]^2b(s)\bigg\},\\
    \end{split}
    \end{equation}
    where $a(s)=\exp(sK_shP_sx^{-\alpha_s})$ and $b(s)=\exp\{-2\pi\lambda_b\int_{R_l}^{r}[1-\exp(sK_shP_sx^{-\alpha_s})]xdx\}$.
    Thus, the expectation of the second moment of received power is
    \begin{equation}
    \begin{split}
    \mathbb{E}[P_{R_0}^2]=&\mathcal{M}''_{P_{R_0}}(0)\\
      =&K_s^2h^2P_s^2\bigg[\frac{\pi\lambda_b}{1-\alpha_s}(R_0^{-2\alpha_s+2}-R_l^{-2\alpha_s+2})+\frac{4\pi^2\lambda_b^2}{(2-\alpha_s)^2}(R_0^{2-\alpha_s}-R_l^{2-\alpha_s})^2\bigg].
    \end{split}
    \end{equation}
    The variance for desired power is
    \begin{equation}
    \begin{split}
    \mathbb{V}[P_{R_0}]    =&K_s^2h^2P_s^2\bigg[\frac{\pi\lambda_b}{1-\alpha_s}(R_0^{-2\alpha_s+2}-R_l^{-2\alpha_s+2})+\frac{4\pi^2\lambda_b^2}{(2-\alpha_s)^2}(R_0^{2-\alpha_s}-R_l^{2-\alpha_s})^2\bigg]\\
    &-\frac{4\pi^2\lambda_b^2 K_s^2 h^2 P_s^2}{(2-\alpha_s)^2}(R_0^{-\alpha_s+2}-R_l^{-\alpha_s+2})^2\\
    =&\frac{\pi\lambda_b K_s^2h^2P_s^2}{1-\alpha_s}(R_0^{-2\alpha_s+2}-R_l^{-2\alpha_s+2}).\\
    \end{split}
    \end{equation}
    In a Rayleigh fading channel, the first and second moments of the MGF of desired power are
    \begin{equation}
    \begin{split}
    \mathcal{M}'_{P_{R_0}}(s)    =&2\pi\lambda_b\int_{R_l}^{R_0}K_sP_sx^{-\alpha_s+1}(1-sK_sP_sx^{-\alpha_s})^{-2}dx\\
    &\times\exp(-2\pi\lambda_b\int_{R_l}^{R_0}(1-(1-sK_sP_sx^{-\alpha_s})^{-1})xdx),\\
    \end{split}
    \end{equation}
     \begin{equation}
    \begin{split}
    \mathcal{M}''_{P_{R_0}}(s)    =&2\pi\lambda_bK_sP_s\Big[\int_{R_l}^{R_0}x^{-2\alpha_s+1}2K_sP_sa(s)^{-3}dxb(s)\\
    &+2\pi\lambda_bK_sP_s\Big(\int_{R_l}^{R_0}x^{-\alpha_s+1}a(s)^{-2}dx\Big)^2b(s)\Big],\\
    \end{split}
    \end{equation}
    where $a(s)=(1-sK_sP_sx^{-\alpha_s})$ and $b(s)=\exp(-2\pi\lambda_b\int_{R_l}^{r}(1-(1-sK_sP_sx^{-\alpha_s})^{-1})xdx)$
    Thus, the expectation of the second moment of received power is
    \begin{equation}
    \begin{split}
    \mathbb{E}[P_{R_0}^2]=&\mathcal{M}''_{P_{R_0}}(0)\\
      =&K_s^2P_s^2\bigg[\frac{2\pi\lambda_b}{1-\alpha_s}(R_0^{-2\alpha_s+2}-R_l^{-2\alpha_s+2})+\frac{4\pi^2\lambda_b^2}{(2-\alpha_s)^2}(R_0^{2-\alpha_s}-R_l^{2-\alpha_s})^2\bigg].
    \end{split}
    \end{equation}
    The variance of the desired power is
    \begin{equation}
    \begin{split}
    \mathbb{V}[P_{R_0}]    =&K_s^2P_s^2\bigg[2\frac{\pi\lambda_b}{1-\alpha_s}(R_0^{-2\alpha_s+2}-R_l^{-2\alpha_s+2})+\frac{4\pi^2\lambda_b^2}{(2-\alpha_s)^2}(R_0^{2-\alpha_s}-R_l^{2-\alpha_s})^2\bigg]\\
    &-\frac{4\pi^2\lambda_b^2 K_s^2 P_s^2}{(2-\alpha_s)^2}(R_0^{-\alpha_s+2}-R_l^{-\alpha_s+2})^2\\
    =&\frac{2\pi\lambda_b K_s^2P_s^2}{1-\alpha_s}(R_0^{-2\alpha_s+2}-R_l^{-2\alpha_s+2}).\\
    \end{split}
    \end{equation}

    In a Nakagami-m fading channel, the result is
    \begin{equation}
    \begin{split}
    \mathbb{V}[P_{R_0}]
     =&K_s^2\Omega^2P_s^2\bigg[\frac{\pi\lambda_b(m+1)}{m(1-\alpha_s)}(R_0^{-2\alpha_s+2}-R_l^{-2\alpha_s+2})+\frac{4\pi^2\lambda_b^2}{(2-\alpha_s)^2}(R_0^{2-\alpha_s}-R_l^{2-\alpha_s})^2\bigg]\\
    &-\frac{4\pi^2\lambda_b^2 K_s^2\Omega^2 P_s^2}{(2-\alpha_s)^2}(R_0^{-\alpha_s+2}-R_l^{-\alpha_s+2})^2\\
    =&\frac{(m+1)\pi\lambda_b K_s^2\Omega^2P_s^2}{m(1-\alpha_s)}(R_0^{-2\alpha_s+2}-R_l^{-2\alpha_s+2}).\\
    \end{split}
    \end{equation}

    Similarly, we can obtain the variance of interfering power. For example, if $h$ is a constant, the variance is
    \begin{equation}
    \begin{split}
    \mathbb{V}[I_{R_0}]=&\frac{\pi\lambda_b K_s^2h^2P_s^2}{1-\alpha_s}(R_m^{-2\alpha_s+2}-R_0^{-2\alpha_s+2}).\\
    \end{split}
    \end{equation}
    In a Rayleigh fading channel, the variance is
    \begin{equation}
    \begin{split}
    \mathbb{V}[I_{R_0}]=&\frac{2\pi\lambda_b K_s^2P_s^2}{1-\alpha_s}(R_m^{-2\alpha_s+2}-R_0^{-2\alpha_s+2}).\\
    \end{split}
    \end{equation}

    In a Nakagami-m fading channel, the variance is
    \begin{equation}
    \begin{split}
    \mathbb{V}[I_{R_0}]=&\frac{(m+1)\pi\lambda_b K_s^2\Omega^2P_s^2}{m(1-\alpha_s)}(R_m^{-2\alpha_s+2}-R_0^{-2\alpha_s+2}).\\
    \end{split}
    \end{equation}

\item \textbf{FPD:} if $h$ is a constant, the first and second derivatives of MGF of the desired power are
    \begin{equation}
    \begin{split}
    \mathcal{M}'_{P_\eta}(s)    =&\int_{R_l}^{R_m}2\pi\lambda_b\int_{R_l}^{r}K_shP_sx^{-\alpha_s+1}\exp(sK_shP_sx^{-\alpha_s})dx\\
    &\times\exp\bigg\{-2\pi\lambda_b\int_{R_l}^{r}\Big[1-\exp(sK_shP_sx^{-\alpha_s})\Big]xdx\bigg\}f_{\eta}(r)dr,\\
    \end{split}
    \end{equation}
     \begin{equation}
    \begin{split}
    \mathcal{M}''_{P_\eta}(s)    =&2\pi\lambda_bK_shP_s\int_{R_l}^{R_m}\Big\{\int_{R_l}^rx^{-2\alpha_s+1}K_shP_sa(s)dxb(s)\\
    &+2\pi\lambda_bK_shP_s\Big[\int_{R_l}^rx^{-\alpha_s+1}a(s)dx\Big]^2b(s)\Big\}f_{\eta}(r)dr,\\
    \end{split}
    \end{equation}
    where $a(s)=\exp(sK_shP_sx^{-\alpha_s})$ and $b(s)=\exp\{-2\pi\lambda_b\int_{R_l}^{r}[1-\exp(sK_shP_sx^{-\alpha_s})]xdx\}$.
    Thus, the expectation of the second moment of received power is
    \begin{equation}
    \begin{split}
    \mathbb{E}[P_\eta^2]=&\mathcal{M}''_{P_\eta}(0)\\
    =&2\pi\lambda_bK_s^2h^2P_s^2\int_{R_l}^{R_m}\bigg[\int_{R_l}^rx^{-2\alpha_s+1}dx+2\pi\lambda_b\Big(\int_{R_l}^rx^{-\alpha_s+1}dx\Big)^2\bigg]f_2(r)dr\\
    =&K_s^2h^2\eta_t^2P_s^2\int_{R_l}^{R_m}\bigg[\frac{2\pi^3\lambda_b^3r^3}{1-\alpha_s}(r^{-2\alpha_s+2}-R_l^{-2\alpha_s+2})\\
    &+\frac{8\pi^4\lambda_b^4r^3}{(2-\alpha_s)^2}(r^{2-\alpha_s}-R_l^{2-\alpha_s})^2\bigg]\exp(-\pi\lambda_b\eta_t^2r^2)dr.
    \end{split}
    \end{equation}
    The variance of the desired power is
    \begin{equation}
    \begin{split}
    \mathbb{V}[P_\eta]=&\mathbb{E}[P_\eta^2]-\mathbb{E}[P_\eta]^2\\
    =&K_s^2h^2\eta_t^2P_s^2\int_{R_l}^{R_m}\bigg[\frac{2\pi^3\lambda_b^3r^3}{1-\alpha_s}(r^{-2\alpha_s+2}-R_l^{-2\alpha_s+2})+\frac{8\pi^4\lambda_b^4r^3}{(2-\alpha_s)^2}(r^{2-\alpha_s}-R_l^{2-\alpha_s})^2\bigg]\\
    &\times\exp(-\pi\lambda_b\eta_t^2r^2)dr-\frac{4K_s^2h^2P_s^2}{(2-\alpha_s)^2}\bigg[\frac{(\pi\lambda_b\eta_t^2)^{\frac{\alpha_s}{2}}}{\eta_t^{2}}\Gamma\Big(\frac{4-\alpha_s}{2},\pi\lambda_b\eta_t^2R_l^2,\pi\lambda_bR_m^2\Big)\\
    &+\pi\lambda_b\Big(\frac{R_l}{\eta_t}\Big)^{2-\alpha_s}\Big(e^{-\pi\lambda_b\eta_t^2R_m^2}-e^{-\pi\lambda_bR_l^2}\Big)\Bigg]^2.\\
    \end{split}
    \end{equation}

    In a Rayleigh fading channel, the variance of the desired power is
    \begin{equation}
    \begin{split}
    \mathbb{V}[P_\eta]=&\mathbb{E}[P_\eta^2]-\mathbb{E}[P_\eta]^2\\
    =&K_s^2\eta_t^2P_s^2\int_{R_l}^{R_m}\bigg[\frac{4\pi^3\lambda_b^3r^3}{1-\alpha_s}(r^{-2\alpha_s+2}-R_l^{-2\alpha_s+2})+\frac{8\pi^4\lambda_b^4r^3}{(2-\alpha_s)^2}(r^{2-\alpha_s}-R_l^{2-\alpha_s})^2\bigg]\\
    &\times\exp(-\pi\lambda_b\eta_t^2r^2)dr-\frac{4K_s^2P_s^2}{(2-\alpha_s)^2}\bigg[\frac{(\pi\lambda_b\eta_t^2)^{\frac{\alpha_s}{2}}}{\eta_t^{2}}\Gamma\Big(\frac{4-\alpha_s}{2},\pi\lambda_b\eta_t^2R_l^2,\pi\lambda_bR_m^2\Big)\\
    &+\pi\lambda_b\Big(\frac{R_l}{\eta_t}\Big)^{2-\alpha_s}\Big(e^{-\pi\lambda_b\eta_t^2R_m^2}-e^{-\pi\lambda_bR_l^2}\Big)\Bigg]^2.\\
    \end{split}
    \end{equation}

    In a Nakagami-m fading channel, the variance of the interfering power is
       \begin{equation}
    \begin{split}
    \mathbb{V}[P_\eta]=&\mathbb{E}[P_\eta^2]-\mathbb{E}[P_\eta]^2\\
    =&K_s^2\Omega^2\eta_t^2P_s^2\int_{R_l}^{R_m}\bigg[\frac{2\pi^3(m+1)\lambda_b^3r^3}{m(1-\alpha_s)}(r^{-2\alpha_s+2}-R_l^{-2\alpha_s+2})+\frac{8\pi^4\lambda_b^4r^3}{(2-\alpha_s)^2}(r^{2-\alpha_s}\\
    &-R_l^{2-\alpha_s})^2\bigg]\exp(-\pi\lambda_b\eta_t^2r^2)dr-\frac{4K_s^2\Omega^2P_s^2}{(2-\alpha_s)^2}\bigg[\frac{(\pi\lambda_b\eta_t^2)^{\frac{\alpha_s}{2}}}{\eta_t^{2}}\Gamma\Big(\frac{4-\alpha_s}{2},\pi\lambda_b\eta_t^2R_l^2,\\
    &\pi\lambda_bR_m^2\Big)+\pi\lambda_b\Big(\frac{R_l}{\eta_t}\Big)^{2-\alpha_s}\Big(e^{-\pi\lambda_b\eta_t^2R_m^2}-e^{-\pi\lambda_bR_l^2}\Big)\Bigg]^2.\\
    \end{split}
    \end{equation}
    Similarly, we can obtain the variance of the interfering power. For example, if $h$ is a constant, the variance of interfering power is
    \begin{equation}
    \begin{split}
    \mathbb{V}[I_\eta]
    =&K_s^2h^2\eta_t^2P_s^2\int_{R_l}^{R_m}\bigg[\frac{2\pi^3\lambda_b^3r^3}{1-\alpha_s}(R_m^{-2\alpha_s+2}-r^{-2\alpha_s+2})+\frac{8\pi^4\lambda_b^4r^3}{(2-\alpha_s)^2}(R_m^{2-\alpha_s}-r^{2-\alpha_s})^2\bigg]\\
    &\times\exp(-\pi\lambda_b\eta_t^2r^2)dr-\frac{4K_s^2h^2P_s^2}{(2-\alpha_s)^2}\bigg[\frac{(\pi\lambda_b\eta_t^2)^{\frac{\alpha_s}{2}}}{\eta_t^{2}}\Gamma\Big(\frac{4-\alpha_s}{2},\pi\lambda_bR_l^2,\pi\lambda_b\eta_t^2R_m^2\Big)\\
    &+\pi\lambda_b\Big(\frac{R_m}{\eta_t}\Big)^{2-\alpha_s}\Big(e^{-\pi\lambda_b\eta_t^2R_m^2}-e^{-\pi\lambda_bR_l^2}\Big)\Bigg]^2.\\
    \end{split}
    \end{equation}

    In a Rayleigh fading channel, the variance of interfering power is
    \begin{equation}
    \begin{split}
    \mathbb{V}[I_\eta]
    =&K_s^2\eta_t^2P_s^2\int_{R_l}^{R_m}\bigg[\frac{4\pi^3\lambda_b^3r^3}{1-\alpha_s}(R_m^{-2\alpha_s+2}-r^{-2\alpha_s+2})+\frac{8\pi^4\lambda_b^4r^3}{(2-\alpha_s)^2}(R_m^{2-\alpha_s}-r^{2-\alpha_s})^2\bigg]\\
    &\times\exp(-\pi\lambda_b\eta_t^2r^2)dr-\frac{4K_s^2P_s^2}{(2-\alpha_s)^2}\bigg[\frac{(\pi\lambda_b\eta_t^2)^{\frac{\alpha_s}{2}}}{\eta_t^{2}}\Gamma\Big(\frac{4-\alpha_s}{2},\pi\lambda_bR_l^2,\pi\lambda_b\eta_t^2R_m^2\Big)\\
    &+\pi\lambda_b\Big(\frac{R_m}{\eta_t}\Big)^{2-\alpha_s}\Big(e^{-\pi\lambda_b\eta_t^2R_m^2}-e^{-\pi\lambda_bR_l^2}\Big)\Bigg]^2.\\
    \end{split}
    \end{equation}

    In a Nakagami-m fading channel, the variance of interfering power is
    \begin{equation}
    \begin{split}
    \mathbb{V}[I_\eta]
    =&K_s^2\Omega^2\eta_t^2P_s^2\int_{R_l}^{R_m}\bigg[\frac{2\pi^3(m+1)\lambda_b^3r^3}{m(1-\alpha_s)}(R_m^{-2\alpha_s+2}-r^{-2\alpha_s+2})+\frac{8\pi^4\lambda_b^4r^3}{(2-\alpha_s)^2}(R_m^{2-\alpha_s}\\
    &-r^{2-\alpha_s})^2\bigg]\exp(-\pi\lambda_b\eta_t^2r^2)dr-\frac{4K_s^2\Omega^2P_s^2}{(2-\alpha_s)^2}\bigg[\frac{(\pi\lambda_b\eta_t^2)^{\frac{\alpha_s}{2}}}{\eta_t^{2}}\Gamma\Big(\frac{4-\alpha_s}{2},\pi\lambda_bR_l^2,\\
    &\pi\lambda_b\eta_t^2R_m^2\Big)+\pi\lambda_b\Big(\frac{R_m}{\eta_t}\Big)^{2-\alpha_s}\Big(e^{-\pi\lambda_b\eta_t^2R_m^2}-e^{-\pi\lambda_bR_l^2}\Big)\Bigg]^2.\\
    \end{split}
    \end{equation}

\end{itemize}
\balance

\vfill

\end{document}